%% file: clear.tex
\documentclass[a4paper,fleqn,english]{cas-sc}

\usepackage[square,numbers,compress,sort]{natbib}
\makeatletter
\renewcommand\@biblabel[1]{#1.}
\makeatother
\usepackage{csquotes}
\usepackage{xcolor}
\usepackage[colorinlistoftodos,prependcaption]{todonotes}
\usepackage[inline]{enumitem}
\usepackage{xspace}
\usepackage{multicol}
\usepackage[utf8]{inputenc}
\usepackage[T1]{fontenc}
\usepackage{wasysym}
\usepackage{float}
\usepackage{color}
\usepackage{acronym} 

\usepackage{subcaption}
\usepackage{longtable}
\usepackage{afterpage}
\usepackage{csquotes}
\usepackage{amssymb}%
\usepackage{rotating}
\usepackage{pifont}
\newcommand{\cmark}{\ding{51}}%
\newcommand{\xmark}{\text{\ding{55}}}
\usepackage[commandnameprefix=ifneeded]{changes}

\def\tsc#1{\csdef{#1}{\textsc{\lowercase{#1}}\xspace}}
\tsc{WGM}
\tsc{QE}
\tsc{EP}
\tsc{PMS}
\tsc{BEC}
\tsc{DE}

\usepackage{float}
\usepackage[usenames,dvipsnames]{pstricks}
\usepackage{pstricks-add}
 \usepackage{epsfig}
 \usepackage{pst-grad} 
 \usepackage{pst-plot} 
 \usepackage[space]{grffile} 
 \usepackage{etoolbox} 
 \makeatletter 
 \patchcmd\Gread@eps{\@inputcheck#1 }{\@inputcheck"#1"\relax}{}{}
 \makeatother
\usepackage{verbatim}
\usepackage{multicol}
\usepackage[ruled,lined,noresetcount ]{algorithm2e}
\usepackage{algpseudocode}
\usepackage{mathtools}

\begin{document}
\let\WriteBookmarks\relax
\def\floatpagepagefraction{1}
\def\textpagefraction{.001}

\shorttitle{Deep Learning Techniques for Hand Vein Biometrics: A Comprehensive Review}

\shortauthors{M. Hemis et~al.}
                   
\title [mode = title]{Deep Learning Techniques for Hand Vein Biometrics: A Comprehensive Review }

\vskip2mm

\author[1]{Mustapha Hemis\corref{cor1}}
\ead{mhemis@usthb.dz}
\cormark[1]
\credit{Conceptualization; Methodology; Resources; Investigation; Writing original draft; Writing, review, and editing}

\author[2]{Hamza Kheddar}
\ead{kheddar.hamza@univ-medea.dz}
\credit{Conceptualization; Methodology; Data Curation; Resources; Investigation; Visualization;  Writing original draft; Writing, review, and editing}

\author[3]{Sami Bourouis}
\ead{s.bourouis@tu.edu.sa}
\credit{Conceptualization; Methodology; Data Curation; Resources; Investigation; Visualization;  Writing original draft; Writing, review, and editing}

\author[4]{Nasir Saleem}
\ead{nasirsaleem@gu.edu.pk}
\credit{Conceptualization; Methodology; Data Curation; Resources; Investigation; Visualization;  Writing original draft; Writing, review, and editing}

\address[1]{LCPTS Laboratory, University of Sciences and Technology Houari Boumediene (USTHB), P.O. Box 32, El-Alia, Bab-Ezzouar, Algiers 16111, Algeria.}

\address[2]{LSEA Laboratory, Department of Electrical Engineering, University of Medea, 26000, Algeria}

\address[3]{Department of Information Technology, College of Computers and Information Technology, Taif University, Taif 21944, Saudi Arabia}

\address[4]{Department of Electrical Engineering, Faculty of Engineering and Technology, Gomal University, Dera Ismail Khan, Pakistan}


\tnotetext[1]{The first is the corresponding author.}

\begin{abstract}
Biometric authentication has garnered significant attention as a secure and efficient method of identity verification. Among the various modalities, hand vein biometrics, including finger vein,  palm vein, and dorsal hand vein recognition, offer unique advantages due to their high accuracy, low susceptibility to forgery, and non-intrusiveness. The vein patterns within the hand are highly complex and distinct for each individual, making them an ideal biometric identifier. Additionally, hand vein recognition is contactless, enhancing user convenience and hygiene compared to other modalities such as fingerprint or iris recognition. Furthermore, the veins are internally located, rendering them less susceptible to damage or alteration, thus enhancing the security and reliability of the biometric system. The combination of these factors makes hand vein biometrics a highly effective and secure method for identity verification.

This review paper delves into the latest advancements in deep learning  techniques applied to {finger vein,  palm vein, and dorsal hand vein} recognition. It encompasses all essential fundamentals of hand vein biometrics, summarizes publicly available datasets, and discusses state-of-the-art metrics used for evaluating the three modes. Moreover, it provides a comprehensive overview of suggested approaches for finger, palm, dorsal, and multimodal vein techniques, offering insights into the best performance achieved, data augmentation techniques, and effective transfer learning methods, along with associated pretrained {deep learning} models. Additionally, the review addresses research challenges faced and outlines future directions and perspectives, encouraging researchers to enhance existing methods and propose innovative techniques.
\end{abstract}




\maketitle


\begin{table}[]
\centering
{\small \section*{Acronyms and Abbreviations}}
\begin{multicols}{3}
\footnotesize
\input{acro_list}
\end{multicols}
\end{table}

\section{Introduction} \label{sec1}

Biometric authentication techniques have become integral in ensuring secure access and identity verification in various domains. These techniques are typically classified into two classes—behavioral and physiological.
Behavioral biometrics encompass characteristics based on an individual's behavior, such as signature,  keystroke dynamics \cite{al2023keystroke}, gait analysis \cite{parashar2023real}, voice recognition \cite{kheddar2023deep,kheddar2024automatic}, and more. On the other hand, physiological biometrics include physical and static characteristics, which  are classified also into two categories: extrinsic and intrinsic. Extrinsic biometrics include features that are external to the body, such as fingerprint, palm print, face, and iris patterns. Intrinsic biometrics, on the other hand, involve features that are inherent to the body, such as vein patterns and retinal scans. 

\begin{figure}[b!]
    \centering
    \includegraphics[scale=0.55]{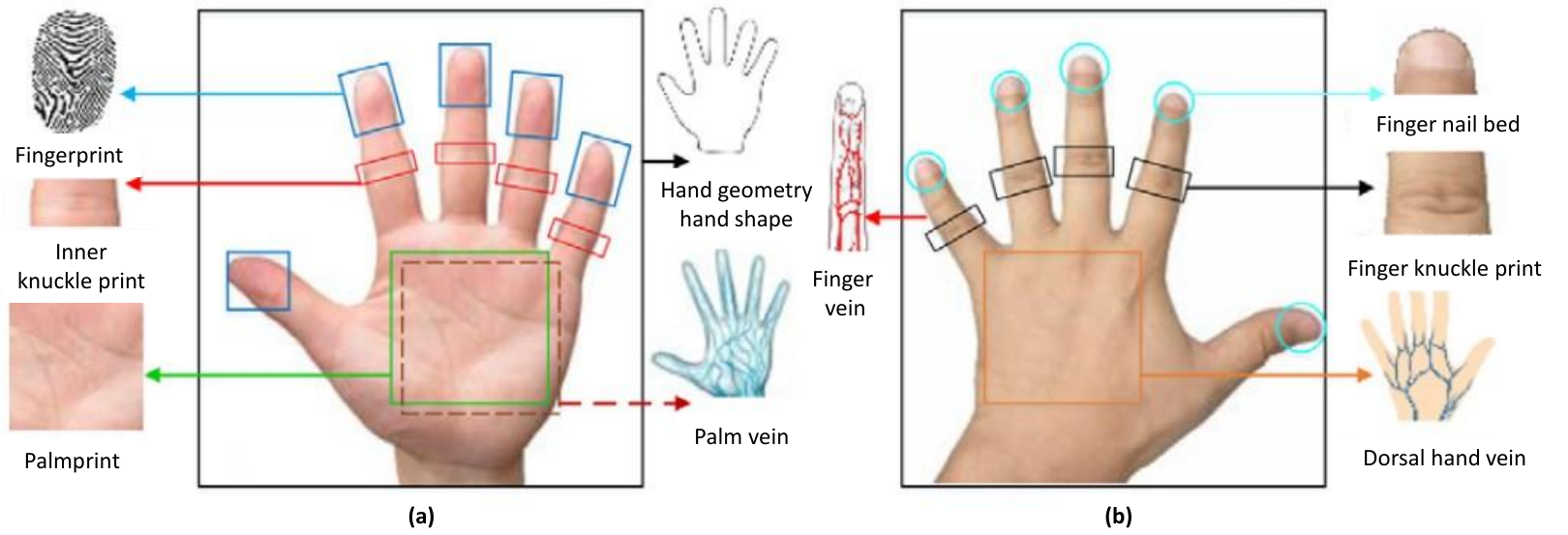}
    \caption{Biometrics of hands \cite{jaswal2016knuckle}. (a) Palm hand features. (b) Dorsal hand features. These characteristics serve as unique biometric identifiers for accurate and reliable personal identification and authentication.}
    \label{fig:handBiometrics}
\end{figure}

From the hand, numerous physiological properties can be extracted for biometric purposes, as illustrated in Figure \ref{fig:handBiometrics}. Among these, hand veins-based biometric method has attracted widespread attention because of several distinct advantages. These biometrics are inherently more secure and difficult to forge, as they are internal to the body and not easily accessible. \ac{FV} biometrics, for instance, have been shown to exhibit a level of diversity similar to that of iris patterns, ensuring high security and suitability for biometric authentication \cite{hou2022finger}. Moreover, capturing vein patterns is only effective in a living body, making it impossible to steal vein patterns from deceased individuals \cite{miura2004feature, kosmala2012human}. Additionally, vein patterns are permanent and do not change in adulthood \cite{syazana2016review}. Furthermore, vein images are typically captured by contactless sensors, ensuring hygiene and convenience \cite{wang2017quality}. Research has also shown that vein recognition is robust against various distortions such as motion blur, defocus, sensor aging, and compression \cite{kauba2015sensor}. 
Despite its advantages, hand vein-based biometrics face challenges. Factors like capture device configuration, ambient light, dust, humidity, and finger misplacement can affect performance  \cite{prommegger2018longitudinal, hou2022finger}. However, research has shown that these issues can be mitigated with optimized settings and processing strategies. Various methods have been developed to overcome those issues.

Since the vein-based biometric system was investigated, many significant achievements have been made in this area as
new systems have been continuously designed and deployed
over the years. \ac{DL} has emerged as powerful tools in enhancing the accuracy and efficiency of vein-based biometric systems. \ac{DL} significantly enhances hand vein recognition by automating feature extraction and improving accuracy. Traditional methods, such as \ac{PCA}, \ac{LDA}, and Wavelet Transform, often rely on manual feature extraction and are sensitive to variations in lighting and hand positioning. In contrast, \ac{DL} models like \acp{CNN} can identify complex patterns and variations in vein structures with minimal human intervention, leading to higher recognition rates and robust performance. Additionally, \ac{DL}'s ability to learn from large datasets enables continuous improvement and adaptation, offering superior security and reliability for biometric authentication.

\subsection{Motivation and contribution}
The motivation behind this comprehensive review stems from the increasing significance of \ac{DL} in the field of vein biometrics. As the demand for robust and secure authentication methods grows, understanding the advancements in \ac{DL} for \ac{FV}, \ac{PV}, and \ac{DHV} recognition becomes pivotal. This review aims to provide a thorough examination of the current \ac{SOTA} in these areas,  spanning
the period from 2015 to 2024, highlighting key contributions, breakthroughs, and challenges encountered in the application of \ac{DL} techniques. To the best of the authors' knowledge, there has been no prior research paper that has explored and critically evaluated contributions in \ac{FV}, \ac{PV} and \ac{DHV} using \ac{DL} until now.   

In recent years, a multitude of survey papers have been published, evaluating various facets of vein biometric models, as summarized in Table \ref{tab:sotareview}. Most of these reviews focus on specific hand vein types such as \ac{FV} \cite{shaheed2018systematic, mohsin2020finger, sidiropoulos2021feature, hou2022finger, shaheed2022recent}, \ac{PV} \cite{wu2020review, salazar2023towards} or \ac{DHV} \cite{jia2021survey}, while others include several physiological biometrics \cite{shaheed2021systematic, shaheed2024deep}. Some of these reviews focus on specific topics such as \ac{PAD} in biometric systems \cite{shaheed2024deep}, feature extraction \cite{sidiropoulos2021feature}, or the generation of synthetic vein images \cite{salazar2023towards}. These reviews primarily cover traditional approaches, with partial inclusion of \ac{DL} approaches or an exclusive focus on \ac{DL}. Also, many of these reviews partially address or miss critical areas such as datasets and metrics reviews, challenges, and future directions.

Unlike previous reviews, our work addresses all three hand vein types—\ac{FV}, \ac{PV}, and \ac{DHV}—as well as multimodal vein recognition, focusing solely on \ac{DL} approaches. Additionally, our review provides an extensive analysis of existing datasets and evaluation metrics. It also examines data augmentation and \ac{DTL} techniques, which are crucial for limited hand vein data. Finally, the review outlines current challenges in the field and future directions for research. The contribution of the manuscript is summarized as follows: 

\begin{itemize}
    \item This paper reviews recent methods for hand vein-based biometrics using \ac{DL}, including preprocessing, feature extraction, and classification. It also covers \ac{PAD} and \ac{TProt}, primarily for \ac{FV} recognition.

    \item The review analyzes \ac{DTL} and data augmentation techniques across all reviewed \ac{DL}-based approaches. It also covers evaluation metrics and  main datasets exploited in hand vein biometrics.
    
    \item The review spans literature from 2017 to 2024, with insights presented through tables and figures to aid understanding \ac{DL}-based hand vein \ac{SOTA} methods.
    
    \item The paper aims to motivate further research in vein-based biometrics by highlighting research gaps, challenges, and proposing future directions.
\end{itemize}

This research suggest paths for further investigation in hand vein biometrics, focusing on various technologies, methodologies, and datasets. By analyzing \acp{IRI}, researchers can accurately identify vein patterns in fingers, hands, and palms. The subsequent section provides a concise overview of these contributions. Table \ref{tab:sotareview} compares the proposed contribution with other hand vein biometric surveys. It is evident that our survey is the most comprehensive.

\begin{table}[h!]
\caption{Comparison of the proposed contribution against other hand vein biometric surveys. The symbols \CIRCLE{}, \LEFTcircle{}, and \Circle{} indicate that the specific field has been addressed, partially addressed, and missed, respectively.
}
\label{tab:sotareview}
\resizebox{\textwidth}{!}{ 
\scriptsize
\begin{tabular}{llm{2.5cm}m{0.8cm}m{0.8cm}m{1cm}m{1.2cm}m{0.5cm}m{0.5cm}m{1cm}m{1cm}m{1cm}m{1cm}}
\hline
Ref. & Year & Description of the survey/review & DL-FV & DL-PV &  DL-DHV & DL-MMV & DTL &DAT & Dataset review & Metrics review & Current \newline challenges & Future \newline directions\\ \hline

\cite{shaheed2018systematic} & 2018 & A systematic review of \ac{FV}
recognition methods  & \LEFTcircle{} & \Circle{} & \Circle{} & \Circle{} & \Circle{} &\Circle{}&\CIRCLE{} & \Circle{} & \Circle{} & \LEFTcircle{}\\

\cite{mohsin2020finger} & 2020 & A review of \ac{FV}
biometric verification systems based software
and hardware components & \LEFTcircle{} & \Circle{} & \Circle{} & \Circle{} &\Circle{} &\Circle{} &\LEFTcircle{} & \Circle{} & \CIRCLE{} & \CIRCLE{}\\

\cite{jia2021survey} & 2021 & A survey of \ac{DHV} biometrics  & \Circle{} & \Circle{}  & \LEFTcircle{} & \CIRCLE{} &\Circle{} &\Circle{} & \CIRCLE{} & \Circle{}  & \CIRCLE{} & \LEFTcircle{} \\

\cite{wu2020review} & 2019 & A review of \ac{PV} recognition methods  & \Circle{} & \Circle{}   &  \Circle{} & \Circle{} &\Circle{} & \Circle{} & \CIRCLE{} & \Circle{}  & \CIRCLE{} & \CIRCLE{} \\

\cite{sidiropoulos2021feature}&2021 & A review of feature extraction methods applied for \ac{FV} recognition  & \LEFTcircle{} & \Circle{} & \Circle{} & \LEFTcircle{} & \Circle{} &\Circle{} & \CIRCLE{} & \Circle{}  & \Circle{} & \Circle{}\\

 \cite{hou2022finger} & 2022 &A review of finger vein
recognition methods  & \CIRCLE{} & \Circle{} & \Circle{} & \Circle{} & \Circle{} & \Circle{}&\LEFTcircle{} & \LEFTcircle{} & \CIRCLE{} & \CIRCLE{} \\

\cite{shaheed2022recent}  & 2022 & A survey of DL, \ac{PAD} and Multimodal based finger vein recognition systems  & \CIRCLE{} & \Circle{} & \Circle{} & \CIRCLE{} &\Circle{} &\Circle{} & \Circle{} & \LEFTcircle{} & \CIRCLE{} & \LEFTcircle{}\\

\cite{salazar2023towards} & 2023 & A review of methods for generating  synthetic images of \ac{PV} patterns & \Circle{} & \CIRCLE{} & \Circle{} & \Circle{} &\Circle{} &\Circle{} & \LEFTcircle{} & \LEFTcircle{} & \CIRCLE{} & \CIRCLE{}\\

\cite{shaheed2021systematic} & 2024 & A systematic review of physiological‑based biometric recognition systems  & \CIRCLE{} & \LEFTcircle{} & \Circle{} & \LEFTcircle{} &\Circle{} & \Circle{} & \LEFTcircle{} & \CIRCLE{} & \CIRCLE{} & \Circle{}\\

\cite{shaheed2024deep} & 2024 & A systematic review of
DL-based \ac{PAD} systems & \LEFTcircle{} & \Circle{} & \Circle{} & \Circle{} &\Circle{} &\Circle{} &\Circle{} & \LEFTcircle{} & \LEFTcircle{} & \Circle{} \\

Our & 2024  & A review of DL-based hand vein recognition systems including \ac{FV}, \ac{PV} and \ac{DHV} & \CIRCLE{} & \CIRCLE{}  &\CIRCLE{} & \CIRCLE{} & \CIRCLE{} &\CIRCLE{} &\CIRCLE{} & \CIRCLE{} & \CIRCLE{} & \CIRCLE{} \\
\hline
\end{tabular}}
\begin{flushleft}
\scriptsize{Abbreviations: DL-based Finger vein (DL-FV),  DL-based Palm vein (DL-PV), DL-based dorsal hand vein (DL-DHV), DL-based mutimodel vein (DL-MMV), Data augmentation techniques (DAT).}   
\end{flushleft}
\end{table}




\begin{figure}
    \centering
    \includegraphics[scale=0.55]{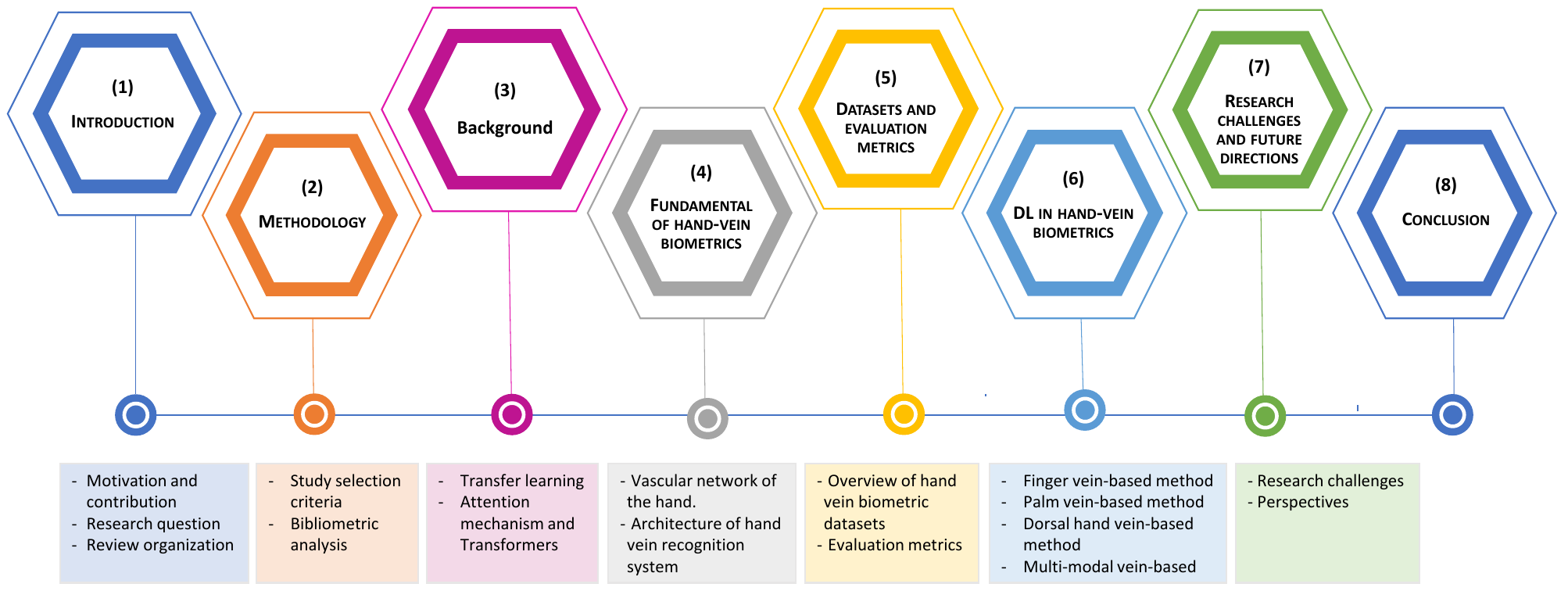}
    \caption{Roadmap of the review, showing main sections of the manuscript.}
    \label{fig:road}
\end{figure}

\subsection{Research questions}



To streamline this systematic mapping review, we outlined ten research questions in Table \ref{tab:research-questions}. Following the study's roadmap, the reader will gain the main insights and understand the study's objectives. The table presents a structured overview of the research questions (RQ) along with their corresponding motivating features. Each row of the table is dedicated to a specific research question, providing a clear understanding of the objectives driving the research in the domain of hand vein identification using automated technologies.

\begin{table}[h]
\centering
\caption{Research Questions and Motivating Features}
\begin{tabular}{|l|m{8cm}|m{8cm}|}
\hline
\textbf{RQ} & \textbf{Research Questions} & \textbf{Motivating Features} \\
\hline
Q1 & Which technologies and approaches are mostly used for automated vein detection in \acp{IRI}? & Objective: To understand the latest innovative methods and the difficulties faced in this endeavor.  \\
\hline
Q2 & How have conventional \ac{ML} methods been used in hand vein identification systems? & Objective: To investigate techniques that use carefully selected attributes/features and established criteria for hand vein identification.  \\
\hline
Q3 & How have \ac{DL} methods been used for hand vein identification systems? & Objective: To explore techniques and architectures that use advanced neural networks to learn features and classifiers from \acp{IRI} data.  \\
\hline
Q4 & Which databases are being employed to assess the effectiveness of hand vein identification and categorization methodologies?  & Objective: To conduct a comprehensive analysis of the quality and availability of current datasets, and to determine shortcomings and the need for additional data sources.  \\
\hline

Q5 & How is the challenge of limited datasets addressed in hand vein biometrics ?  & Objective: To investigate the strategies and techniques used to overcome the limitations of small datasets in hand vein biometric research, including data augmentation, synthetic data generation, and transfer learning.   \\
\hline

Q6 & Which measures have been used to evaluate the performance of hand vein identification techniques?  & Objective: To better understand the requirements and benchmarks used in assessing the precision and effectiveness of various methodologies.  \\\hline

Q7 & Which \ac{DL} model is most commonly utilized in hand vein recognition ?  & Objective:  To identify the predominant \ac{DL} architectures used in hand vein biometrics and to analyze their effectiveness and popularity among researchers.  \\
\hline

Q8 & Which types of hand vein biometrics are most extensively  studied by researchers ?  & Objective: To explore the focus areas within hand vein biometrics research, identifying whether \ac{FV}, \ac{PV}, or \ac{DHV} receive the most attention from the scientific community.  \\ \hline

Q9 & How can \ac{DL} enhance the security of hand vein recognition systems to mitigate presentation attacks and protect biometric templates ? & Objective: To explore \ac{DL} techniques that can detect and prevent presentation attacks, and to develop methods for securing biometric templates against potential threats. \\ \hline

Q10 & What are the research challenges and future directions in hand vein biometrics ?
 & Objective: To identify the current limitations and obstacles in the field, to highlight areas requiring further research, and to propose potential future research directions to advance hand vein biometrics. \\ \hline 

\hline
\end{tabular}
\label{tab:research-questions}
\end{table}

\subsection{Review organization}
To achieve a structured and insightful exploration, this review is organized into several key sections. The methodology in Section \ref{sec2} outlines the criteria used for selecting studies and employs bibliometric statistics to present an overview of the research landscape. Following this,  Section \ref{sec3} delves into the fundamentals of hand vein biometrics, covering the vascular network of the hand, its characteristics, and the architecture of hand vein recognition systems. Section \ref{sec:back} offers background knowledge essential for understanding the proposed ML/DL-based hand vein identification approaches, including \ac{DTL}, attention mechanisms, and Transformers. 
Moving forward, Section \ref{sec4} provides an overview of available datasets and discusses evaluation metrics relevant to \ac{DL} models in the hand vein biometrics domain. Subsequently, the \ac{DL} in hand vein biometrics,  Section \ref{sec5}, explores most suggested \ac{SOTA} \ac{DL}-based vien techniques, including \ac{FV}-based,  \ac{PV}-based, \ac{DHV}-based, and multi-modal vein-based. As the review progresses, research challenges and future directions, thoroughly detailed in Section \ref{sec6}, highlight the existing hurdles in the field, offering perspectives on potential solutions and paving the way for future advancements. The conclusion  in Section \ref{sec7} summarizes the key findings, emphasizing the importance of \ac{DL} in advancing the reliability and security of vein-based biometric systems. Figure \ref{fig:road} illustrates the roadmap and detailed organization of the paper.

\section{Methodology} \label{sec2}
The research methodology is crucial for conducting any type of research or survey. This section covers the research approach we utilized to conduct a thorough survey. It outlines our criteria for selecting studies and provides bibliometric statistics to offer an overview of the research landscape.

\subsection{Study selection criteria}
\begin{figure}
    \centering
    \includegraphics[scale=0.5]{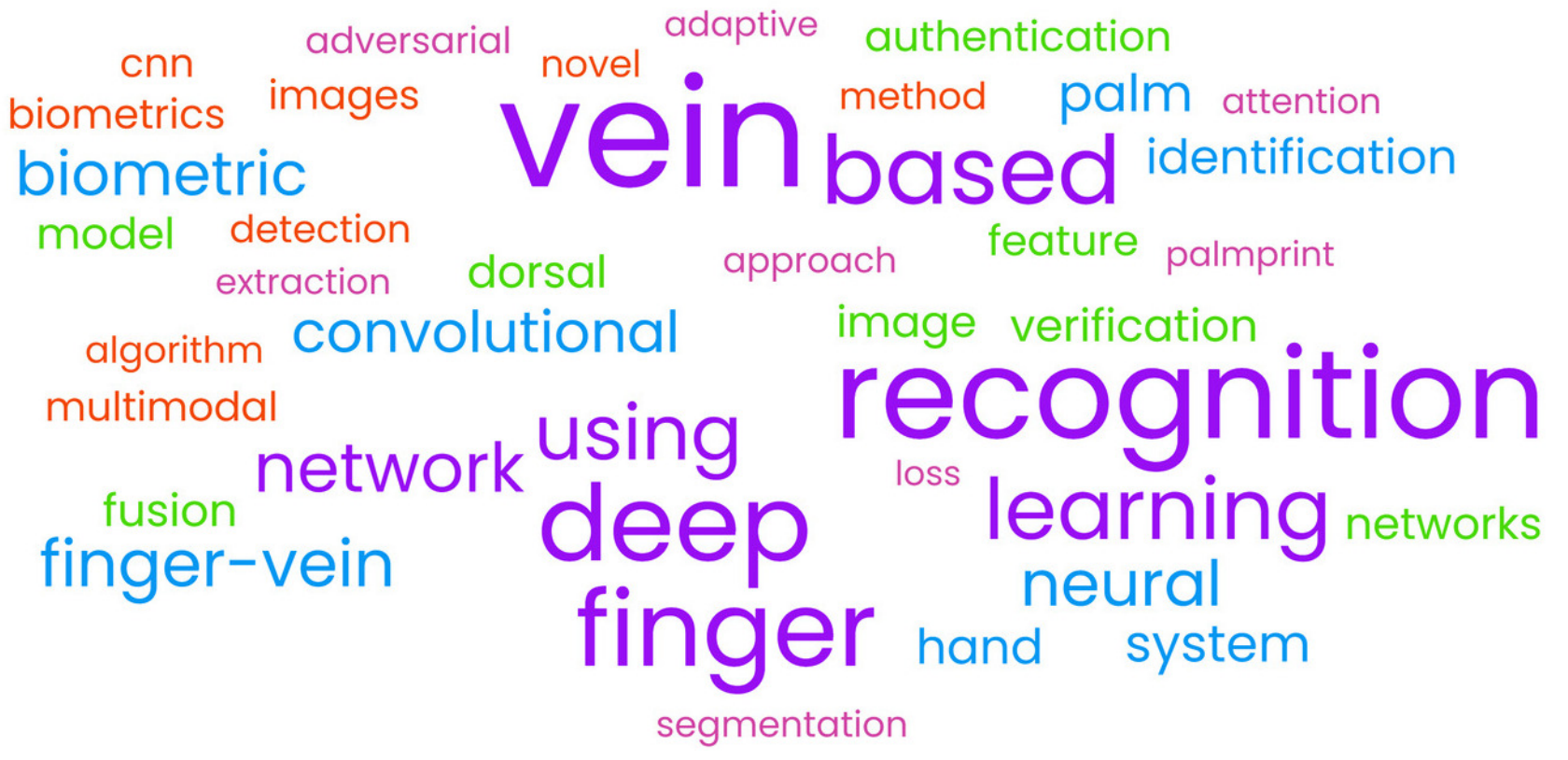}
    \caption{Word cloud of the most essential terms in the field of hand vein biometrics.}
    \label{fig:cloud}
\end{figure}

 To identify and review existing studies on \ac{DL}-based hand vein biometrics, a comprehensive search was conducted across several leading publication databases renowned for their high-quality scientific research written in English. The primary search was carried out in Scopus, which systematically includes databases such as Web of Science, Elsevier, IEEE, among others. Articles published in the last ten years were prioritized. However, older publications were also considered when necessary to provide fundamentals, datasets, metrics, etc. Additionally, high-quality pre-prints from arXiv, SSRN, among others, were selected. Figure \ref{fig:cloud} summarizes the most frequently used keywords by authors. Keywords with larger font sizes indicate higher usage, while smaller font sizes indicate less frequent usage.

\subsection{Bibliometric analysis}

\begin{figure}[h!]
    \centering
    \includegraphics[width=0.9\linewidth]{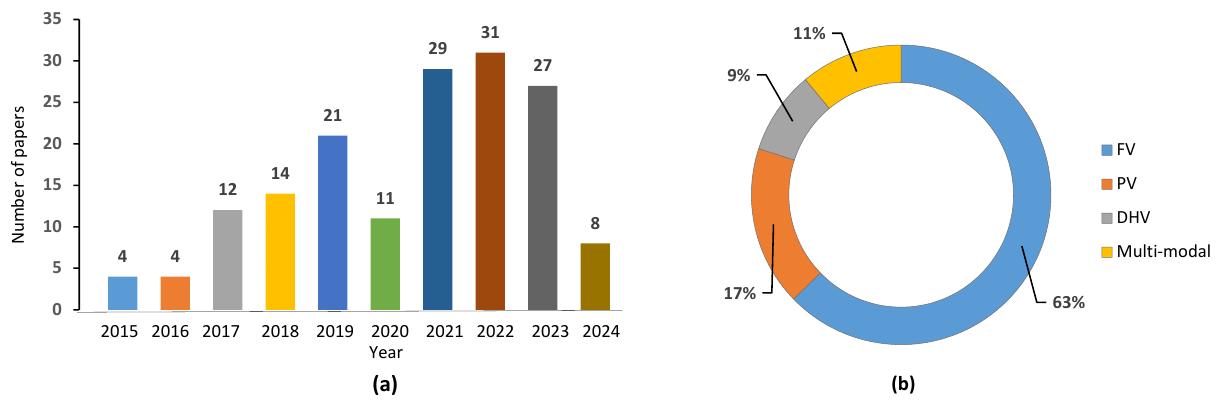}
    \caption{Improved caption:
Bibliography Statistics: (a) Annual publications on \ac{DL}-based hand vein research; (b) Percentage distribution of publications across different domains: \ac{FV}, \ac{PV}, \ac{DHV}, and multi-modal. The data clearly indicates a predominant research focus on \ac{DL}-based \ac{FV}. }
    \label{fig:stats}
\end{figure}

A total of 174 papers published between 2015 and 2024 were involved in the main sections discussing fundamentals and \ac{DL}-based hand vein biometrics. Figure  \ref{fig:stats}(a) illustrates the distribution of these papers over the years. Additionally, 145 out of the 174 papers discussed in Section 6 belong to one of the following domains: \ac{FV}, \ac{PV}, \ac{DHV}, and multi-modal. Figure \ref{fig:stats}(b) illustrates the distribution of these papers across these vein biometrics domains. 
The remaining papers in this review, some of which fall outside the 2015-2024 range, include  review papers used for comparison,  reviews and studies related to \ac{DL} techniques, \ac{DTL}, Transformers,  preprocessing, metrics, datasets, and more. Additionally, other papers were included to provide a comprehensive foundation for the introduction, address the challenges faced, and offer insights into future directions.

\section{Background}\label{sec:back}

According to the reviewed \ac{SOTA} work, some background knowledge is necessary to deeply understand the proposed \ac{ML}/\ac{DL}-
based hand vein identification approaches. {For a comprehensive understanding of various \ac{ML} and \ac{DL} techniques, researchers are encouraged to consult our previous work in \cite{kheddar2024deep}. This publication covers a broad spectrum of methodologies, including \ac{DBN}, \ac{GNN}, \ac{DNN}, \ac{RNN}, \ac{LSTM} networks, \acp{AE}, \ac{RBM}, and \ac{CNN}. These techniques are valid and adaptable across multiple fields of research. In contrast, the background section of this review is specifically focused on advanced \ac{DL} techniques employed in hand vein recognition research.} For instance, \ac{DTL}, particularly fine-tuning and \ac{DA}, is widely
adopted. Additionally, attention layers and Transformers, which are advanced versions of attention mechanisms, are also extensively
used. {By concentrating on these newer and more complex concepts, this review
aims to equip readers with the necessary foundation to comprehend current trends and innovations in the field. While fundamental
concepts like \acp{CNN} are crucial to many approaches, they are well-established in the literature and thus are not detailed in this section.}

\subsection{Deep transfer learning}

\ac{DTL} is a \ac{ML} technique where a pre-trained model, which was initially trained on a large dataset for a specific task, is reused or fine-tuned for a different but related task. Instead of training a model from scratch, \ac{DTL} leverages the patterns and knowledge learned by the pre-trained model to improve performance on the new task, often with less training data \cite{himeur2023video}.

\begin{figure}
    \centering
    \includegraphics[scale=0.75]{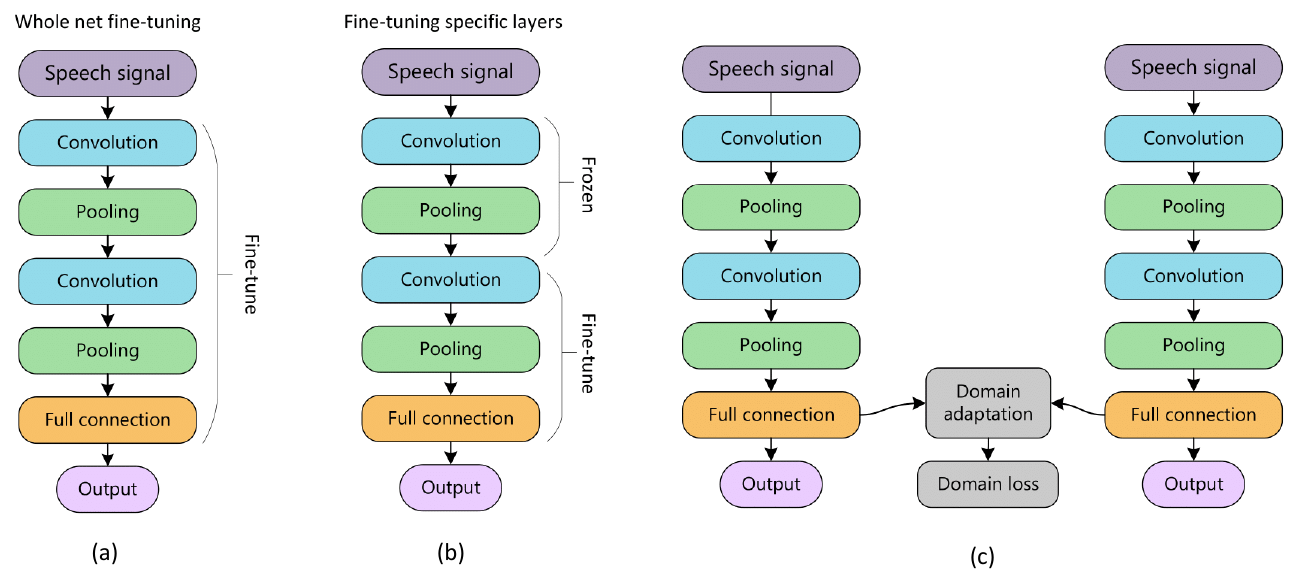}
    \caption{\ac{DTL} principles. (a) Full fine-tuning; (b) Partial fine-tuning; (c) \ac{DA}.}
    \label{fig:TL}
\end{figure}

\noindent \textbf{(a) Fine-tuning all layers in a model:} When researcher fine-tune all layers of a pre-trained model, he adjust the weights of the entire network on your new dataset (Figure \ref{fig:TL} (a)). This approach is beneficial when the new task is similar but has some significant differences from the original task \cite{kheddar2023deep}.

\noindent \textbf{(b) Fine-tuning some layers in a model:} In this approach, the researcher only fine-tune certain layers of the pre-trained model while keeping others frozen, i.e. their weights remain unchanged (Figure \ref{fig:TL} (b)). This is useful when the new task is very similar to the original task \cite{kheddar2023deep}. 

\noindent \textbf{(c) \ac{DA}:} Is a subfield of \ac{DTL} where the goal is to adapt a model trained on a source domain to perform well on a target domain, despite the differences between the domains (Figure \ref{fig:TL} (c)). This involves adjusting the model, by feature adaptation or fusion, to bridge the gap between the source and target domains, which may have different distributions or feature spaces \cite{himeur2023video}.

For example, in the context of \ac{FV} biometric recognition, fine-tuning all layers in a model might involve starting with a \ac{CNN} pre-trained on ImageNet and then training it entirely on a \ac{FV} dataset to capture the unique patterns of veins across different individuals. Fine-tuning some layers could involve freezing the initial layers of the same pre-trained \ac{CNN}, which capture general features, and only training the later layers on the \ac{FV} dataset to adapt to vein-specific details. For \ac{DA}, a model pre-trained on a similar biometric dataset, such as \ac{PV} images, could be adapted to \ac{FV} images by using techniques like adversarial training to align the feature distributions between the two domains, ensuring robust performance despite the differences in image characteristics. Thus, \ac{DTL} and \ac{DA} can be extremely useful due to the limited availability of hand-vein image datasets.

\subsection{Attention mechanism and Transformers}

In recent years, the fields of \ac{DL} and \ac{ML} have been revolutionized by the concepts of Attention mechanisms and Transformers. These innovations have significantly improved the performance of models in tasks such as language translation, text generation, and image processing.

\noindent \textbf{(a) Attention mechanism:}  The attention mechanism is a concept that enables models to focus on specific parts of the input data when making predictions. It was introduced to address the limitations of traditional sequence models, such as \acp{RNN} and \acp{LSTM}, which often struggle with long-range dependencies and information retention over long sequences. Attention works by assigning different weights to different parts of the input sequence. This allows the model to prioritize important information and ignore irrelevant details. The key components of the attention mechanism include \cite{kheddar2024automatic}:

\begin{itemize}
    \item \textbf{Query (Q)}: Represents the current state or the part of the sequence that is being processed.
    \item \textbf{Key (K)}: Represents all parts of the input sequence.
    \item \textbf{Value (V)}: Also represents all parts of the input sequence, but these are the actual values being processed.
\end{itemize}

The attention score is calculated as a similarity measure between the query and the keys, and this score is used to weigh the values. This weighted sum then forms the attention output.\\

\noindent \textbf{(b) Transformer model:}  The Transformer model, introduced by Vaswani et al. in \cite{vaswani2017attention}, is built entirely on attention mechanisms, eschewing the use of recurrence altogether. Transformers have become the foundation for \ac{SOTA} models in NLP, such as \ac{BERT}, \ac{GPT}, among others \cite{djeffal2023automatic}. Key components of the Transformer model include:

\begin{itemize}
    \item \textbf{Multi-head self-Attention}: Allows the model to focus on different parts of the input sequence simultaneously by employing multiple attention mechanisms (heads). Each head captures different aspects of the input.
    \item \textbf{Positional encoding}: Since transformers do not process sequences in a sequential manner, positional encoding is added to the input embeddings to give the model information about the position of each word in the sequence.
    \item \textbf{Feed-forward neural Networks}: Applied to the attention outputs, adding non-linearity and improving the model's expressive power.
    \item \textbf{Layer normalization and residual connections}: Help in stabilizing and improving the training of deep models by normalizing the inputs and adding skip connections.
\end{itemize}

Transformers consist of an encoder and a decoder. The encoder processes the input sequence and generates a set of attention-based representations. The decoder takes these representations and generates the output sequence, often using a similar attention mechanism to focus on different parts of the input.

\section{Fundamentals of hand vein biometrics}\label{sec3}

\subsection{Vascular network of the hand}
\begin{figure}
    \centering
    \includegraphics[scale=0.5]{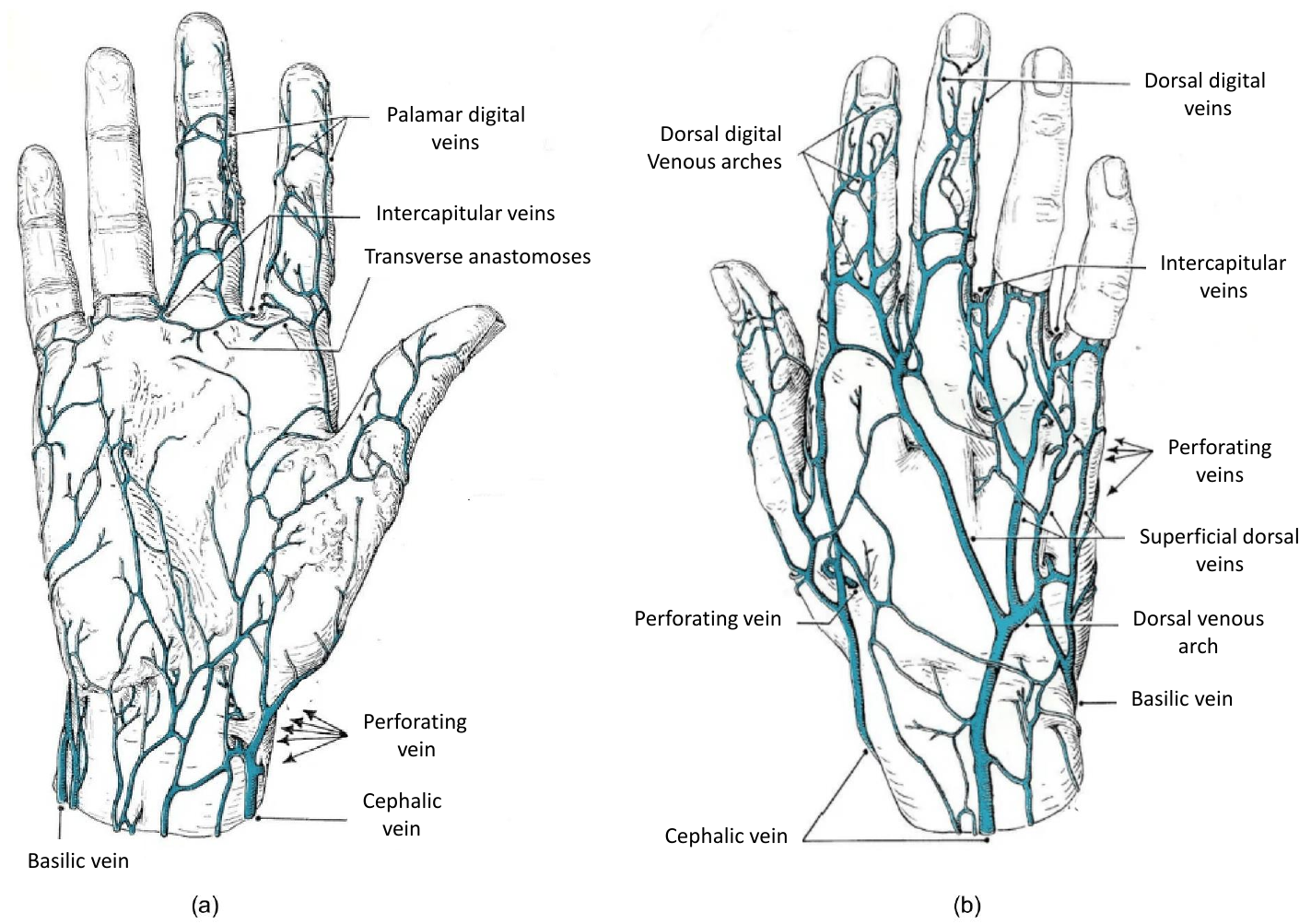}
    \caption{Illustration of the hand's vascular network \cite{HandVein}. (a) Palmar view; (b) Dorsal view.}
    \label{fig:HandVeins}
\end{figure}

Hand vein biometrics utilize the vascular patterns present in various parts of the hand, including the fingers, palms, and the dorsal hand (Figure \ref{fig:HandVeins}). The hand's vascular network consists of veins and arteries. Veins carry deoxygenated hemoglobin (Hb), while arteries transport oxygenated hemoglobin (HbO2) \cite{salazar2023towards}. The hand's vascular structure displays intricate patterns that facilitate unobstructed blood flow during hand movements. These specific patterns enable the vascular network to serve as a biometric identifier, with superficial networks being primarily utilized due to the optical limitations of acquisition systems \cite{crisan2008hand}.  Each individual's vein pattern is unique, akin to fingerprints, due to the random development of vascular structures influenced by genetic and environmental factors. The following are the types and characteristics of hand veins:
\begin{itemize}
    \item \textbf{\acp{FV}:} The vascular network in the fingers is composed of several small veins running parallel to the bones and joints. These veins are typically arranged in a branching pattern that varies significantly from person to person. These patterns offer high resolution due to the smaller captured area, revealing rich details. However, the smaller area limits the amount of data captured compared to \acp{PV}.
    \item \textbf{\acp{PV}:} The palm's vein structure includes a more complex and denser network of veins, owing to the larger surface area. The palm's veins are situated beneath the skin and muscle, providing a unique and stable pattern for each individual. Because the palm has no hair, it is easier to photograph its vascular pattern.  The palm also has no significant variations in skin color compared with fingers or the dorsal hand, where the color can darken in certain areas.  
    The larger surface area offers more data for recognition, but recognition accuracy might be affected by wrinkles or scars present on the palm.
    \item \textbf{\acp{DHV}:} The veins on the back of the hand are more superficial and visible compared to those in the palm. They form distinctive patterns that can be easily captured with appropriate imaging techniques. They are more convenient for non-intrusive capture (easier to position the hand). 
\end{itemize}

\subsection{Architecture of hand vein recognition system}
\begin{figure}
    \centering
    \includegraphics[scale=0.9]{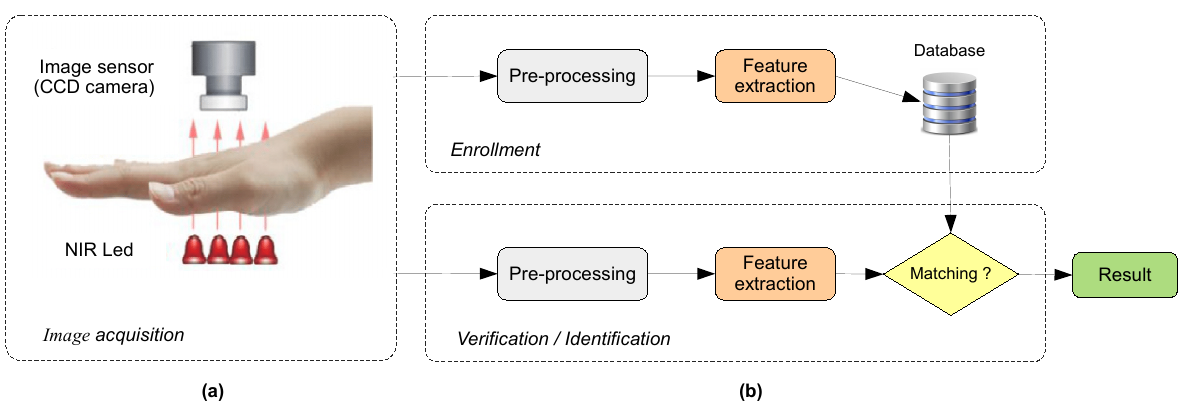}
    \caption{Framework of hand vein verification system. (a) hand vein image acquisition, specifically the \ac{PV} using transmission mode; (b) Enrollment and identification phases. }
    \label{fig:VeinVerificationDiagram}
\end{figure}

A hand vein recognition system operates through a three-step process: acquisition, enrollment, and verification (or identification), as illustrated in Figure  \ref{fig:VeinVerificationDiagram}.  
During the enrollment phase, individual images are collected and undergo preprocessing to remove noise. Feature extraction algorithms then identify unique characteristics of the vein network. These features are compiled into a secure digital template specific to the user and stored in a database.
In the identification phase, the user's image is captured, and the system repeats the preprocessing and feature extraction steps. The extracted features are then compared with the database templates to either determine the user's identity in identification mode or verify the user's identity in verification mode \cite{kolivand2023finger}.

\subsubsection{Image acquisition}
The acquisition of hand vein images is carried out usually using \ac{NIR} devices that interact with oxidized hemoglobin (HbO2) and deoxidized hemoglobin (Hb) in the vascular network \cite{wu2020review, wang2007infrared}. 
When the light wavelength is between 720 nm and
760 nm, the radiation is strongly absorbed by Hb, producing a shadow
corresponding to the vein pattern. At 790 nm, there is an intersection
point where Hb and HbO2 present the same absorption, which allows
the visualization of veins and arteries. Meanwhile, for higher spectral
ranges of the optical window HbO2 presents a slight increase compared
to Hb. Thus, the vascular pattern inside the hand can be rendered visible with the help of an \ac{NIR} light source in combination with an \ac{NIR}-sensitive image sensor. Several studies highlighted that wavelength range of 875nm-890nm is the best choice for hand vein recognition \cite{zhang2016dorsal, chen2011band, walus2017impact}. 

In \ac{NIR}-based hand vein image acquisition (Figure  \ref{fig:VeinVerificationDiagram} (a)), the device initially uses an \ac{LED} array to emit \ac{NIR} light onto the hand. A \ac{CCD} camera, which is sensitive to \ac{NIR} light, then captures the detailed hand vein image. Several acquisition  modes can be found:


\begin{itemize}
\item  Based on \ac{NIR} imaging technique, hand vein image acquisition can be categorized into transmission mode and reflection mode \cite{kauba2018shedding}. In transmission mode, \ac{NIR} light passes through the hand, and an \ac{NIR}-sensitive camera captures the light on the opposite side, creating a clear vein pattern due to the absorption properties of hemoglobin. In reflection mode, \ac{NIR} light is emitted onto the hand's surface, and the reflected light is captured by an \ac{NIR}-sensitive camera placed on the same side as the light source. This mode is more practical for various applications due to its ease of use and less restrictive positioning.

\item Based on contact with the capture device, the acquisition modes can be divided into two categories: touch-based acquisition \cite{wang2015automatic, zhu2015near} and touchless-based acquisition \cite{joardar2016real}. In the first mode, the hand is in contact with the device, ensuring stable positioning and consistent imaging conditions. The second mode does not require physical contact, enhancing hygiene and user comfort. It is particularly suitable for public or high-traffic environments but may require more advanced image processing to handle variations in hand positioning.

\item  Based on the illumination type, it can be divided into four categories: top-light illumination \cite{kumar2011human}, side-light illumination \cite{ton2013high}, bottom-light illumination \cite{bazrafkan2016finger}, and both-light (side and top) illumination \cite{ramachandra2019design}. In top-light illumination, \ac{NIR} light is directed from above the hand, which highlights vein patterns with minimal interference from ambient light. In side-light illumination, \ac{NIR} light is directed from the side, providing contrast that can enhance the visibility of veins  through light reflection.  In bottom-light illumination, \ac{NIR} light is directed from below the hand, which can be effective in transmission mode setups.
Both-light illumination combines side and top lighting, utilizing both reflection and transmission for comprehensive lighting and improved image quality.

\item Based on hand pose for dorsal hand imaging, the acquisition modes can be classified into two categories: open hand mode \cite{wang2008minutiae}, where the hand is positioned open and flat, making it easier to capture the dorsal vein patterns, and clenched hand mode \cite{huang2014hand}, where the hand is clenched, which might be used to highlight different aspects of the dorsal vein network.

\end{itemize}

These various acquisition modes can be selected based on the specific needs and constraints of the hand vein recognition system, optimizing for factors such as image quality, user convenience, and application environment.


\subsubsection{Image preprocessing}
Image preprocessing plays a crucial role in hand vein recognition systems by ensuring that captured images are suitable for feature extraction and of high quality. Various factors, such as ambient light, sensor cleanliness, humidity, temperature, illumination, and user behavior, can affect the acquisition of hand vein images. Neglecting these factors can lead to low-quality images and decreased recognition accuracy. To address these issues, preprocessing involves several steps to improve vein pattern visibility and eliminate artifacts or noise that might hinder recognition. These steps include image quality assessment, image restoration and enhancement, and \ac{ROI} extraction.
\begin{itemize}
    \item \textbf{Image quality assessment:} Image quality assessment is a crucial initial step in the preprocessing of hand vein images. This process involves evaluating the overall quality of the captured images to determine if they meet the required standards for further processing. Factors such as focus, contrast, brightness, and overall clarity are assessed to ensure that the vein patterns are clearly visible and suitable for feature extraction. Images that do not meet the desired quality criteria may be discarded or undergo further enhancement to improve their suitability for subsequent processing steps. 

    Based on the human visual system's sensitivity and prior knowledge, numerous handcrafted descriptor-based methods have been developed for image quality assessment \cite{qin2018finger,hsia2017new, nino2023palm} . However, these handcrafted descriptor-based methods face several challenges. Firstly, some features directly derived from human visual system characteristics and prior knowledge may not adequately represent the quality of hand vein images. Secondly, single-feature methods are insufficient for comprehensive evaluation, while multi-feature methods are complex and time-consuming \cite{hou2022finger}. Integrating \ac{DL} into image quality processes can significantly enhance the performance and efficiency of hand vein recognition systems. Unlike traditional handcrafted descriptor-based methods, \ac{DL} models can automatically learn and extract relevant features from images without the need for explicit human knowledge or sensitivity to the visual system. \ac{DL} models are capable of handling the complex and diverse characteristics of hand vein images, making them more effective at evaluating image quality. By utilizing large amounts of training data, \ac{DL} models can establish an accurate mapping between image features and quality, enabling them to effectively assess the quality of new images. Several approaches have been developed, primarily focusing on \ac{FV} recognition. An overview of these \ac{DL}-based approaches is presented in section \ref{DLPrep-based}.   
    
    \item \textbf{Image restoration and enhancement:}     
    Image restoration and enhancement processes are designed to improve the quality and clarity of captured images, making them more suitable for feature extraction and subsequent analysis. Image restoration focuses on recovering the original vein pattern from degraded images caused by factors such as sensor noise, motion blur, and compression artifacts. Several traditional image restoration techniques, primarily focusing on removing scattering in \ac{FV} images, include the point spread function (PSF) \cite{lee2009restoration, lee2011image} and the biological optical model (BOM) \cite{yang2012scattering, yang2014towards}.

    On the other hand, image enhancement techniques are principally used for denoising and contrast enhancement to further improve the visibility of vein patterns. These techniques are crucial for ensuring the accuracy and reliability of hand vein recognition systems by providing high-quality input images for subsequent processing stages. Common denoising methods include median filters, Gaussian smoothing filters, Wiener filters, and wavelet transforms. For contrast enhancement, widely used techniques include \ac{HE}, \ac{CLHE}, \ac{AHE}, \ac{CLAHE} and Gabor filter. An example of image enhancement applied to \ac{FV} images using these methods is illustrated in Figure \ref{fig:ImageEnhancement}. These tools are primarily integrated with handcrafted feature extraction approaches but are also widely combined with \ac{DL} methods.

    \begin{figure}
        \centering
        \includegraphics[scale=0.75]{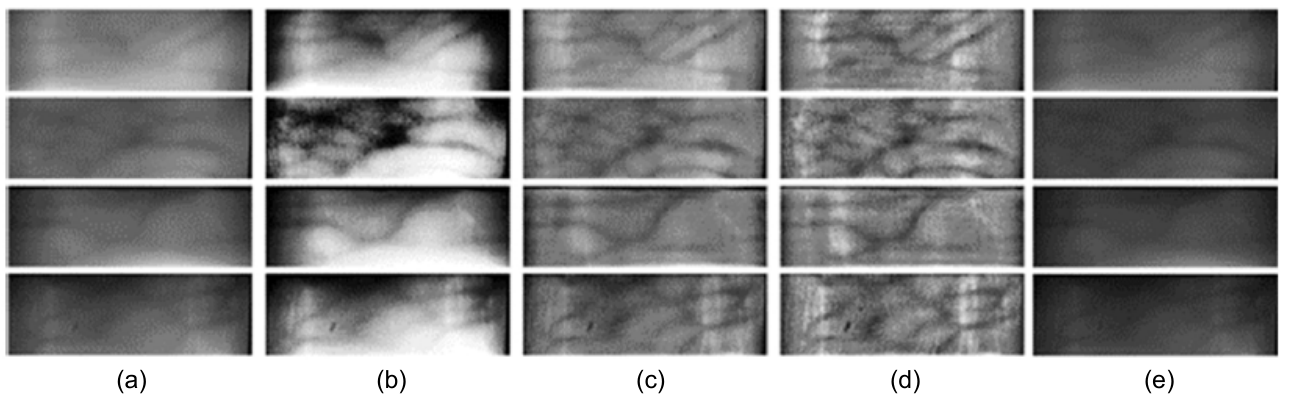}
        \caption{Examples of image enhancement methods applied to \ac{FV} images (adopted from \cite{hou2022finger}). (a) Orginal images, (b) \ac{HE} enhanced images, (c) \ac{AHE} enhanced images, (d) \ac{CLAHE} enhanced images, (e) Gabor enhanced images. }
        \label{fig:ImageEnhancement}
    \end{figure}

Both traditional and \ac{DL}-based image restoration and enhancement techniques have their strengths and limitations. On one hand, traditional methods often rely on predefined mathematical models or filters, such as Gaussian filters for noise reduction, Gabor filters for edge detection to enhance vein patterns, and \ac{HE} for contrast enhancement. These techniques have well-established mathematical foundations, making them computationally efficient and easy to implement without the need for training data. However, they may struggle to capture the full complexity of hand vein images across varied imaging conditions, such as changes in lighting, skin texture, or vein visibility. On the other hand, \ac{DL} approaches, while potentially more powerful in learning complex image features, indeed rely on the predefined network architecture and training data. This reliance can limit their generalizability to different imaging conditions or noise types not represented in the training set. Therefore, successful \ac{DL} models often require training data encompassing a wide range of image conditions and noise types to ensure robust performance and transferability. When properly trained on diverse datasets, they can often outperform traditional methods in specific tasks, particularly in handling complex, non-linear image degradations that are difficult to model mathematically.
\color{black}
    

    \item \textbf{Image \ac{ROI} extraction:} 
    The \ac{ROI} extraction process is crucial for isolating the specific regions of the hand that contain vein patterns, enhancing the accuracy of feature extraction, and reducing computational load by focusing only on relevant portions of the image. This process typically includes steps such as boundary selection and \ac{ROI} localization. Image alignment techniques are also primarily applied to \ac{FV} images during the \ac{ROI} extraction process to reduce the effects of finger movement and rotation. This ensures that the extracted region is consistent and accurately represents the vein patterns, despite any variations in finger positioning during image capture \cite{hou2022finger}. \ac{ROI} localization relies principally on detecting key points in the hand or finger. Due to the differences in vein pattern positions among the finger, palm, and dorsal hand, this process varies for each type. For \acp{FV}, the process often involves identifying the phalangeal joints, which are typically used to locate the region in the middle of the phalanx of a finger, ensuring that the central vein patterns are accurately captured \cite{hou2022finger}. For the palm and dorsal hand, the key points include the contours and reference (generally, the mid-point of the wrist) points, helping to define a consistent and precise \ac{ROI} that encompasses the dense vein patterns \cite{jia2021survey, wu2020review}. Several approaches have been explored for this step, each tailored to the specific characteristics of the vein patterns and the anatomical structure of the finger, palm, or dorsal hand. Figures \ref{fig:ROIFingerVein} and \ref{fig:ROIDHVVein} illustrate the \ac{ROI} localization process for the finger \cite{yang2012finger} and dorsal hand \cite{wang2007infrared}. The process for \ac{PV} \ac{ROI} localization is quite similar to that of the dorsal hand.
    
    \begin{figure}
        \centering
        \includegraphics[scale=0.65]{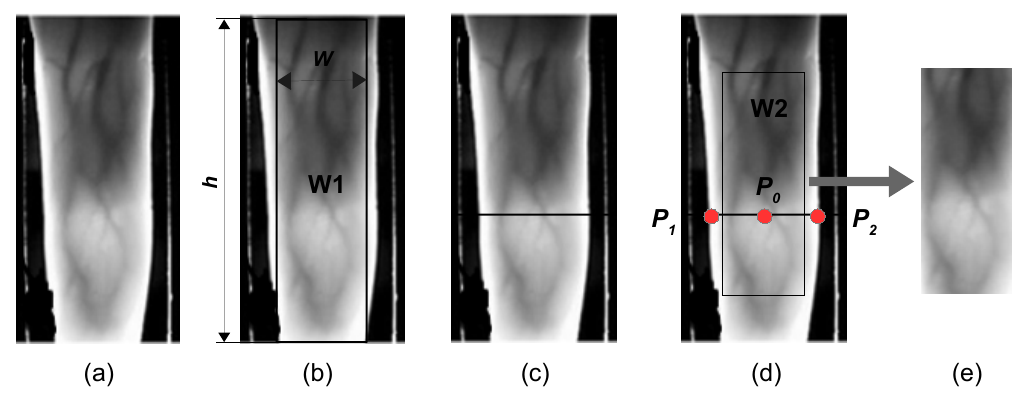}
        \caption{\ac{FV} \ac{ROI} extraction process \cite{yang2012finger}. (a) \ac{FV} image; (b) A candidate region W1 centered in the  w $\times$ h; (c) Inter-phalangeal joint position; (d) A \ac{ROI} region denoted by W2 extracted after locating three key points P$_0$, P$_1$ and P$_2$ points; (e) A \ac{FV} \ac{ROI} image.}
        \label{fig:ROIFingerVein}
    \end{figure}

    \begin{figure}
        \centering
        \includegraphics[scale=0.75]{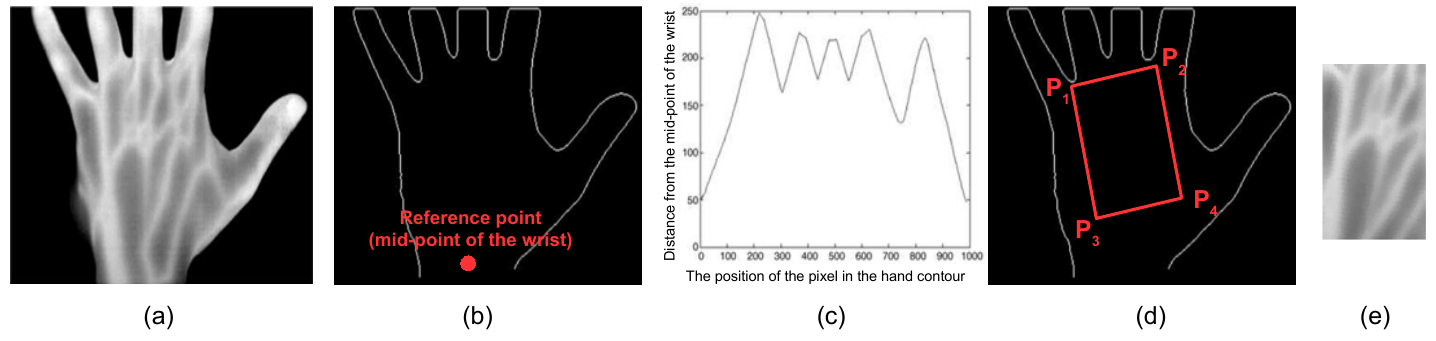}
        \caption{\ac{DHV} \ac{ROI} extraction process \cite{wang2007infrared}. (a) \ac{DHV} image, (b) the boundary of the hand, (c) the distance profile between the reference point and the contour points, (d) \ac{ROI} localization after locating P$_1$, P$_2$, P$_3$ and P$_4$. (e) A \ac{DHV} \ac{ROI} image.}
        \label{fig:ROIDHVVein}
    \end{figure}
   
\end{itemize}

\subsubsection{Feature extraction}
After preprocessing, Feature extraction is a crucial step in hand vein recognition systems, as it involves identifying and quantifying the unique patterns in vein images that can be used for accurate identification and authentication. This process translates the raw, preprocessed images into a set of numerical features that represent the essential characteristics of the vein patterns, facilitating the subsequent matching and classification stages. Traditional methods for feature extraction from vein images often rely on handcrafted techniques that utilize domain-specific knowledge. While traditional methods have been successful to some extent, they often require careful tuning and may not generalize well across different imaging conditions and populations. Handcrafted features are sensitive to noise and do not perform well on low-quality images. Additionally, setting standard parameters for these methods can be challenging, requiring extensive experimentation to determine suitable parameters for different databases. The choice of matching strategy after feature extraction can also significantly affect recognition performance \cite{hou2022finger}.

\ac{DL} methods have revolutionized feature extraction in recent years, offering a more automated and robust approach compared to traditional methods.  \ac{DL} can learn hierarchical feature representations directly from raw images without the need for handcrafted descriptors. These methods can simultaneously extract features, reduce data dimensionality, and classify within a single network structure. Compared to traditional methods, a significant advantage of \ac{DL} is its ability to maintain good and stable recognition performance regardless of image quality and hand/finger placement. \ac{DL} methods also reduce the need for extensive parameter tuning and can handle the complex and varied characteristics of hand vein images more effectively. 
Most proposed \ac{DL} approaches in hand vein recognition focus on feature extraction tasks, as reviewed in section \ref{sec5}.

\subsubsection{Matching}
After feature extraction, the final stage involves comparing the acquired vein image to a stored template. This comparison generates a matching score that indicates how similar the two images are. A high score suggests the input image belongs to the enrolled individual, while a low score implies a potential imposter. Two main methods are used in this phase: distance-based and classification-based \cite{hou2022finger, al2022vein}. The distance-based method directly compares the extracted feature vectors using a distance metric. The smaller the distance, the more similar the images are. On the other hand, the classification-based method utilizes \ac{ML} algorithms to classify the input image as either a match or a non-match based on the extracted features.

In the context of \ac{DL} approaches, classification methods often integrate feature extraction and classification into a single model. This integration allows the model to learn the most relevant features for matching during the training process, leading to more accurate and efficient recognition performance.


\section{Datasets and evaluation metrics}\label{sec4}
This section outlines the datasets utilized in this study, detailing their composition, diversity, and relevance to the task of \ac{FV}, \ac{PV} and \ac{DHV} recognition. Additionally, the section discusses the evaluation metrics employed to rigorously measure the effectiveness of our proposed \ac{DL} approaches. These metrics provide a comprehensive understanding of model accuracy, robustness, and real-world applicability.

\subsection{Overview of hand vein biometric datasets}
Vein biometric datasets are essential for the development, testing, and evaluation of hand vein recognition systems. These datasets provide the necessary data to train and validate algorithms, ensuring they perform accurately and reliably under various conditions. In this section, the paper review the most commonly used datasets in \ac{DL}-based approaches, focusing on \ac{FV}, \ac{PV}, and \ac{DHV} images, as summarized in Table \ref{tab:VeinDatasets}. These datasets are developed by the research community and are generally collected using self-designed and low-cost devices and different collection protocols. Consequently, they encompass a range of  image parameters,  including illumination, contrast, and performance results. Most of these datasets are publicly available on the internet or can be obtained by requesting them from the principal investigator.

For \ac{FV}, THUMVFV-3V and MultiView-FV are two are two datasets  specifically designed for multi-view recognition problems, offering data from various angles to enhance recognition accuracy in rotation and translation in practical applications. Additionally, the SCUT-SFVD, ISPR, and VERA finger vein datasets are utilized for \ac{PAD}. These datasets include fake images created by printing real \ac{FV} images, making them valuable for developing and testing anti-spoofing techniques.

\ac{DL} models developed for hand vein recognition face significant challenges related to the small number of samples, which limits the training data for vein recognition tasks and, consequently, reduces the generalization performance of the models,  and hinder their effectiveness.  To address these challenges, researchers have explored two main solutions: (i) the use of pre-trained models, which leverage knowledge from large, diverse datasets, improving performance even with limited hand vein data, and (ii) the application of data augmentation techniques, which artificially increase the dataset size by creating variations of the existing data, enhancing the model's ability to generalize. Given the importance of these techniques in \ac{DL}, this paper systematically analyzed their usage across all the reviewed articles in our study, as indicated in the Tables \ref{tabNoTL}, \ref{tabTL}, \ref{tab:PADFV}, \ref{tab:PalmVein}, \ref{tab:DorsalmVein}, and \ref{tab:MultiMod}. 

Additionally, some researchers have adopted the generation of synthetic databases, such as S-PVDB and NS-PVDB for \ac{PV} recognition \cite{kilgore2017palm,salazar2021automatic}. Synthetic images can be highly effective for evaluating the performance of image processing algorithms. However, their use in biometric applications remains controversial. Synthetic databases have the advantage of avoiding the time-consuming process of real data collection and can facilitate the validation of proposed approaches on large-scale databases. Despite these benefits, synthetic datasets cannot entirely replace validation with real images. Therefore, while they help in large-scale testing, real-world data remains crucial for final validation and ensuring the robustness and reliability of hand vein recognition systems. The research conducted by Salazar-Jurado et al.  {\cite{salazar2023towards}} provides an overview of cutting-edge methods in synthetic \ac{PV} imaging for biometric purposes, examining their strengths and weaknesses.

\begin{table}[]
\caption{Summary of publicly available hand vein datasets. The quantity of subjects and samples is denoted as subjects $\times$ hands (or fingers) and samples $\times$ acquisition sessions, respectively.}
\scriptsize
\label{tab:VeinDatasets}
\resizebox{\textwidth}{!}{ 
\begin{tabular}{m{0.3cm}m{2cm}m{1.5cm}m{1.2cm}m{1.2cm}m{1.5cm}m{1.5cm}m{0.9cm}}
\hline
VM & Dataset & Total images & Subjects & Samples & Images sizes &  SIs & Ref\\ \hline
\multirow{16}{*}{\rotatebox{90}{\ac{FV}}}
& FV-USM & 5904 & 123 $\times$ 4 & 6 $\times$ 2 & 640 $\times$ 480 & >15 days & \cite{asaari2014fusion} \\ \cline{2-8}
&HKPU-FI & 6264 & 156 $\times$ 2 & 6 $\times$ 2 & 513 $\times$ 256& 1 to 6 months & \cite{kumar2011human} \\ \cline{2-8}
& MMCBNU\_6000  &  6000 &  100 $\times$ 6   & 10 $\times$ 1 & 480 $\times$ 640 &  -- & \cite{lu2013available} \\ \cline{2-8}
& SDUMLA-HMT & 3816 & 106 $\times$ 6  & 6 $\times$ 1  & 320 $\times$ 240 &  -- & \cite{yin2011sdumla} \\ \cline{2-8}
& UTFVP  & 1,440 & 60 $\times$ 4 & 2 $\times$ 2 & 672 $\times$ 380 &   15 days & \cite{ton2013high} \\ \cline{2-8}
& THU-FVDT1 & 1760 & 220 $\times$ 1 & 4 $\times$ 2 & 720 $\times$ 576 &  12 sec & \cite{yang2014comparative}  \\ \cline{2-8}
& THU-FVFDT2 & 1220 & 610 & 1 $\times$ 2 &  720 $\times$ 576 &  3 to 7 days & \cite{yang2014comparative} \\ \cline{2-8}
& THU-FVFDT3 & 9760 & 610 & 16 $\times$ 2 & 720 $\times$ 576 &   & \cite{yang2014comparative} \\ \cline{2-8}
&THUMVFV-3V & 23,760 & 180 $\times$ 4  & 6 $\times$ 2 &  & 30-106 days  & \cite{zhao2024vpcformer} \\ \cline{2-8}
&MultiView-FV& 6480 & 135 $\times$ 4  & 12 $\times$ 2 & 1280 $\times$ 1024 & 30-106 days  & \cite{lin2021finger} \\ \cline{2-8}
& NUPT-FPV* & 16800 & 140 $\times$ 6 & 10 $\times$ 2 & 300 $\times$ 400 &  -- & \cite{ren2022dataset} \\ \cline{2-8}
& PLUSVein-FV3 & 7200 & 60 $\times$ 6 & 20 $\times$ 1 & 1280 $\times$ 1024 &  -- & \cite{kauba2018focussing} \\ \cline{2-8}
& SCUT-FV & 61344 & 568 $\times$ 1 & 108 $\times$ 1 & 640 $\times$ 288 &  -- & \cite{tang2019finger} \\ \cline{2-8}
& SCUT-SFVD & 7200\(^{a}\)  & 14 $\times$ 6 & 6 $\times$ 1 & 158 $\times$ 467 &   & \cite{qiu2018finger} \\ \cline{2-8}
& ISPR & 5820\(^{b}\)& 33 $\times$ 10 & 10 $\times$ 1 & 640 $\times$ 480&  -- & \cite{nguyen2013fake} \\ \cline{2-8}
& VERA FingerVein  & 880\(^{c}\)  & 110 $\times$ 2 & 2 $\times$ 1  & 250 $\times$ 665 &  -- & \cite{tome20151st, tome2014vulnerability} \\ \hline

\multirow{9}{*}{\rotatebox{90}{\ac{PV} }} & CASIA & 7,200 & 100 $\times$ 2 &  3 $\times$ 2 & 768 $\times$ 576 &  30 days & \cite{hao2008multispectral}  \\ \cline{2-8}
 & VERA PalmVein & 2,200 & 110 $\times$ 2 & 5 $\times$ 2 & 480 $\times$ 680 & 5 minuts & \cite{tome2015palm}  \\ \cline{2-8}
& PUT & 1,200 & 50 $\times$ 2 & 4 $\times$ 3 & 1,280 $\times$ 960 &  7 days & \cite{kabacinski2011vein}   \\ \cline{2-8}
& PolyU & 6,000 & 250 $\times$ 2  & 6 $\times$ 2 & 352 $\times$ 288 &  9 days &  \cite{zhang2009online} \\ \cline{2-8}
& Tonji & 12,000 & 300 $\times$ 2 & 10 $\times$ 2 & 800 $\times$ 600&  61 days & \cite{zhang2018palmprint} \\ \cline{2-8}
& IITI & 2,220 & 185 $\times$ 2 & 6 $\times$ 1 & 2,592 $\times$ 1944 &  -- & \cite{bhilare2018single} \\ \cline{2-8}
& FYO* & 640 & 160 $\times$ 2 & 1 $\times$ 2 & 800 $\times$ 600 &  10 minutes & \cite{toygar2020fyo} \\ \cline{2-8}
& S-PVDB & 12,000 & 10,000 $\times$ 1 & 6 $\times$ 1 & 128 $\times$ 128 &  -- & \cite{kilgore2017palm} \\ \cline{2-8}
& NS-PVDB & 12,000 & 2,000 $\times$ 1 & 6 $\times$ 1  & 128 $\times$ 128 &  -- & \cite{salazar2021automatic} \\ \cline{2-8}
\hline
\multirow{4}{*}{\rotatebox{90}{\ac{DHV}}}
 & NCUT & 2040 & 102 $\times$ 2 & 10 $\times$ 1 &  640 $\times$ 480&  -- & \cite{wang2009gradient} \\ \cline{2-8}
 & Bosphorus & 1575 & 100 $\times$ 1 & 3 $\times$ 4 & 300 $\times$ 240 &  -- & \cite{yuksel2011hand} \\ \cline{2-8}
 & Badawi's dataset & 5000 & 500 $\times$ 2 & 5 $\times$ 1 &  320 $\times$ 240 &  -- & \cite{badawi2006hand} \\ \cline{2-8}
 & JLU & 3680 & 736 $\times$ 1 & 5 $\times$ 1  & 320 $\times$ 240&  -- & \cite{liu2020recognition} \\ \cline{2-8}
 & DHVI & 1782 & 251 $\times$ 2 & 4/3 $\times$ 2 & 752 $\times$ 560 & 2 months  & \cite{wilches2020database}\\
\hline
\end{tabular}
}
\begin{flushleft}
Abbreviation: Vein mode (VM), session intervals (SIs), (*) : Means dataset used in many modes.

\(^{a}\)(3600 real,  3600 fake): Fake images are generated from each real image by printing them onto two aligned and stacked overhead projector films.

\(^{b}\)(3300 real, 2520 fake): Fake images are created by printing 56 real images using three printers, each set at three different z-distances.

\(^{c}\)(440 real, 440 fake): Fake images are generated by printing real images from 50 subjects using a laser printer.
\end{flushleft}
\end{table}

\subsection{Evaluation metrics}

A variety of evaluation metrics have been employed in assessing biometric recognition systems focusing on finger, hand, palm, and multi-modal approaches. While some metrics, such as \ac{PSNR}, \ac{SSIM}, \ac{AUC}, \ac{Acc}, \ac{Rec}, \ac{Pre}, and \ac{F1}, are commonly used for general image assessment and \ac{DL} tasks, their definitions can be found in \cite{kheddar2024deep,kheddar2023deep,habchi2023ai}. Other specific metrics tailored to hand vein-based \ac{DL} assessment are summarized in Table \ref{tab:metrics}.

\begin{longtable}{m{30mm} m{50mm} m{90mm}}
\caption{An overview of the evaluation measures utilized for appraising \ac{DL} techniques applied to hand vein recognition.}  \label{tab:metrics} \\
\hline
Metric & Formula & Description   \\ 
\hline
\endfirsthead
\multicolumn{3}{c}{Table \thetable\ (continued)} \\
\hline
Metric & Formula & Description  \\\hline 
\endhead
\hline
\endfoot
\hline \hline
\endlastfoot
\Ac{FAR}  &  $\mathrm{\frac{{FP}}{FP + TN}}.100$  &   This refers to instances where unauthorized individuals are mistakenly allowed access. Adjusting the system's threshold gradually shifts security levels from high to low, causing \ac{FAR} to increase from 0 to 1. \Ac{FAR} is called also \ac{FMR}. \\ \hline

\Ac{FRR}  & $\mathrm{\frac{{FN}}{TP + FN}}.100$    &  It measures legitimate users denied access due to unrecognized vein patterns. Crucial for system reliability, it influences user experience and security by evaluating hand vein biometric authentication effectiveness. When \ac{FAR} increase from 0 to 1, \ac{FRR} decrease from 1 to 0. \ac{FRR} also referred to as \ac{FNMR}.\\ \hline

\Ac{EER} & $\mathrm{\frac{{FAR + FRR}}{2}}.100$ & Acts as a measure of the equilibrium between security and convenience within a biometric authentication setup. It serves as the primary safety gauge for the verification system. It also quantifies the overall error rate when referred to as \ac{HTER}. \\\hline

\Ac{GAR} & $\mathrm{\frac{TP}{TP + FN}}.100$ & It represents the rate at which legitimate users are correctly granted access by the biometric authentication system. It could also be expressed as $\mathrm{GAR = 1 - FRR}$. \Ac{GAR} also referred to as \Ac{CIR}.  \\ \hline

 
\Ac{FDR} & $\mathrm{\frac{FP}{FP + TP}}.100$ &  Refers to the rate at which the system incorrectly identifies a non-matching vein pattern as a match.  \\\hline

\Ac{APCER} & \(\displaystyle  \mathrm{1 - \Big(\frac{1}{I_{fake}}\Big)\sum_{i=1}^{I_{fake}} \text{Res}_i}
 \) & The \ac{APCER} denotes the rate of error in misclassifying a spoof attack image (a fake image) as a genuine one (a real image). $I_{fake}$ is the number of fake images. ${Res}_i$ are the responses from authentication system. ${Res}_i$ equals 0 if fake image wrongly classified as real, 1 if correctly classified as fake for input fake image $i$. \\\hline
\Ac{BPCER} &  \(\displaystyle \mathrm{\Big(\frac{1}{I_{real}}\Big)\sum_{i=1}^{I_{real}} \text{Res}_i}
 \) & This refers to the error rate of misclassifying a genuine image as a fake one. Also known as \ac{NPCER}. ${Res}_i$ becomes 1 if real image wrongly classified as fake, 0 if correctly classified as real. \Ac{ACER} represents the mean of both \ac{APCER} and \ac{BPCER}.\\\hline
\Ac{FID} & \(\displaystyle  \lVert \mu_X - \mu_Y \rVert^2 + \newline \text{Tr}\Big(\Sigma_X + \Sigma_Y - 2.(\Sigma_X \Sigma_Y)^{\frac{1}{2}}\Big)
\) & Evaluate the extent of contrast between the GAN-generated image and the authentic real image. $X$ and  $Y$ are the feature distribution of real and synthetic vein patterns, respectively,  
$\mu_X$  and  $\mu_Y$  are the mean feature vectors of  $X$  and  $Y$ respectively, 
$\Sigma_X$ and   $\Sigma_Y$  are the covariance matrices of  $X$  and  $Y$  respectively,
$Tr$  represents the trace operator, and  $\lVert \cdot \rVert^2$ denotes the squared Euclidean norm.
 \\\hline
\Ac{WD} & $\Big( \inf\limits_{\substack{\pi \in \Pi(P;Q)}}  \int_{\mathbb{R}^d \times \mathbb{R}^d} \lvert X - Y \rvert^p \, d\pi \Big)^{\frac{1}{p}}
$ & In the context of biometric vein recognition, it can be used to quantify the dissimilarity between two sets of vein patterns. \(P\) and \(Q\) represent the probability distributions of vein patterns from two different individuals, and $X$ and $Y$ represent feature vectors extracted from vein patterns of individuals. \\\hline


\text{Similarity}($\mathbf{x}, \mathbf{y}$) & $\cos(\theta) = \frac{\mathbf{x}^T \cdot \mathbf{y}}{\lVert \mathbf{x} \rVert \cdot \lVert \mathbf{y} \rVert}$ &  The cosine similarity score is crucial for calculating the matching score between feature vectors \( x \) and \( y \), distinguished by angle \( \theta \). It facilitates hand vein pattern recognition by comparing feature vectors' angles \cite{obayya2020contactless}. \\ \hline
\end{longtable}

\section{DL in hand vein biometrics} \label{sec5}
This section reviews \ac{SOTA} \ac{DL} approaches in hand vein biometrics recognition, focusing on \ac{FV}, \ac{PV}, and \ac{DHV} identification. It examines recent advancements and methodologies proposed for enhancing biometric accuracy and robustness. By exploring various neural network architectures and training techniques, this section aims to highlight significant contributions and breakthroughs in the field.

\subsection{\ac{FV}-based methods}

\ac{FV} recognition has gained significant attention through the use of \ac{DL}. Several approaches have been proposed in this field, with a focus on applying \ac{DL} in preprocessing steps to enhance the quality of input images or in feature extraction to automatically learn discriminative features. Additionally, \ac{DL} methods have been employed to protect biometric template and  to detect  presentation attacks, such as spoofing with fake \acp{FV}.

\subsubsection{Preprocessing-based methods}\label{DLPrep-based}
In the preprocessing step, \ac{DL} models have been developed to automatically assess the quality of \ac{FV} images, enabling the selection of the most reliable samples for feature extraction. Furthermore, image enhancement and restoration techniques based on \ac{DL} have been proposed to improve the clarity and contrast of \ac{FV} images, enhancing the overall recognition accuracy.  Tables \ref{tabNoTL} and \ref{tabTL} summarize proposed schemes identified by their \textbf{\textit{objective}} fields, including preprocessing techniques for image quality assessment, enhancement and restoration. 

\begin{enumerate}[leftmargin=12pt,label=\arabic*)]
    \item \textbf{Image quality assessment:}
Qin and El-Yacoubi \cite{qin2015finger, qin2017deep} were the first to apply \ac{DL} for \ac{FV} quality assessment. They proposed training a \ac{CNN} on binary images to predict vein quality, aiming to minimize verification errors. Low and high-quality images are automatically labeled, assuming low-quality images are false rejected \ac{FV} images. Experimental outcomes show the proposed method's efficacy in achieving high performance in a public database.
Zeng et al. \cite{zeng2018finger} utilized a light-\ac{CNN} approach for assessing image quality, specifically to determine if it contains well-defined and consistent vein patterns. They segmented the image into multiple blocks and calculated average results from these block images to assess overall quality. This light-\ac{CNN} architecture reduces computation time while also extracting strong features to represent the quality of \ac{FV} images.
In \cite{ren2022high}, a \ac{CNN}-based method for evaluating \ac{FV} image quality was introduced to improve recognition performance and reduce false rejection of low-quality images. The method uses statistical analysis to compare different finger samples, automatically determining and classifying image quality. A lightweight \ac{CNN} is then used to train images classified as high or low quality based on the evaluation criteria. By identifying common attributes among low-quality images, the system can efficiently assess image quality in practical settings. Experimental results demonstrate the method's excellent performance on three public datasets, showing its effectiveness in enhancing the original recognition system's performance.

\item \textbf{Image restoration and enhancement:}
Guo et al. \cite{guo2019image} proposed an image restoration approach to enhance the integrity of venous networks. The method involves an adaptive threshold method to detect the unknown incomplete regions in \ac{FV} images. Subsequently, a \ac{CAE} network model is employed to restore the \ac{FV} images. Experimental results demonstrate the effectiveness of the approach in improving vein network integrity in \ac{FV} images. \ac{CAE} was also used for \ac{FV} image enhancement \cite{bros2021vein}, where a residual \ac{CAE}  architecture is trained in a supervised manner to enhance vein patterns in near-\acp{IRI}. The method has been evaluated on several databases, showing promising results on the UTFVP database as a main outcome. In \cite{choi2020modified}, authors proposed a method using a modified conditional \ac{GAN} to restore optically blurred \ac{FV} images. The \ac{GAN} includes a generator with an encoder and decoder, along with a discriminator. The restored images were then used for \ac{FV} recognition using a deep \ac{CNN}, leading to improved recognition performance. Similarly, a \ac{GAN}-based restoration model was proposed in \cite{yang2020finger} to recover missed patterns in \ac{FV} images. The results show that the proposed method restores missed vein patterns and reduces the \ac{EER} of the \ac{FV} verification system. Jiang et al. \cite{jiang2022finger} proposed a finger
vein restoration algorithm based on neighbor binary Wasserstein \ac{GAN} (NB-WGAN). The method uses texture loss as part of the loss function of the
generator to recover more vein texture details.
In recent work \cite{hong2024deep}, \acp{GAN} were used to restore \ac{FV} images by addressing multiple degradation factors like non-uniform illumination and noise. The approach classified the degradation factor using a task adaptor and adaptively restored the image based on this factor through task channel-wise attention. Gated fusion was employed to merge features from four points in the generator, simultaneously considering feature characteristics from low to high levels. \Ac{DRL} approach enables automatic decision-making by considering the environment's state, the desired actions, and the rewards obtained, is used with \ac{GAN} for \ac{FV} image restoration \cite{gao2023drl}. This model trains an agent to select restoration behaviors based on the image state, enabling continuous restoration. The tasks are divided into deblurring, defect restoration, and denoising/enhancement. For deblurring, a \ac{FV} deblurring \ac{GAN} based on DeblurGAN-v2 with an Inception-ResNet-v2 backbone is used. For defect restoration, a \ac{FV} feature-guided restoration network is proposed with two stages: feature image restoration and original image restoration. Experimental results show reduced \ac{EER} for single and multiple image restoration problems.

Lei et al. \cite{lei2019finger} proposed a model based on the \ac{PCNN} to enhance the quality of \ac{FV} images. They developed a parameter setting scheme to automatically adjust the parameters in the \ac{PCNN} model, eliminating the need for empirical correlations or training. The results demonstrate that the \ac{PCNN} can produce enhanced \ac{FV} images with higher quality, thereby improving the efficiency of \ac{FV} image recognition.
Du et al. \cite{du2021fvsr} proposed  FVSR-Net architecture, an end-to-end \ac{CNN} designed for restoring \ac{FV} scattering images. The network utilizes a multi-scale \ac{CNN} for the task of \ac{FV} scattering removal, aiming to extract clear vein backbone features. Experimental results demonstrate that the proposed method achieves better visual and recognition performance.

\end{enumerate}

\subsubsection{Features extraction-based methods}

\ac{DL} methods for feature extraction have demonstrated significant advancements in capturing discriminative features from \ac{FV} images. These approaches utilize \ac{DL} models to learn hierarchical representations of \ac{FV} patterns, enhancing their potential for matching and identification tasks. Deep features extracted through these methods have exhibited superior performance compared to traditional handcrafted features, particularly in terms of recognition accuracy. The \ac{DL} approaches discussed in this section focus either on feature extraction alone or on both feature extraction and matching processes.

Due to the robust feature learning capabilities of \ac{DL}, a majority of the proposed approaches for \ac{FV} analysis fall into this category, accounting for 67 articles in our case. To enhance organization and readability, these approaches are categorized based on the specific \ac{DL} techniques employed.  Tables \ref{tabNoTL} and \ref{tabTL} summarize proposed schemes for feature extraction, without and with use of \ac{DTL}, respectively.


\scriptsize
\begin{center}
\begin{longtable}[t]{llm{2cm}m{2cm}m{3cm}m{4cm}m{4cm}}

\caption{Summary of proposed works in preprocessing, feature extraction and matching-based \ac{DL} for \ac{FV} recognition without \ac{DTL}. Some studies have employed specific data augmentation techniques to achieve the reported results. In cases where multiple scenarios are examined, only the top-performing outcome is mentioned.}
\label{tabNoTL}\\
\hline
Ref. & Year & Obj. & DL approach & Data augmentation & Dataset & Results \\ \hline

\endfirsthead
\multicolumn{6}{c}{Table \thetable\ (Continue)} \\
\hline
Ref. & Year & Obj. & DL approach  & Data augmentation & Dataset & Results \\ \hline
\endhead
\hline
\endfoot
\hline \hline
\endlastfoot

\cite{qin2015finger} & 2015  &  & CNN+P-SVM & -- & HKPU-FI , FV-USM & Acc=88.99\%, 74.98\% \\

\cite{zeng2018finger} & 2018  & Image quality assessment & Lightweight CNN & -- & MMCBNU\_6000, SDUMLA-HMT & Acc=71.95\%, 74.63\% \\

\cite{ren2022high} & 2022  &  & CNN & Low quality images generation by applying cropping and mirroring & SDUMLA-HMT, MMCBNU\_6000, FV-USM & Acc=81.97\%, 75.27\%, 73.85\% \\ \hline

\cite{guo2019image} & 2019  &  & CAE & -- & Self-built dataset & EER = 0.0016\% \\

\cite{bros2021vein} & 2021  &  & CAE & Images random horizontal flip, rotation, translation and shear & UTFVP, SDUMLA-HMT & HTER=1.0\%, 9.8\% \\

\cite{choi2020modified} & 2020  &  & GAN+CNN & Images shifting by 1-3 pixel (up, down, left, right directions), applying Gaussian blur & SDUMLA-HMT, HKPU-FI & EER= 3.934\%, 2.746\% \\

\cite{yang2020finger} & 2020  & Image restoration and enhancement & GAN & -- & MMCBNU\_6000, FV-USM & EER = 5.66\%, 2.37\% \\

\cite{jiang2022finger} & 2022  &  & NB-WGAN & -- & Self-built dataset & PSNR= 52.085dB; FRR= 30.77\% (when FAR= 0\%) \\

\cite{gao2023drl} & 2023  &  & DeblurGAN-v2+DRL & -- & MMCBNU\_6000, SDUMLA-HMT, FV-USM, UTFVP & PSNR=41.18dB, 44.26dB, 53.49dB, 49.85dB; EER\(\simeq\) 6.5\%, 4.3\%, 4.8\%, 3.7\%  \\

\cite{lei2019finger} & 2019  &  & PCNN & -- & Self-built (NJUST-FV) \& SDUMLA-HMT, THU-FV, HKPU-FI & EER \(\simeq\) 4.7\%, 5.1\%, 0.2\%, 5.9\%  \\

\cite{du2021fvsr} & 2021  &  & CNN & -- & Self-built dataset, SDUMLA-HMT & PSNR= 15.95dB, 13.59dB; EER= 0.11\%, 8.39\% \\

 \hline


\cite{qin2017deep2}&  2017 & & CNN &-- &HKPU-FI, FV-USM & EER = 2.70\%, 1.42\% \\

\cite{das2018convolutional} & 2018 &  & CNN  & -- & HKPU-FI, FV-USM, SDUMLA-HMT, UTFVP   & Acc=96.55\%, 98.58\%, 97.48\%, 95.56\%  \\

\cite{lian2023fedfv}& 2023 &  & CNN + FL  & -- & MMCBNU\_6000, HKPU-FI, PLUSVein-FV3, SDUMLA-HMT, THU-FVFDT3, FV-USM, UTFVP, VERA FingerVein,  SCUT-FV
   & EER=0.35\%, 0.48\%, 0.76\%, 1.52\%, 2.09\%, 0.07\%, 1.62\%, 4.54\%, 0.82\%  \\

\cite{mu2024pafedfv}& 2024 &  & CNN + FL & -- & FV-USM, HKPU-FI, NUPT-FPV,
SDUMLA-HMT, UTFVP, VERA FingerVein  & EER=0.68\%, 0.97\%, 1.10\%, 3.73\%, 1.62\%, 9.09\%  \\

\cite{zhang2019adaptive} & 2019 &  & Gabor CNN &-- & MMCBNU\_6000, FV-USM, SDUMLA-HMT  & EER=0.11\%, 0.57\%, 1.09\% \\

\cite{chang2023design}& 2023 & 
& Lighweight CNN  &Images shifting and rotating  &Self-built dataset, SDUMLA-HMT & Acc=95.82\%; EER=4.17\%; Inference time=0.356 s \\

\cite{liu2023deep}& 2023 & & CNN &  -- & HKPU-FI, FV-USM & Acc=92.11\%, 94.17\%\\

\cite{huang2017deepvein} & 2017 & &CNN & -- & Self-built dataset, MMCBNU\_6000,  SDUMLA-HMT, FV-USM& EER= 2.04\%, 0.17\%, 0.87\%, 0.50\%   \\

\cite{liu2021finger}& 2021 & &CNN & -- &  SDUMLA-HMT, FV-USM,  MMCBNU\_6000 & EER= 2.29\%, 0.47\% \\

\cite{boucherit2022finger}&2021& & Merge CNN & --& FV\_USM, SDUMLA-HMT, THU-FVFDT2 & Acc=96.15\%, 99.48\%, 99.56\% \\

\cite{fang2018novel} & 2018 &  & Lightweight CNN & --&MMCBNU\_6000, SDUMLA-HMT & EER=0.10\%, 0.47\% \\

\cite{XIE2019148} & 2019 &  & Lightweight CNN & -- & HKPU-FI & EER=0.097\%\\

\cite{shen2021finger}& 2021 &  & Lightweight CNN & --&  SDUMLA-HMT, PKU-FVD & Acc= 99.3\%, 99.6\%;  EER=1.13\%, 0.67\%\\

\cite{zhao2020finger} & 2020 & & Lightweight CNN &-- & MMCBNU\_6000, FV-USM & Acc=99.05\%, 97.95\%; EER= 0.503\%, 1.070\%\\

\cite{zhang2023convolutional}& 2023 &  & Lightweight CNN (LModel) &-- & HKPU-FI &Acc=99.13\%\\ 

\cite{wang2023residual}&2023&  & Residual Gabor CNN  & Grayscale normalization and linear mixing of \ac{ROI} fine FVIs image&MMCBNU\_6000, FV\_USM, UTFVP, SDUMLA-HMT, self-built dataset&Acc=99.67\%, 100\%, 97.22\%, 98.98\%, 100\%; EER=0.08\%, 0.00\%, 0.55\%, 0.31\%, 0.0003\% \\

\cite{li2019finger}&2019& & CNN&  & Self-built dataset, SDUMLA-HMT & Acc=100\%, 98.78\%; EER=0.21\%, 2.82\%\\

\cite{wang2019multi}&2019&  & CNN& &   SDUMLA-HMT & EER=0.30\%\\

\cite{liu2021improved}& 2018 & &CNN + RAB  &--  &FV-USM, MMCBNU\_6000& Acc= 98.58\%, 97.54\%   \\
\cite{sulaiman2022attention}&2022&  & CNN+Attention  & --&SDUMLA-HMT& Acc=100\%\\

\cite{huang2021joint}&2022&   & CNN+Attention  &-- &  SDUMLA-HMT, MMCBNU\_6000, FV\_USM, Self-built dataset & EER=0.35\%, 0.08\%, 0.34\%,  0.49\% \\

\cite{zhang2022convolutional}&2022&  & Lightweight CNN+CBAM & --& FV\_USM, HKPU-FI & Acc=100\%, 100\% \\

\cite{liu2023mmran}&2023& & CNN+Residual AM &Luminance
transformation and gaussian noise&SDUMLA-HMT, FV\_USM, MMCBNU\_6000, UTFVP, HKPU-FI & EER=0.44\%, 0.05\%, 0.01\%, 1.21\%, 0.18\%\\

\cite{huang2023axially}&2023& Feature extraction  & CNN+Attention  & --& SDUMLA-HMT, HKPU-FI, NUPT-FPV   & Acc=94.50\%, 97.86\%, 98.92\%; EER=0.53\%, 0.32\%, 0.14\%\\

\cite{huang2023fvfsnet}&2023&  & CNN+Attention  &  Random color jitter, translation and perspective& FV\_USM, MMCBNU\_6000, THU-FVDT2, SDUMLA-HMT, HKPU-FI, PLUSVein-FV3, UTFVP, VERA FingerVein, SCUT-FV&EER=0.20\%, 0.18\%, 2.15\%, 1.10\%, 0.81\%, 1.32\%, 2.08\%, 6.82\%, 0.83\%\\

\cite{lu2021novel}&2021&& ViT &--  & FV\_USM, SDUMLA-HMT, MMCBNU\_6000 & Acc=93.50\%, 91.75\%, 91.84\% \\

\cite{huang2022fvt}&2022&  & ViT &  --& FV\_USM, MMCBNU\_6000, SDUMLA-HMT, HKPU-FI & EER=0.44\%, 0.92\%, 1.50\%, 2.37\% \\

\cite{li2023fv}&2022&  & ViT & Images cropping, rotation and dropout &  FV\_USM, SDUMLA-HMT & Acc=99.73\%, 92.77\%; EER= 0.042\%, 1.033\% \\

\cite{lu2022finger}&2022&  & T2T-ViT &--& FV\_USM, SDUMLA-HMT & Acc=94.85\%, 99.48\%; EER= 2.46\%, 0.94\% \\

\cite{zhao2024vpcformer}&2022&  & ViT & -- & Self-built dataset (THUMVFV-3V)& Acc=99.79\%\\

\cite{chen2016finger} & 2016 & 
 & DBN & -- & Self-built dataset & Acc=96.9\% \\

\cite{hou2019convolutional} & 2020 &  & CAE+SVM &  Image flip, rotation, shift,
shear, and zoom & FV-USM, SDUMLA-HMT &  Acc=99.95\%, 99.78\%; EER=0.12\%, 0.21\% \\

\cite{jalilian2018finger}& 2018 &  & CAE & -- &  UTFVP, SDUMLA-HMT &  Acc=99.95\%, 99.78\%; EER=2.17\%, 2.87\% \\

\cite{hou2018convolutional} &2018&  & CAE+CNN & --& FV-USM & Acc=99.49\%; EER=0.16\% \\

\cite{qin2019finger} &2019&  & SCNN+LSTM &--& HKPU-FI & EER=2.38\% \\

\cite{zhang2019gan}&2019&  & FCGAN+CNN & GAN & HKPU-FI & EER=0.87\% \\

\cite{hou2022triplet}&2019&  & GAN+CNN & GAN & Self-built dataset, FV\_USM, SDUMLA-HMT, HKPU-FI & EER=0.06\%, 0.03\%, 0.05\%, 0.15\% \\

\cite{li2022vit}&2022& & ViT+CapsNet&-- & MMCBNU\_6000, SDUMLA-HMT, FV\_USM, HKPU-FI & Acc=97.52\%, 93.24\%, 98.68\%, 95.61\%; EER=0.63\%, 1.3\%, 0.28\%, 1.66\% \\

\cite{kamaruddin2019new}&2019&  & PCANet & --&  FV\_USM, SDUMLA-HMT, THU-FVFDT2 & Acc=99.49\%, 98.19\%, 100\% \\

\cite{yang2019fv}&2018&  & CycleGAN & -- & THU-FVFDT2, SDUMLA-HMT  & EER=1.12\%, 0.94\% \\

\cite{muthusamy2022steepest}&2022& & SDBCCML & -- &SDUMLA-HMT & Acc=94\%; F-Score=96\%; Time complexity=66ms \\

\cite{muthusamy2022trilateral} &2022&  & TFHFT-DPFNN & --&SDUMLA-HMT & PSNR=58.58; Acc=98\%; Time complexity=60ms \\

\cite{song2022eifnet}&2023&  & FNN &Modified Mixup Method \cite{zhang2017mixup}& FV\_USM, SDUMLA-HMT, MMCBNU\_6000 & EER=0.10\%, 0.06\%, 0.24; Time complexity=32.24ms \\

\cite{wan2022optimization}&2022&  & Deep Forest &-- &SDUMLA-HMT& Acc=98.40\%; EER=2.8\% \\

\cite{qin2023ag}&2024& & GRU+attention +NAS& --& Self-built dataset, HKPU-FI, MMCBNU\_6000& Acc=97.33\%, 97.92\%, 98.29\%; EER=1.25\%, 0.99\%, 1.00\%\\

\hline
\end{longtable}
\end{center}

\begin{center}
\begin{longtable}{llm{1.5cm}m{1.8cm}m{2.5cm}m{2.5cm}m{3cm}m{3cm}}
\caption{Summary of proposed works in preprocessing, feature extraction and matching-based \ac{DL} for \ac{FV} recognition with \ac{DTL}. Some studies have employed specific data augmentation techniques to achieve the reported results. In cases where multiple scenarios are examined, only the top-performing outcome is mentioned.}

\label{tabTL}\\

\hline
Ref. & Year & Obj. & DL approach & DTL & Data augmentation & Dataset & Results \\ \hline

\endfirsthead
\multicolumn{6}{c}{Table \thetable\ (Continue)} \\
\hline
Ref. & Year & Obj. & DL approach & DTL & Data augmentation & Dataset & Results \\ \hline
\endhead
\hline
\endfoot
\hline \hline
\endlastfoot

\cite{qin2017deep} & 2018  & Image quality assessment &  CNN+P-SVM & Pretraining and fine-tuning between datasets& -- & FV-USM \& HKPU-FI \& MMCBNU\_6000  \& SDUMLA-HMT & Acc=73.57\%, 87.08\% \\ \hline

\cite{hong2024deep} & 2024  & Image restoration and enhancement & GAN+CNN & Domains adaptation with DenseNet-161, and  ConvNeXt-small & Images Cropping and 4-way translation & SDUMLA-HMT, HKPU-FI & FID= 15.09, 26.63; WD=2.89, 3.22; EER = 4.72\%, 1.73\% \\ \hline

{\cite{huang2024towards}}&2024&  & VGG Net-16+ELM&DTL form source dataset to  target dataset using ELM & Images cropping, scaling, brightness and
contrast transformation, masking, and rotation &SDUMLA-HMT, THU-FVFDT2, HKPU-FI   & Acc= 99.21\%, 99.51\%, 98.62\%, 99.88\\

\cite{zhang2023finger}&2023&   & ResNet+Attention  & Domains adaptation & FV\_USM, SDUMLA-HMT, THU-FVFDT3   & Acc=99.90\%, 100\%, 100\%\\

\cite{ma2023finger}&2023&  & EfficientNetV2+ Attention & Domains adaptation & --& FV\_USM, SDUMLA-HMT&Acc=98.96\%, 97.29\%\\

\cite{lin2021finger}&2021&  & CAE+Siamese CNN&Pretrained weights from
ImageNet dataset  & --& MultiView-FV, SDUMLA-HMT and MMCBNU\_6000 & EER=1.69\%, 0.47\%, 0.1\% \\

\cite{tang2019finger}& 2019 & & Lighweight CNN & Knowledge distillation from pretrained-weights ResNet-50 & Images gamma transformation, shear, translation,
rotation, enlargement and channel color shifting &Self-built dataset (SCUT-FV),  SDUMLA-HMT, FV-USM, HKPU-FI & Acc= 97.90\%, 99.25\%, 99.79\%, 97.90\%;  EER= 3.97\%, 1.53\%, 0.25\%, 1.30\%   \\

\cite{song2019finger}& 2019 &  & DensNet & Fine-tuning pretrained DensNet-161 & Images translation& SDUMLA-HMT, HKPU-FI & EER = 2.35\%, 0.33\% \\

\cite{tao2021dglfv} & 2021 &  & InceptionResNetV2 & Pretrained weights from COCO dataset &-- & SDUMLA-HMT, Self-built dataset   & Acc=99.25\%, 99.08\% \\

\cite{tran2023anti}& 2023 & 
 & Densenet-161 & Pretrained weights from ImageNet dataset &-- &FV-USM, SDUMLA-HMT, THU-FVFDT2 & Acc=97.66\%, 99.94\%, 88.19\%; EER=2.03\%, 0.24\%, 12.61\%  \\

\cite{deshmukh2024optimized}& 2024 & 
& ResNet152V2 and MobileNetV3 & Domains adaptation & &  &
Acc=99.6\%, Precision=99.42\%\\

\cite{chai2022shape} & 2022 &  & Lightweight CNN & Fine-tuning pre-trained MobileNetV2& -- & HKPU-FI, FV-USM, SDUMLA-HMT &
Acc=93.05\%, 94.67\%, 96.61\%, EER=2.47\%, 1.37\%,
0.99\%\\

\cite{hsia2022new} & 2022 & & AlexNet & Domains adaptation &-- & FV-USM, SDUMLA-HMT & Acc= 99.1\%, 98.1\%; EER = 0.002\%, 0.003\% \\

\cite{hou2021arcvein}& 2021 & Feature extraction &ECA-ResNet & Domains adaptation  & Image
flip, rotation, shift, shear, and zoom & SDUMLA-HMT, FV-USM,  MMCBNU\_6000 & EER= 1.20\%, 0.76\%, 0.30\%   \\

\cite{chai2023vascular} & 2023 & & Lightweight CNN & Fine-tune first pretrained layers of MobileNetV2 on CIFAR dataset & --& HKPU-FI, FV-USM, SDUMLA-HMT, UTFVP   &Acc=91.03\%, 96.17\%, 95.07\%, 92.50\%;  EER=4.25\%,  1.21\%,\%,  1.27\%,  1.93\%\\

\cite{hu2018fv} & 2018 & &CNN & Fine tuning pretrained VGGFace-Net &Images projective transform, rotation and cropping & Self-built dataset,  SDUMLA-HMT, FV-USM, HKPU-FI & Acc= 97.90\%, 99.25\%, 99.79\%, 97.90\%;  EER= 3.97\%, 1.53\%, 0.25\%, 1.30\%   \\

\cite{zheng2020new} & 2020 & & Lightweight CNN (ShuffleNet V2) &Pretrained   weights from ImageNet dataset &Images random brightness, cropping, rotating and erasing&SDUMLA-HMT, FV-USM, and MMCBNU\_6000& EER= 0.37\%, 0.31\%, 0.05\%\\

\cite{ou2021fusion}&2021&  & ResNet & Pretrained   weights from ImageNet dataset & Inter-class augmentation via vertical flip and intra-class augmentation via geometric and color transformations & FV\_USM, MMCBNU\_6000, HKPU-FI & EER=0.48\%, 0.21\%, 1.90\% \\

\cite{hong2017convolutional} &  2017 &  & CNN &Fine tuning pretrained VGG Net-16 &Image translation and cropping & 2 self-built datasets, SDUMLA-HMT & EER = 0.804\%, 2.967\%, 6.115\% \\

\cite{lu2019exploring}&2019&  & CNN&Pre-trained convolutional filters from ImageNet dataset & -- & MMCBNU\_6000, SDUMLA-HMT & EER=0.74\%, 2.37\%\\

\cite{shaheed2022ds}&2022&  & CNN&Fine-tuning 
pre-trained Xception on ImageNet dataset & Image rotating, width
shifting, zooming, horizontal flipping and shearing &SDUMLA-HMT, THU-FVFDT2, FV\_USM,   & Acc=98.5\%, 90\%\\


\cite{qin2022local}&2022&  & Local attention Transformer &Fine-tuning several pretrained models & -- & Self-built dataset,  FV\_USM& Acc=98.44\% (12 views, 0$^{\circ}$  rotation), 99.80\%\\

\cite{li2024fv}&2024& & MobileViT & \ac{DA} & Mixup method \cite{zhang2017mixup} & SDUMLA-HMT, FV\_USM& Acc=99.53\%, 100.00\%; EER=0.47\%, 0.02\%; Time
complexity=10m\\

\cite{huang2021robust}&2021&  & CAE+CNN  & Fine-tuning pretrained U-Net and ResNet on ImageNet dataset& Images cropping& SDUMLA-HMT, THU-FVFDT2  & Acc=99.53\%, 98.64\% \\

\hline
\end{longtable}
\end{center}

\normalsize

\begin{enumerate}[leftmargin=12pt,label=\arabic*)]

\item \textbf{CNN:} Qin and Yacoubi \cite{qin2017deep2} introduced a segmentation model for \ac{FV} verification based on \ac{DL}. They utilized a \ac{CNN} to predict the probabilty of pixels belonging to veins or background by learning deep feature representations. Additionally, they presented a method for recovering missing \ac{FV} patterns using a  fully convolutional network to enhance recognition performance.
Das et al. \cite{das2018convolutional} introduced a \ac{DL} approach utilizing \ac{CNN} to extract robust features, achieving high accuracy regardless of image quality. Experimental evaluations on four widely used databases validated the method's robustness in feature extraction and classification under varying image qualities. 
Song et al. \cite{song2019finger} employed a deep densely connected \ac{CNN}, DensNet, for \ac{FV} recognition, utilizing composite images as input data. This approach exhibited robustness against noise and misalignment. 
Hsia et al. \cite{hsia2022new} introduced an enhanced edge detection method designed to adapt to the rotational and translational movements of detected fingers, as well as to mitigate interference from external light and other environmental factors. They proposed integrating this approach, which utilizes the \ac{CNN} AlexNet, into carputer systems.
Tran et al. \cite{tran2023anti} introduced a \ac{FV} recognition system designed for the virtual reality human-robot equipment in the Metaverse to prevent misappropriation. Their approach utilizes a \ac{CNN} and anti-aliasing technique, yielding promising results across three public datasets. Recently, \ac{FL} technique is introduced in \ac{FV} to solve the problem of small sample size and category diversity while protecting user privacy \cite{lian2023fedfv, mu2024pafedfv}. 

To enhance the capacity for learning features, metric learning, which is proposed to measure the distance between sample features, is introduced in \ac{DL}-based \ac{FV} systems \cite{huang2017deepvein, tang2019finger,XIE2019148, shen2021finger, zhao2020finger, zheng2020new, hou2021arcvein, ou2021fusion, liu2021finger}. 
Huang et al. \cite{huang2017deepvein} applied contrastive loss in a deep \ac{CNN} for \ac{FV} identification. They paired intra-class and inter-class samples as input, aiming to maximize inter-class distance and minimize intra-class distance during training. Their method used metric learning to enhance the network's ability to distinguish between different \ac{FV} patterns. Later, Tang et al. \cite{tang2019finger} further modified the contrastive loss to extract more discriminative features, pairing intra-class and inter-class samples to achieve the same goal of maximizing inter-class distance and minimizing intra-class distance.
Hou and Yan \cite{hou2021arcvein} presented the Arc-vein method, which incorporates a novel arccosine center loss into a \ac{CNN} to improve its ability to extract and recognize \ac{FV} features.  {Figure \ref{fig:arcvein} illustrates this Arc-vein method, providing a prominent example of metric learning applied to \ac{FV} verification. The diagram effectively demonstrates the complete process, from Region of \ac{ROI} extraction to feature extraction using the efficient channel attention residual network (ECA-ResNet), and finally, feature identification using the combined softmax and arccosine center loss.}
In  \cite{liu2021finger}, the authors proposed a shallow \ac{CNN} with improved interval-based loss function for efficient real-time application in both closed-set and open-set architectures of \ac{FV} recognition. 
Several approaches \cite{XIE2019148, shen2021finger, zhao2020finger} have employed the triplet loss function{, which has shown promising results. Figure \ref{fig:TripletLoss} illustrates the general mechanism of this widely used metric learning approach. It clearly shows how the function learns effective embeddings for FV images, ensuring that images from the same individual are grouped closely together in the embedding space, while images from different individuals are separated.}

\begin{figure}
    \centering
    \includegraphics[scale= 0.6]{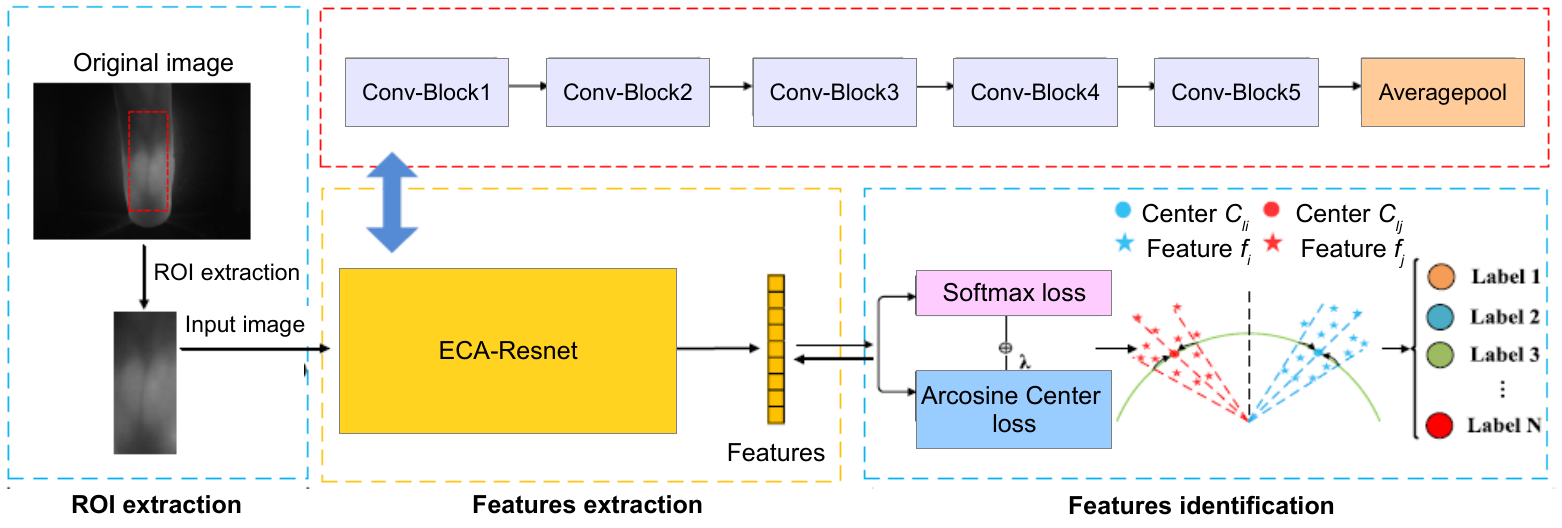}
    \caption{Arc-Vein \ac{FV} image verification framework. This approach combines the softmax loss with the newly designed arccosine center loss during model training, with the goal of enhancing \ac{FV} image verification performance. The Arc-vein framework, which includes \ac{ROI} image extraction, feature extraction using efficient channel attention residual network (ECA-ResNet), and feature identification with arccosine loss and softmax loss  \cite{hou2021arcvein}.}
    \label{fig:arcvein}
\end{figure}

\begin{figure}
    \centering
    \includegraphics[scale= 0.6]{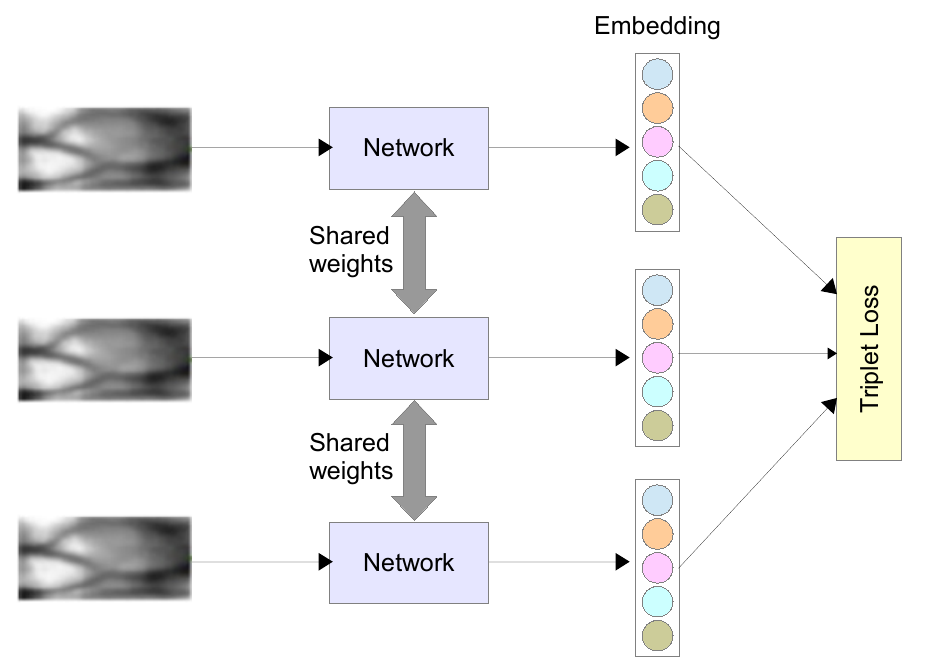}
    \caption{Mechanism of action of Triplet loss function.  The triplet loss in \ac{FV} recognition aims to learn effective embeddings for each \ac{FV} image. In the embedding space, \ac{FV} images from the same individual should be closely grouped together, forming distinct clusters. The objective of the triplet loss is to ensure that examples with the same label have their embeddings positioned closely together, while examples with different labels have their embeddings placed far apart.}
    \label{fig:TripletLoss}
\end{figure}

The increasing depth of \ac{CNN}s raises computation costs, posing challenges for deploying network architectures efficiently on various hardware platforms. Training a customized \ac{CNN} model from scratch demands substantial computing resources in terms of time, energy, finances, and environmental impact. Research in \ac{DL} aims to miniaturize neural networks, reducing model depth and parameter count. Lightweight networks focus on minimizing the size of \ac{CNN}-based \ac{FV} recognition methods while maintaining satisfactory accuracy. Several work have been proposed in this context \cite{fang2018novel, XIE2019148, tang2019finger,  shen2021finger, zheng2020new, zhao2020finger, zhang2022convolutional, chai2023vascular, zhang2023convolutional}.  For instance,  Fang et al. introduced a lightweight \ac{CNN} framework for \ac{FV} verification \cite{fang2018novel}, which overcomes the high computational demands and training sample requirements of traditional \acp{CNN}. Their approach involves a two-stream model integrating the original image with a mini-\ac{ROI} image, addressing displacement and data limitations. Leveraging features extracted from this two-stream network, they achieved \ac{SOTA} results with reduced computational costs. Similarly,  in their study, Xie et al. \cite{XIE2019148} introduced a light \ac{CNN} framework to improve \ac{FV} matching performance using enhanced images. 
Their experiments demonstrated that, when combined with a triplet similarity loss function {(as illustrated in Figure \ref{fig:TripletLoss})} and incorporating supervised discrete hashing, superior accuracy was achieved while also reducing the template size. Moreover, Shen et al.  \cite{shen2021finger}
proposed algorithm using a lightweight \ac{CNN} model in the backbone network and employs a triplet loss function to train the model. While the feature and pattern extraction is based on Min-\ac{ROI} and Gaussian filter optimization, recognition and matching is based on lightweight \ac{CNN}. 
On the other hand, Zhao et al. \cite{zhao2020finger} proposed a lightweight linear \ac{CNN} model similar to AlexNet. This model achieved exceptional performance in \ac{FV} verification by implementing dynamic regularization and center loss. {To illustrate the simplified structure of a lightweight \ac{CNN}, the architecture adopted in \cite{XIE2019148} is depicted in Figure \ref{fig:LighCNN} as an example. This architecture, with its reduced depth and complexity compared to traditional models, effectively balances accuracy and efficiency.}

\begin{figure}[h!]
    \centering
    \includegraphics[scale=0.65]{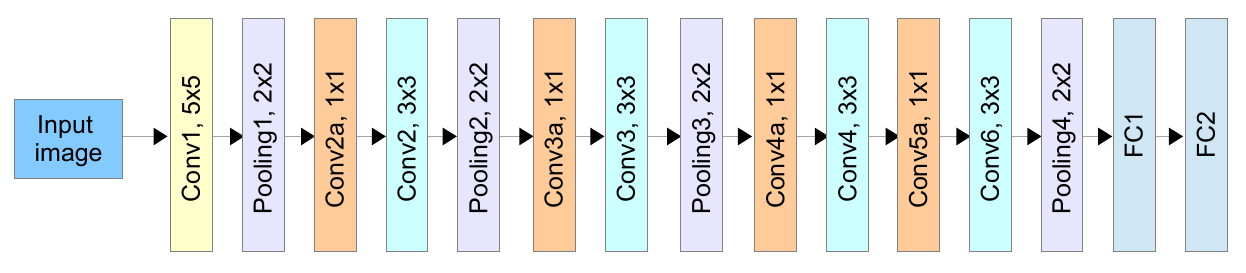}
    \caption{Example of lightweight \ac{CNN} architecture used in \cite{XIE2019148}. }
    \label{fig:LighCNN}
\end{figure}

Data augmentation technique have been introduced \ac{DL}-based \ac{FV} methods to improve the feature learning ability when the amount of \ac{FV} database is limited \cite{hong2017convolutional, hu2018fv,song2019finger, tang2019finger, zheng2020new, ou2021fusion, hou2021arcvein, shaheed2022ds, wang2023residual, chang2023design, liu2023mmran, huang2023fvfsnet, huang2024towards}. Hu et al. \cite{hu2018fv} introduced the \ac{FV} Network (FV-Net), a \ac{CNN} model to learn discriminative and robust features representative of \acp{FV}. They first applied sample augmentation strategies to expand the training set and transplanted a pre-trained model to prevent overfitting. Additionally, they designed a network structure to extract features with spatial information and proposed a matching strategy to address misalignment issues caused by translations and rotations in vein pattern recognition. Their approach demonstrated effectiveness on three publicly available databases. Zheng et al. \cite{zheng2020new} presented a lightweight ShuffleNet V2 \ac{CNN} model for \ac{FV} verification. They devised a data augmentation strategy to address the issue of insufficient training samples with brightness variations, partial cropping, or rotation. Additionally, they incorporated label smoothing and a joint loss function for enhancing the learning of discriminative features. Ou et al. \cite{ou2021fusion} introduced a framework for \ac{FV} recognition, integrating intensive data augmentation, robust loss function design, and network architecture selection. This approach integrates interclass data augmentation with traditional intra-class data augmentation, effectively addressing the data shortage challenge in \ac{FV} recognition.  Figure  \ref{fig:dataAugm} provides a visual example of how this approach is applied to an actual \ac{FV} sample. {By showcasing the most widely used transformations, including cropping, rotation, perspective distortion, and color jittering, the figure offers a clear and intuitive understanding of how data augmentation artificially expands the training dataset and introduces variability.}
Wang et al. \cite{wang2023residual} presented a \ac{FV} data augmentation strategy named FV-Mix, which involves grayscale normalization and linear mixing of fine \ac{FV} \ac{ROI}. They also introduced a residual Gabor convolutional network for \ac{FV} recognition, enhancing scale and directional information using Gabor filters. Additionally, they proposed a dense semantic analysis module to assist in classifying and recognizing \ac{FV} images.

\begin{figure}
    \centering
    \includegraphics[scale=0.61]{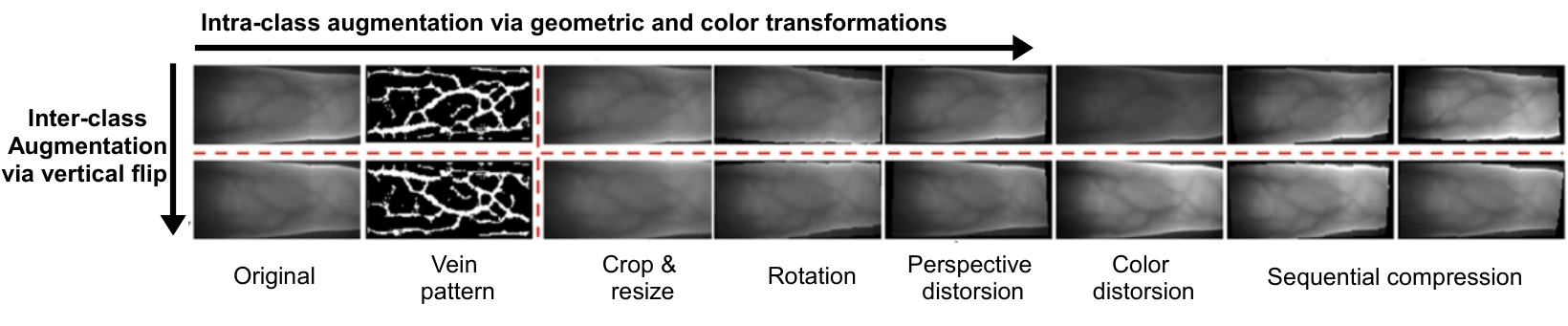}
    \caption{Demonstration of data augmentation using an actual \ac{FV} example (top left), as depicted in \cite{ou2021fusion}. Four types of intra-class transformations were empirically chosen and applied sequentially to the image: random crop and resize, rotation, perspective distortion, and color jittering. For inter-class, vertical flipping of \ac{FV} images is utilized, significantly altering the vein pattern to create new identity semantics, thus defining a new label for such synthetic samples. It should be noted that all data augmentations in this approach are applied solely to the training set.}
    \label{fig:dataAugm}
\end{figure}

To achieve better accuracy and reduce the training time complexity, trasfer learning  has
been adopted in \ac{FV} recognition \cite{hong2017convolutional, hu2018fv, tang2019finger, lu2019exploring, song2019finger, li2019finger, wang2019multi, zheng2020new, tao2021dglfv, ou2021fusion,  shaheed2022ds, chai2022shape, chai2023vascular, tran2023anti, huang2024towards}.  
For instance, Hong et al. \cite{hong2017convolutional} proposed a \ac{CNN}-based approach for \ac{FV} recognition, aiming to address misalignment and shading issues. They trained the \ac{CNN} using three databases with varying characteristics from a vein recognition system's camera, to make it robust to environmental changes. The proposed method simplified the \ac{CNN} structure by using one difference image as input, eliminating the need for complex structures and post-processing steps. Results showed that fine-tuning VGG Net-16 on difference images, as proposed in the research, achieved higher accuracy than existing methods and other \ac{CNN} structures across all three databases.
In \cite{lu2019exploring}, \ac{CNN} competitive, a pre-trained \ac{CNN} model from ImageNet is utilized for feature extraction. Specifically, \ac{CNN} filters similar to Gabor filters, which capture line structures, are selected. A local descriptor leveraging these selected \ac{CNN} filters is then proposed for \ac{FV} recognition. This approach eliminates the need to develop a specific training strategy for \ac{CNN} models using limited \ac{FV} datasets.
Shaheed et al. \cite{shaheed2022ds} introduced the Xception model, a pre-trained \ac{CNN} based on depth-wise separable \acp{CNN} with residual connections, which is deemed to be a more efficient and less complex neural network for extracting robust features.
Huang and Guo {\cite{huang2024towards}} proposed a cross-dataset \ac{FV} recognition solution using a single-source dataset for training. Their approach includes an instance alignment transformation to reduce dataset gaps, a feature alignment and clustering learning algorithm based on VGG Net-16 for learning invariant features, and a modified ELM-based classifier for fast \ac{DTL}. Experimental results on four public datasets show that their model transfers efficiently to target datasets, achieving recognition performance comparable to \ac{SOTA} models trained end-to-end on the target data.

\item \textbf{{Attention and  Transformers:}}
{This section explores two significant advancements in \ac{DL} architectures for \ac{FV} recognition, extending beyond traditional \acp{CNN}: (i) the integration of attention mechanisms—fundamental components of Transformers—into \ac{CNN}-based models, and (ii) the advent of Transformer-based models, particularly \ac{ViT}. These developments are presented together as they epitomize a continuum in the evolution of attention-based techniques within image processing. The integration of attention mechanisms in \ac{CNN} aims to selectively focus on the most pertinent parts of the input, thereby enhancing feature extraction and improving recognition performance.} 
\begin{itemize}
\item \textbf{{CNN-Attention-based approaches:}} 
Attention mechanisms have been increasingly {integrated into \ac{CNN} architectures to enhance feature extraction and improve recognition performance in \ac{FV} recognition \cite{liu2021improved, sulaiman2022attention, 
huang2021joint, hou2021arcvein, zhang2022convolutional, liu2023mmran, zhang2023finger, huang2023axially, huang2023fvfsnet, ma2023finger}}. These mechanisms allow models to focus on specific parts of the input data that are most relevant to the task at hand, effectively reducing noise and improving the overall robustness of the system. 
For instance, Liu et al. \cite{liu2021improved} introduced a  \ac{CNN} model with a modified \ac{RAB} and incorporated the Inception structure to capture features across various scales. The results demonstrate the model's strong recognition performance with reduced parameters.
Huang et al. \cite{huang2021joint} introduced an attention mechanism called the joint Attention, JA module, to extract discriminative features from low-contrast images. This module dynamically adjusts and aggregates information in the spatial and channel dimensions of feature maps, emphasizing fine-grained details and enhancing the contribution of vein patterns for identity feature extraction. Their approach achieved competitive results on various datasets. Zhang and Wang \cite{zhang2022convolutional} proposed a lightweight \ac{CNN} with a \ac{CBAM} for \ac{FV} recognition, aiming to capture visual structures more accurately.  In \cite{zhang2023finger}, the authors introduced a ResNet model enhanced with self-attention. By combining the global focusing ability of self-attention with the local feature extraction capability of \ac{CNN}, this approach achieves higher accuracy in \ac{FV} recognition. Huang et al. \cite{huang2023axially} introduce the axially enhanced local attention network, called ALA Net, to enhance the extraction of long-distance vessels and local textures in vein images. They propose a feasible variable ALA to improve identity discrimination of multi-scale features, inferring attention using a given vascular topological distribution. Additionally, feature amplification via low-cost grouped convolution generates convolutional feature mappings directly, reducing the parameter size for miniaturized device deployment. Experimental results demonstrate ALA Net's significant accuracy improvement over baseline networks without attention blocks. Huang et al. \cite{huang2023fvfsnet} developed FVFSNet model for \ac{FV} authentication, processing spatial and frequency domains. The frequency domain module uses transformation and convolution layers, the spatial module employs efficient convolution layers, and the coupling module combines features with channel
and spatial attention mechanisms. Experimental results demonstrate FVFSNet's effectiveness, particularly the frequency domain network, in \ac{FV} authentication. The majority of \ac{DL}-based \ac{FV} approaches have traditionally relied on manually designed architectures, which can be time-consuming and prone to errors. Recently, Qin et al. \cite{qin2023ag} introduced AG-NAS, a neural architecture search method based on attention-gated recurrent units. AG-NAS automatically discovers the optimal network architecture, leading to improved recognition performance. By integrating self-attention and gated recurrent units in an attention \ac{GRU} module, AG-NAS acts as a controller to generate architectural hyperparameters. Additionally, a parameter-sharing supernet policy is used to reduce the search space and computational costs. Experimental results on multiple datasets demonstrate that AG-NAS outperforms existing methods, achieving \ac{SOTA} accuracy.

\item \textbf{{\ac{ViT}:}}
{The Transformer architecture, particularly the \ac{ViT}, represents a significant shift from traditional \acp{CNN}.}  Unlike traditional \acp{CNN}, or \acp{CNN}-attention-based models, \ac{ViT} divides an image into patches and processes them using a Transformer encoder with multi-head attention mechanisms. This approach captures long-range dependencies and contextual information across the image more effectively. {A comparative analysis of traditional CNN, CNN-Attention, and \ac{ViT}-based models, highlighting their characteristics, advantages, and disadvantages, is provided in Table \ref{tab:comparison}.}

\begin{scriptsize}
\begin{table}[H!]
\centering

\caption{{Comparison of CNN-only, \ac{CNN}-Attention, and \ac{ViT}  models for hand vein recognition.}}
\label{tab:comparison}

\begin{tabular}{|m{1cm}|m{5cm}|m{5cm}|m{4.5cm}|}
\hline
\textbf{Model Type} & \textbf{Characteristics} & \textbf{Advantages} & \textbf{Disadvantages} \\ \hline
\textbf{CNN-only} & 
- Utilizes multiple convolutional layers to extract features from hand vein images, focusing primarily on local spatial relationships.\newline
- Does not incorporate attention mechanisms, relying solely on stacked convolutions for feature hierarchy. & 
- Highly efficient with moderate to small-sized datasets due to efficient parameter usage and fewer computational requirements.\newline
- Simpler architectural design facilitates easier optimization and faster training times. & 
- Limited in capturing long-range dependencies between vein structures without the help of attention mechanisms.\newline
- May not perform as well with very complex vein patterns or large-scale datasets compared to models incorporating attention. \\ \hline

\textbf{\ac{CNN}-Atten.} & 
- Combines \ac{CNN} to capture local vein patterns with Transformer layers to model long-range dependencies between vein structures.\newline
- Leverages \ac{CNN}’s spatial hierarchies and the Transformer's attention mechanism for a more holistic understanding of vein networks. & 
- Effective at capturing both local vein details and global vein structures.\newline
- Performs well on tasks involving complex vein patterns, even with moderate datasets.\newline
- More efficient on smaller hand vein datasets compared to \ac{ViT}, due to \ac{CNN}’s feature extraction capabilities. & 
- Higher architectural complexity, requiring careful tuning of \ac{CNN}-Transformer balance.\newline
- Potentially more computationally intensive than pure \ac{CNN} models, depending on the depth of Transformer layers. \\ \hline

\textbf{\ac{ViT}} & 
- Divides hand vein images into patches and processes them as sequences, similar to language tokens, using multi-head attention mechanisms to extract features.\newline
- No convolution layers are used, purely attention-based for global context understanding. & 
- Capable of capturing long-range dependencies and structural relationships between distant vein patterns.\newline
- Highly scalable and can generalize well with large hand vein datasets. & 
- Requires large amounts of labeled hand vein data for optimal performance and pre-training.\newline
- Struggles with small hand vein datasets without extensive pre-training.\newline
- Computationally expensive, demanding high \ac{GPU}/memory resources. \\ \hline

\end{tabular}
\color{black}
\end{table}
\end{scriptsize}

{The general architecture of the ViT, which serves as a foundation for various approaches in \ac{FV} recognition \cite{lu2021novel,huang2022fvt,li2023fv,lu2022finger}, is illustrated in Figure \ref{fig:ViT}}. For example, Lu et al. \cite{lu2021novel} proposed a \ac{FV} recognition method using the \ac{ViT}. They split \ac{FV} images into patches, linearly embed each patch, and feed the sequence into the Transformer encoder. The model employs multi-head attention mechanisms to capture long-range contextual information without explicitly considering spatial distances, showing advantages over other \acp{CNN}, particularly with smaller data categories. Huang et al. \cite{huang2022fvt} proposed using \ac{ViT} for \ac{FV} authentication. Their model utilizes a pyramid structure to improve multiscale feature extraction and introduces conditional position embedding to aggregate local information. They also proposed an efficient \ac{MLP} module and adopted the local \ac{FFN} to improve the local information extraction. Experimental results on benchmark datasets show competitive performance with \ac{SOTA} methods. Later, Li and Zhang \cite{li2023fv} enhance the recognition capability of the \ac{ViT} in \ac{FV} recognition. Their approach, FV-ViT, achieves remarkable performance by applying rigorous regularization only to the \acp{MLP} head without altering the \ac{ViT} backbone architecture, surpassing other \ac{SOTA} methods. Moving on, in their work \cite{lu2022finger}, the authors propose a \ac{FV} feature extraction model based on the \ac{T2T} \ac{ViT} (T2T-ViT). The \ac{T2T} module is utilized for its ability to integrate local information from vein patterns.  

\begin{figure}
    \centering
    \includegraphics[scale=0.55]{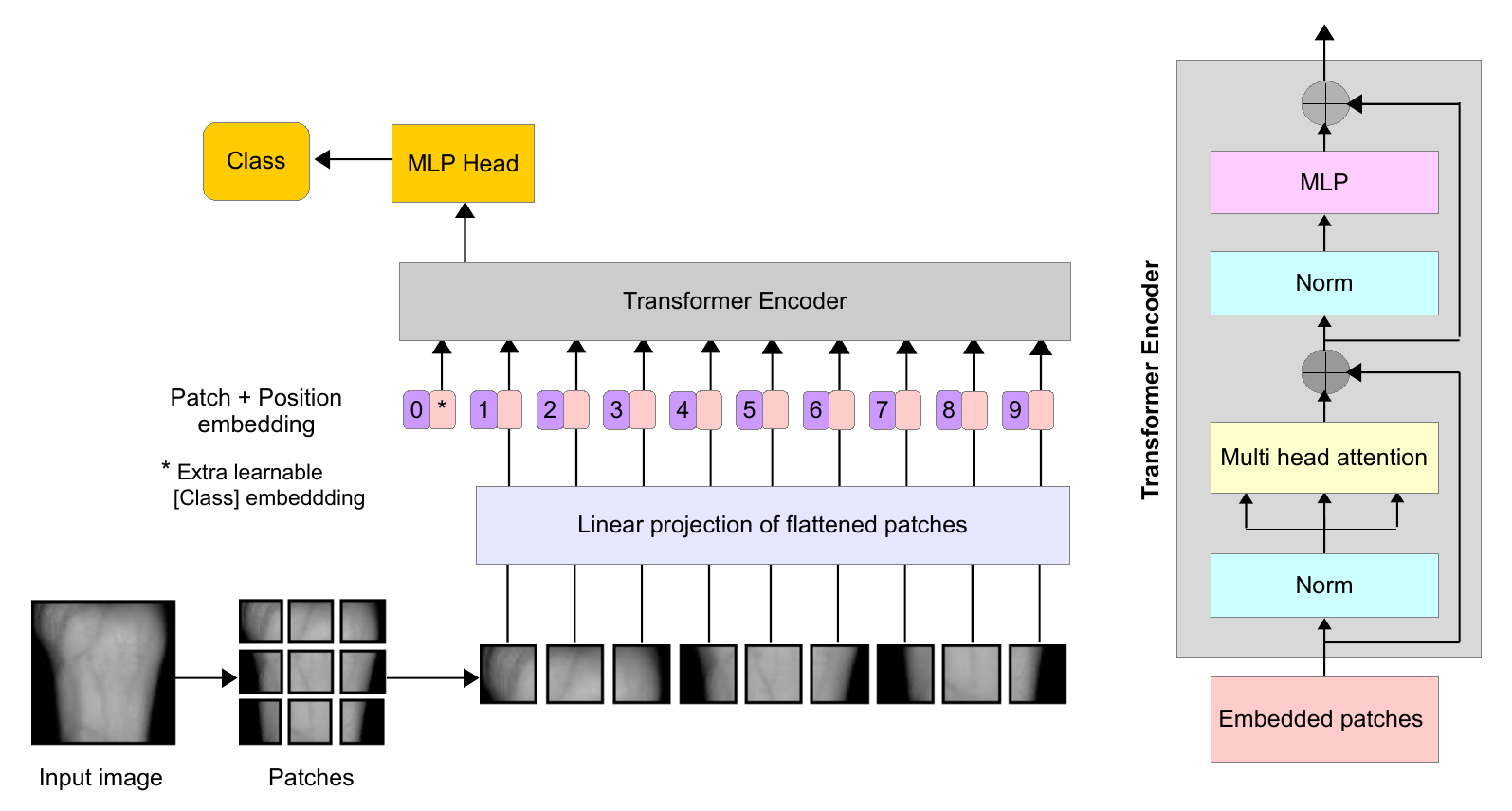}
    \caption{\ac{ViT} architecture for \ac{FV} image.  The \ac{ViT} applies transformer models, originally designed for NLP, to image processing by treating image patches as sequence tokens with positional embeddings, enabling efficient handling of visual data. \ac{MLP} head performing classification or regression tasks based on the learned features.}
    \label{fig:ViT}
\end{figure}

FV-LT, introduced in \cite{qin2022local}, addresses limitations in multi-view \ac{FV} recognition by employing a local Attention Transformer for robust 3D vein vessel feature representation, reducing sensitivity to finger positional variations from roll movements, and mitigating costs and space constraints associated with multiple cameras. The approach involves designing a contactless, low-cost image capturing device using a single camera for acquiring full-view \ac{FV} images. Additionally, a local attention transformer-based identification method is proposed, considering tokens in adjacent nodes for dependency feature learning. Experimental results on public databases demonstrate that FV-LT outperforms existing 2D/multi-view vein recognition approaches, particularly in overcoming variations caused by finger roll, achieving superior performance for full-view \ac{FV} recognition.  Similarly, Zhao et al. \cite{zhao2024vpcformer} proposed a ViT-based model, VPCFormer, for multi-view \ac{FV} feature extraction, effectively capturing global intra- and inter-view correlations with the constraint of vein patterns. VPCFormer achieves excellent performance in multi-view \ac{FV} recognition.
Recently, Li et al. \cite{li2024fv} proposed a MobileViT model for \ac{FV} recognition, addressing high parameter counts and limited training samples. The model includes the Mul-MV2 block, for multi-scale convolutions and the enhanced MobileViT block, for separable self-attention, reducing parameter counts and feature extraction times. They also introduced a soft target center cross-entropy loss function. Experimental results show high accuracy and low computational time.
\end{itemize}
\item \textbf{\Ac{AE}:} 
\acp{AE} are neural networks designed to learn efficient data representations, primarily for dimensionality reduction or feature learning. They consist of an encoder that compresses the input data and a decoder that reconstructs it. In \ac{FV} recognition, \acp{AE} are used to extract meaningful features from vein images for improved accuracy in verification tasks \cite{hou2019convolutional, jalilian2018finger}. As an example, Hou et al. \cite{hou2019convolutional} proposed a \ac{DL} method that combines a \ac{CAE} with a \ac{SVM} for \ac{FV} verification. The \ac{CAE} learns features from \ac{FV} images, with an encoder extracting high-level features and a decoder reconstructing images. The \ac{SVM} classifies the feature codes. Results show that this approach outperforms traditional methods, indicating its potential in \ac{FV} verification. One more example,  Jalilian et al. in their study \cite{jalilian2018finger} assessed the effectiveness of three classical \ac{CAE} architectures—U-net, RefineNet, and SegNet—in extracting \ac{FV} patterns. Their experiments revealed that the RefineNet \ac{AE} architecture, when paired with automatically generated labels, exhibited the best performance and remained consistent regardless of the quantity of training labels.

\item \textbf{Hybrid:}
Hybrid approaches in \ac{FV} recognition represent a cutting-edge strategy that leverages the strengths of different \ac{DL} techniques to achieve superior performance. These approaches are designed to address the challenges inherent in \ac{FV} recognition, such as robustness to varying lighting conditions, accurate feature extraction, and efficient classification. Several approaches in the literature have explored this concept \cite{chen2016finger, hou2018convolutional, qin2019finger, zhang2019gan, zhang2019gan, hou2022triplet, huang2021robust, lin2021finger, li2022vit}.

For exemple, Chen et al. \cite{chen2016finger} combined feature block fusion and \ac{DBN}  for \ac{FV} recognition to address issues of template matching and whole feature recognition methods, which are prone to low robustness due to light instability in acquisition equipment. FBF-DBN utilizes the \ac{DNN}'s nonlinear learning ability to recognize \ac{FV} features. The algorithm improves the deep network's input by using a feature points set in vein images, reducing learning and detection time. Experimental results show that FBF-DBN achieves good recognition performance and faster speed. 
In \cite{hou2018convolutional}, the author employed a combination of \ac{CAE} and  \ac{CNN} to verify vein patterns in images. The \ac{CAE} was utilized to learn the extracted features, while the \ac{CNN} was applied to classify \ac{FV} images based on the extracted feature code. The results demonstrated an enhancement in \ac{FV} verification performance.
Qin et al. \cite{qin2019finger} proposed a \ac{DL} model that combines the stacked \ac{CNN} and \ac{LSTM} models to extract vein features. The SCNN-\ac{LSTM} model outputs represent the vein patterns, and a P-\ac{SVM} is used to predict pixel probabilities for vein networks. The \ac{CNN} learns robust features of \ac{FV} images, while the \ac{LSTM} captures complex spatial relationships of vein patterns. Experimental results on a public \ac{FV} database show significant improvement in verification accuracy. 
In \cite{zhang2019gan}, a combination of \ac{GAN} and \ac{CNN} is proposed for \ac{FV} data augmentation and recognition. They introduce FCGAN, a fully convolutional GAN-based architecture, to generate high-quality and diverse \ac{FV} images. Subsequently, a \ac{CNN} is employed for feature extraction and classification. Experimental results demonstrate that the generated \ac{FV} images improves classification performance compared to classic sample augmentation methods. Similarly, Hou et al. \cite{hou2022triplet} utilized a combination of \ac{GAN} and \ac{CNN} in their work. In contrast to traditional \ac{GAN} methods, their triplet-classifier \ac{GAN} utilizes generated "fake" data to enhance the learning capability of the triplet loss-based \ac{CNN} classifier. This approach expands the training data and improves the \ac{CNN}'s discriminative ability. Experimental results demonstrate the superior performance of the proposed triplet-classifier \ac{GAN} in \ac{FV} verification. 
Huang and Guo \cite{huang2021robust} introduced a \ac{CNN}-based method for \ac{FV} recognition that incorporates bias field correction and a spatial attention mechanism. The bias field correction aims to eliminate unbalanced bias fields from original images using a two-dimensional polynomial fitting algorithm, while the spatial attention module employs a U-Net to extract robust vein patterns. Classification is performed using a ResNet \ac{CNN} model. Experimental results demonstrate the effectiveness of the approach on two public databases.
Liu et al. \cite{lin2021finger} designed an end to-
end deep \ac{CNN} for robust \ac{FV} recognition, incorporating an intrinsic feature learning module with a \ac{CAE} network and an extrinsic feature learning module using a Siamese \ac{CNN}. The intrinsic module estimates intra-class \ac{FV} features under different offsets and rotations, while the extrinsic module learns inter-class feature representations. The experimental results demonstrated that the proposed method achieved comparable \ac{EER} with existing methods on public datasets.
Li et al. \cite{li2022vit} introduced a \ac{FV} recognition model that combines a CapsNet with a \ac{ViT}. This approach leverages CapsNet's strength in processing visual hierarchies and \ac{ViT}'s ability to handle relationships between visual elements and objects. By addressing CapsNet's limitation in encoding long-range dependencies and selectively focusing on important features, the proposed model achieved improved classification accuracy for \ac{FV} recognition.

\item \textbf{Other models: } 
In \cite{kamaruddin2019new}, the authors proposed a method that utilizes PCANet to efficiently extract features from \ac{FV} images. PCANet combines cascaded \ac{PCA}, binary hashing, and block-wise histogram techniques. The method focuses on the vein line feature as an essential part of the filter to extract optimal vein features for classification. The approach uses the original grayscale image in combination with the vein line image using the \ac{CCA} method to generate the filters. This method achieved high accuracy in three public datasets. To resolve the issue of low-level ground-truth pattern maps for training \acp{CNN} due to outlier and vessel break problems, Yang et al. \cite{yang2019fv} developed a model based on cycle-consistent (CycleGAN). Their approach aimed to extract \ac{FV} patterns and outperformed previous \ac{CNN} models in performance. Muthusamy et Rakkimuthu, proposed, the steepest deep bipolar cascade correlative \ac{ML}, called  SDBCCML technique \cite{muthusamy2022steepest}, and trilateral filterative hermitian feature transformation based deep perceptive
fuzzy neural network, named TFHFT-DPFNN technique \cite{muthusamy2022trilateral},  to
efficiently perform the \ac{FV} verification with minimum time.

To enhance \ac{FV} recognition accuracy and reduce verification time while addressing limited training data, Wan et al. \cite{wan2022optimization} introduce a deep forest approach. This method utilizes the deep forest algorithm to identify feature points and the oriented FAST and rotated BRIEF  algorithm, called also ORB, to match these features, extracting angular information from matched pairs. Identity determination is based on the sparse distribution of angles. Compared to traditional \ac{ML} models, this approach maximizes the feature representation capacity of deep forest, particularly in scenarios with limited samples. 

\end{enumerate}

\subsubsection{\Ac{PAD}-based methods}

\begin{figure}[h]
    \centering
    \includegraphics[scale=0.7]{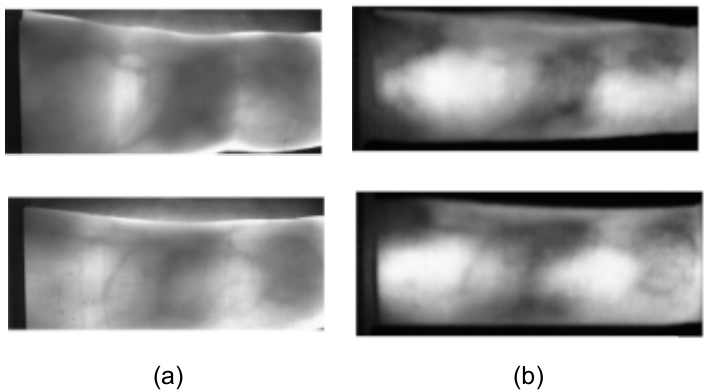}
    \caption{Example of real and presentation attack \ac{FV} images in VERA FingerVein database. (a) Real images; (b) Presentation attack (fake) images.}
    \label{fig:PADImages}
\end{figure}

\begin{figure}[h]
    \centering
    \includegraphics[scale=0.7]{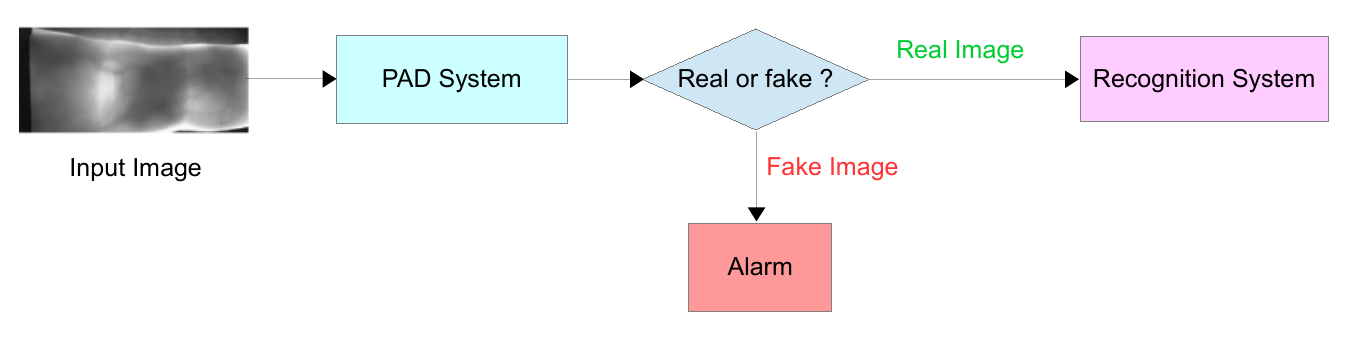}
    \caption{{Integration of \ac{PAD} System with \ac{FV} Recognition System.}}
    \label{fig:pad-fvr}
\end{figure}

Even though \ac{FV} {recognition} system offers advantages and attractive characteristics, it also presents a vulnerability to presentation attacks where fake \ac{FV} images are used to deceive a recognition system, posing a serious threat to its reliability. Figure \ref{fig:PADImages} shows examples of real and presentation attack \ac{FV} images taken from the VERA  database.
Therefore, incorporating \ac{PAD} mechanisms into these systems is essential to maintain their reliability and security. {The integration of a PAD system with a recognition system is presented in Figure \ref{fig:pad-fvr}. The PAD system is used initially to classify whether an input image is real or fake. If fake, an alarm is triggered; if real, the recognition system proceeds.}

Despite the increasing popularity of \ac{FV} recognition technology, research on {\ac{PAD} for \ac{FV} images remains limited}. Traditional \ac{PAD} methods have been proposed but are limited by their feature extraction methods. These methods typically use handcrafted feature extractors designed to distinguish between real and presentation attack \ac{FV} images based on observations of differences in spatial and/or frequency domains. This approach results in limited detection accuracy. Recently, \ac{DL}-based methods have shown promise in improving the detection accuracy of PAD systems. {While most proposed \ac{DL} methods focus solely on designing and testing PAD systems independently of the recognition system \cite{nguyen2017spoof, qiu2017finger, raghavendra2017transferable, kim2022spoof, shaheed2022finger, mu2024federated}, there is a gap in exploiting \ac{DL} in overall systems that combine PAD with recognition \cite{yang2020fvras}.}

One of the first \ac{DL}-based \ac{PAD} methods for \ac{FV} was proposed by Nguyen et al. \cite{nguyen2017spoof}. They introduced a \ac{CNN} approach with \ac{DTL} to combat overfitting issues caused by limited training data or complex \ac{CNN} architectures. \ac{DTL} was applied using pre-trained models like AlexNet and VGGNet-16 from the ImageNet database. To further improve detection, post-processing steps involving \ac{PCA} for feature space dimensionality reduction and \ac{SVM} for classification were employed. Experimental results demonstrated the effectiveness of \ac{DTL} in addressing overfitting and highlighted the suitability of \ac{CNN}-based methods for {\ac{FV} \ac{PAD}}. Qiu et al.  \cite{qiu2017finger} developed FPNet, a shallow \ac{CNN}, for \ac{FV} \ac{PAD}. FPNet was trained without pre-trained models using patches extracted from vein images for dataset augmentation. Mixing training patches from two databases improved model generalizability and robustness, achieving 100\% accuracy on both test datasets, surpassing \ac{SOTA} {\ac{PAD}} methods. 
\ac{DTL} was also applied {for \ac{FV} \ac{PAD} system} in \cite{raghavendra2017transferable}. The authors augmented the pre-trained AlexNet architecture with seven additional layers and fine-tuned the modified \ac{CNN} using a \ac{FV} artefact database. This approach aimed to adapt the pre-trained and modified network to the specific application of \ac{FV} \ac{PAD}. Experiments were conducted using two different \ac{FV} presentation attack databases, with two different \ac{FV} artefact species generated using two different kinds of printers such as Laser print artefact
and Inkjet print artefact. Shaheed et al. \cite{shaheed2022finger} developed a {\ac{FV}} \ac{PAD} method  by combining residual connections and a depth-wise separable \ac{CNN} with a linear \ac{SVM} technique. It aims to address challenges like limited datasets, high computational complexity, and the lack of efficient feature descriptors. The depth-wise separable \ac{CNN} automatically extracts features from spoof and genuine \ac{FV} images, which are then classified by the linear \ac{SVM}. Experimental results demonstrate that this method outperforms existing \ac{DL} models in both performance and computational cost efficiency.

Previous \ac{PAD} studies have primarily concentrated on methods for identifying presentation attacks that involve the use of printed or replayed fake \ac{FV} images. Kim et al. \cite{kim2022spoof} introduced a novel \ac{PAD} method focusing on fake \ac{FV} images generated through a \ac{GAN}. They utilized CycleGAN to create presentation \ac{FV} images aimed at compromising traditional \ac{FV} recognition systems. Their proposed \ac{PAD} method utilizes a score fusion technique that leverages ensemble networks. These networks are based on \ac{SVM} and optimized using both Adam optimization and a sharpness-aware minimization optimizer, known as SAM. Experimental evaluations conducted on public datasets demonstrate the efficacy of this method in detecting GAN-based fake \ac{FV} images.

Recently, Mu et al. {\cite{mu2024federated}} proposed the use of \ac{FL} for \ac{FV} \ac{PAD} across various clients. This approach addresses the challenges faced by terminal clients and enhances model generalization for institutional clients. The framework accounts for differences in data volume and computational power among clients, categorizing \ac{FL} clients into two groups: institutional and terminal. For institutional clients, an enhanced triplet training mode combined with \ac{FL} is suggested to boost the generalization of the \ac{FV} \ac{PAD} model. For terminal clients, an unsupervised learning module is designed to enhance performance in terminal devices.

While the aforementioned methods concentrate exclusively on developing standalone PAD systems, Yang et al. \cite{yang2020fvras} developed a comprehensive \ac{FV} recognition system that integrates both recognition and \ac{PAD} tasks using a unified \ac{CNN} model, leveraging a \ac{MTL} approach. Additionally, they adopted a multi-intensity illumination strategy to improve the captured image quality, obtaining vein images with the best quality. Experimental results demonstrate that the proposed system achieves outstanding performance in both recognition and \ac{PAD} tasks on public datasets, as well as a Self-built dataset containing images depicting axial rotation.
\color{black}

\begin{table}
\caption{Summary of proposed works in \ac{PAD} and \ac{TProt}-based \ac{DL} for \ac{FV} recognition. In cases where multiple scenarios are examined, only the top-performing outcome for each dataset used in the test is mentioned.}
\scriptsize
\label{tab:PADFV}
\resizebox{\textwidth}{!}{ 
\begin{tabular}{m{0.5cm}m{0.5cm}m{0.4cm}m{1.5cm}m{2cm}m{2.3cm}m{3cm}m{3cm}}
\hline
Ref. & Year & Obj. & DL approach & DTL &Data augmentation & Dataset &  Results \\ \hline

\cite{nguyen2017spoof}& 2017 &  & AlexNet, VGGNet-16 &Pretrained weights from
ImageNet dataset & Images shifting and cropping& ISPR, VERA FingerVein & APCER=0\%; BPCER =0\%; ACER=0\%, 0.0311\%   \\

\cite{qiu2017finger}& 2017 &  & Shalow CNN  & --&Patches extraction with stride s &VERA FingerVein, SCUT-SFVD & ACER=0.00\% \\

\cite{raghavendra2017transferable}& 2017 & & AlexNet  & Fine-tuning & --& Two self-built datasets & 
APCER=0\%,0.4\%; BPCER=0\%,0\% (Laserjet printed artefact species)\\

\cite{yang2020fvras}& 2020 &  \multirow{7}{*}{\rotatebox{90}{\ac{PAD}}}& Lighweight \ac{CNN}  & Fine-tuning between
datasets  & Images rotating, cropping and blurring& PAD (VERA FingerVein, SCUT-SFVD), Recognition (FV\_USM, SDUMLA-HMT MMCBNU\_6000, Idiap, Self-built dataset)  & PAD (HTER=0.00\%, 0.00\%); Recognition (EER=0.95\%, 1.71\%, 1.11\%, 3.25\%, 2.02\%)  \\

\cite{kim2022spoof}& 2022 & & GAN, DensNet-161, DensNet-169, and SVM &Fine-tuning  DensNet-161 and DensNet-169 on ImageNet dataset& --& ISPR, VERA FingerVein& FID=20.57, 28.85; WD=5.95, 8.30;  APCER=0.64\%, 0.45\%; BPCER=0.00\%, 0.00\%; ACER= 0.32\%, 0.23\% \\ 

\cite{shaheed2022finger}& 2022 & 
& Xception+LSVM & Fine-tuning 
Xception on ImageNet
dataset & Image rotating, 
shifting, zooming, horizontal flipping and 
shearing &VERA FingerVein, SCUT-SFVD & APCER, BPCER, and ACER all at 0.00\%, and Acc. of 100\%\\

\cite{mu2024federated}& 2024 & & MobileNetV2 and MobileViT with \ac{FV} &Fine-tuning& --& VERA FingerVein,  SCUT-SFVD, Self-built dataset  & ACER= 0.00\%;  EER=0.00\%; BPCER=0.00\% \\ \hline

\cite{yang2019securing}& 2018 & \multirow{3}{*}{\rotatebox{90}{\begin{tabular}{c}\ac{TProt}
\end{tabular}}} & ML-ELM &-- &-- &SDUMLA-HMT, MMCBNU\_6000, UTFVP & EER= 7.04\%; CIR=93.09\%, 98.70\%, 98.61\%   \\

\cite{liu2018finger}& 2018 &     & DBN &-- &-- & Self-built dataset & GAR= 91.2\%; FAR=0.3\%  \\

\cite{shahreza2021deep}& 2021 & & CAE & --&Images rotating, shifting and zooming & UTFVP & FMR= 0\%; FMNR=0\%; HTER= 0\%\\
\hline
\end{tabular}}
\begin{flushleft}   
\end{flushleft}
\end{table}

\subsubsection{Template protection-based methods}
Biometric systems face a critical issue regarding privacy, as once a biometric template is compromised, it cannot be changed or revoked. To tackle this challenge, various \ac{BTP} schemes have been developed. However, while these schemes enhance privacy, they can also impact recognition performance. Several \ac{DL} approaches have been proposed to mitigate this issue. Yang et al. \cite{yang2019securing} devised a novel \ac{BTP} algorithm using binary decision diagrams for \ac{DL}-based \ac{FV} biometric systems. This algorithm creates a noninvertible version of the original \ac{FV} template, which is then combined with an artificial neural network, specifically the\ac{ML-ELM}, to form a privacy-preserving \ac{FV} recognition system, called BDD-ML-ELM. This approach provides an advantage over existing \ac{ML}/\ac{DL} biometric systems, as the raw biometric templates are protected from inversion attacks on the artificial neural network. Liu et al. \cite{liu2018finger} introduced a secure \ac{FV} recognition method, FVR-DLRP, using \ac{DBN} and random projections. The \ac{DBN}, comprising two \acp{RBM}, generates the biometric template. Experimental results show that the proposed approach maintains identification accuracy and enhances security for \ac{FV} authentication.
In \cite{shahreza2021deep}, a CAE-based method for secure \ac{FV} recognition is proposed. The approach utilizes a \ac{CAE} neural network with a multi-term loss function, incorporating both \ac{AE} and triplet loss functions for embedding features. Biohashing is then applied to these deep features to generate protected templates.






\subsection{\ac{PV}-based methods}
Moving from \ac{FV} to \ac{PV} recognition for biometric identification offers several advantages. \ac{PV} recognition captures a larger and more intricate vascular pattern than \ac{FV} (Figure \ref{fig:handBiometrics} (a)), providing higher accuracy and security. The palm's wider surface area allows for more data points, improving matching reliability and reducing false acceptance/rejection rates. Additionally, \ac{PV} scanning is less invasive and more user-friendly, as it doesn't require precise finger placement. This technology is also less affected by external factors like skin conditions, cuts, or dryness, enhancing its robustness. Furthermore, the contactless nature of \ac{PV} scanning promotes hygiene, an important consideration in public or shared environments. Overall, \ac{PV} recognition's enhanced security, reliability, user experience, and hygiene make it a superior choice for efficient biometric identification. Numerous \ac{ML}/\ac{DL}-based \ac{PV} recognition methods have been proposed in the literature. These methods are discussed and classified below. Table \ref{tab:PalmVein} summarizes the proposed schemes for \ac{PV} recognition.


\subsubsection{\ac{DL}-based methods}

\ac{PV} characteristics, including high recognition rates, stable features, easy localization, and superior image quality, have garnered significant attention from researchers. This feature makes it suitable for individual identification.  For example, the contactless \ac{PV} recognition system proposed in \cite{chen2021contactless} was developed based on prior research and \ac{CNN} technology. It comprises two main components: training and testing. Initially, the \ac{ROI} method was employed to identify the relevant area, followed by convolution calculations using the \ac{ROI} image and Gabor filter to extract features. During the deep network training phase, input fusion was executed between the original \ac{ROI} image and Gabor features. Subsequently, in each subsequent deep network training batch, triplet loss and cross-entropy were computed to optimize the network weights via backpropagation, thus ensuring the capture of sufficient \ac{PV} data. In the same manner, Obayya et al. \cite{obayya2020contactless} , the \ac{CNN} operates as a feature extraction algorithm, and accomplished through the utilization of a Bayesian optimization algorithm, aimed at reaching the optimal network structure and training parameters within the search space of potential solutions. When the \ac{ROI} underwent processing through a Jerman vessel enhancement filter, the process notably reduced the \ac{EER}. The authors in \cite{ma2023focal}  introduce a \ac{CNN} based method for \ac{PV} authentication, so-called FCPVN,  integrating label information into self-supervised contrastive learning. They propose a focal contrastive loss to prioritize hard examples, enhancing model focus. FCPVN demonstrates competitive performance across five \ac{PV} databases, compared to  existing \ac{SOTA} methods.

\begin{figure}[h!]
    \centering
    \includegraphics[scale=0.7]{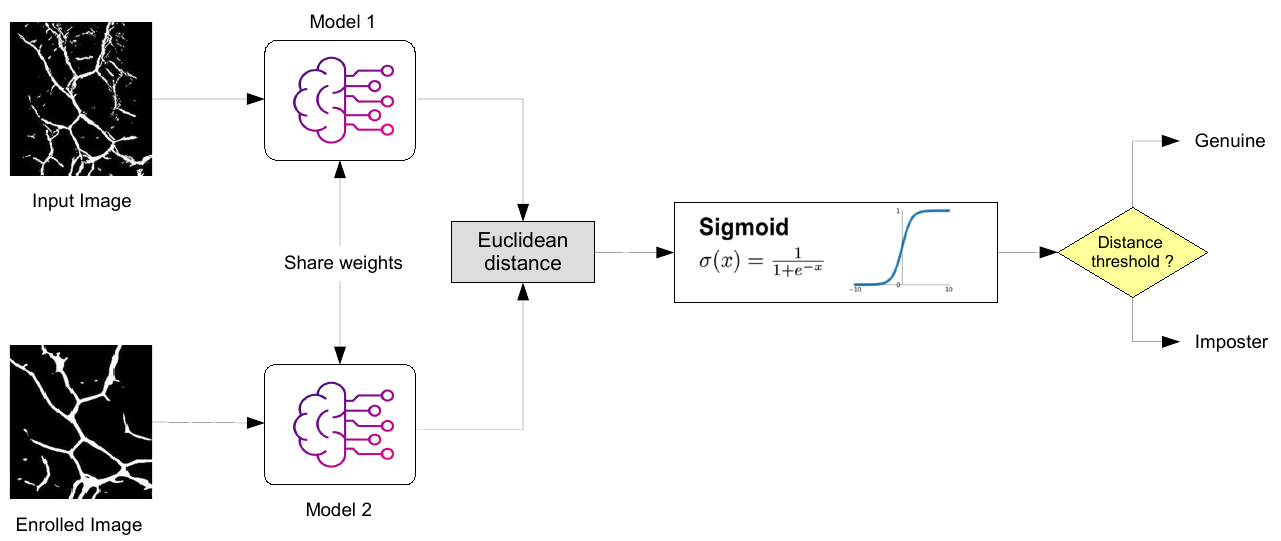}
    \caption{A standard Siamese neural network architecture used in vein biometric systems involves training two identical subnetworks simultaneously, fine-tuning them on two datasets to compare input pairs. It uses Euclidean distance to measure similarity and a sigmoid function to output the likelihood of the inputs being similar.}
    \label{fig:siamese}
\end{figure}

Recent advances in hand vein identification via \acp{DNN} rely on extensive training data, hindering robust feature extraction from single images. Addressing this, Qin et al. in \cite{qin2021multi} propose single-sample-per-person \ac{PV} identification. The proposed method,  combines multi-scale and multi-direction \acp{GAN} for data augmentation with a \ac{CNN}. The scheme generates diverse samples, improving \ac{CNN} stability in \ac{PV} recognition, as demonstrated on public databases,  CASIA and PolyU. The authors later introduced an enhanced multi-scale \ac{PV} identification method based on Transformer in \cite{qin2023label}. The method employs convolutional and self-attention blocks to capture local and scale information. It enhances vein classification via a \ac{GCNLE} approach, which leverages label correlations to improve feature representation in a multiscale vein Transformer. Horng et al. \cite{horng2021recognizing} introduces a cost-effective \ac{PV} recognition system for smartphones, utilizing RGB images. It enhances vein patterns using the saturation channel, instead of red channel, and introduces a novel method for \ac{ROI} extraction based on convex hulls and key vectors. Additionally, a lightweight \ac{DL} model, called MPSNet,  optimized for smartphones is proposed, incorporating convolutional layers, depthwise separable convolution, inverted residual bottleneck, and \ac{SPP}modules, enhancing accuracy through fusion strategy. Moving forward, in their work  \cite{thapar2019pvsnet}, Thapar et al. introduced a \ac{CNN} architecture, called  Siamese framework (The same principle is illustrated in Figure  \ref{fig:siamese}, {  a widely used architecture in image recognition, face verification, and few-shot learning due to its unique twin networks that compare pairs of inputs to determine similarity}), coupled with a triplet loss function. This latter is derived from the disparities among (i) anchor, (ii) positive, and (iii) hard-negative embeddings. Initially, an encoder-decoder network is employed to capture domain-specific traits, followed by the assembly of the inception layers to produce \ac{PV} characteristics. Augmentation techniques are utilized to address the scarcity of positive samples for a specific subject and to boost learning, and a smaller batch size is chosen to expedite convergence, while dropouts are incorporated to prevent overfitting. {Another method, \cite{qin2019iterative}, presents an iterative \ac{DBN} specifically designed for hand vein verification. It initially extracts vein features from labeled data and then iteratively refines them during training. The \ac{DBN} is a probabilistic generative model formed by stacking \acp{DBN} using a layer-wise pre-training approach. This study trains the \ac{DBN} to estimate the likelihood of a pixel belonging to a vein pattern. To accomplish this, it utilizes established handcrafted image segmentation techniques for vein extraction, creates a training dataset based on labeled pixels, and employs the \ac{DBN} for training. This approach effectively addresses the challenges of vein segmentation and leads to improved verification performance.}

Regrettably, recent studies \cite{ma2021understanding,kheddar2023deep} have revealed the vulnerability of \ac{DL} algorithms to attacks involving the addition of carefully crafted adversarial perturbations to authentic samples as input. These perturbations are often subtle and imperceptible to human eyes, yet they can result in significant misclassification of genuine samples during inference.  In this context, adversarial attacks pose a threat to the security of \ac{DL}-based biometric models. Currently, numerous attacking techniques have been devised to manipulate network classifiers into altering their predictions. For instance, Goodfellow et al. suggested the \ac{FGSM}, which introduces adversarial perturbations to create adversarial examples \cite{ma2021understanding}, among others. For that reason, researchers highlight the urgent need for approaches to safeguard \ac{DL}-based \ac{PV} models. In light of the preceding details, Li et al. \cite{li2023transformer} proposed VeinGuard to defend deep neural network-based vein classifiers from adversarial attacks. It incorporates a \ac{LTGAN} and a purifier. The \ac{LTGAN} models the unperturbed image distribution using conditional position embedding and local depth-wise convolution in Swin Transformer blocks. The purifier removes perturbations by identifying the nearest sample to the given perturbed image. It comprises a pre-trained generator and a trainable residual network, fine-tuned during testing for vein recognition. \ac{LTGAN} outperforms \ac{SOTA} methods under black-box \ac{HSJA}.

\subsubsection{\ac{DL}-based hashing}

\begin{figure}[h!]
    \centering
    \includegraphics[scale=0.8]{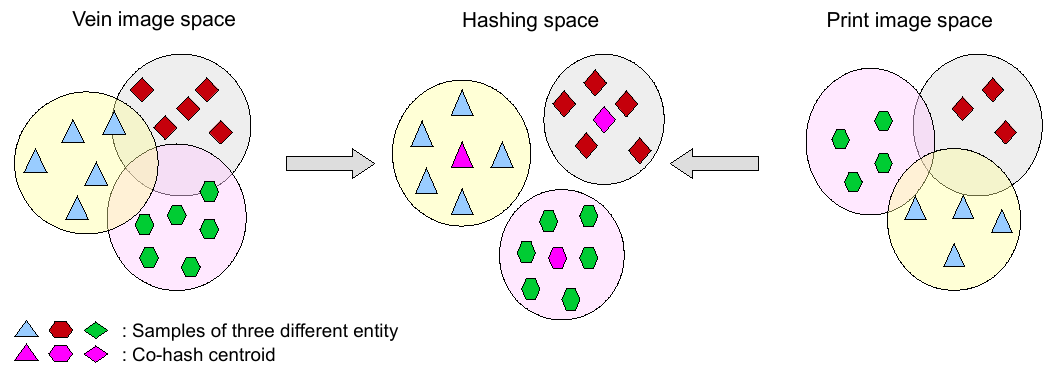}
    \caption{Principle of mapping \ac{PV} and palm print samples to their respective co-hash centroids linked with each identity. This ensures that the hashing space is maximally utilized, as these co-hash centroids are mutually orthogonal.}
    \label{fig:crossHashPVein}
\end{figure}

Hashing methods offer impressive efficiency in data storage, retrieval speed, and enhanced security, which are vital for \ac{IoT} services. Zhong et al. \cite{zhong2018palm} put forth an end-to-end \ac{CNN}-based hashing network specifically designed for \ac{PV} recognition. They employ a hashing code, generated using $tanh$
and $sign$ functions, to encode image features into a binary code of consistent length. Through comparing the Hamming distances between the binary codes of various \ac{PV} images, the proposed method can discern whether they share the same category. In \cite{wu2021palmprint}, hash codes representing palm print and \ac{PV} are derived through a deep hashing network. A spatial transformer network is employed to mitigate issues related to rotation and displacement. The methodology incorporates both image-level fusion and score-level fusion techniques. However, the inability to perform cross-modality matching limits the extent of performance enhancement in this study.  The work proposed by Dong et al. \cite{dong2022co} introduces PalmCohashNet, a biometric hashing network designed for \ac{IoT} applications, which combines palm print and \ac{PV} data. PalmCohashNet consists of two separate hashing subnetworks, one for each palm modality, trained together to produce shared hash codes (co-hash codes) for each modality. {This approach is depicted in Figure \ref{fig:crossHashPVein}, a method widely adopted across all \ac{DL}-based \ac{PV} implementations that utilize hashing techniques.} A novel \ac{CMH} loss function is developed to ensure that co-hash codes from the same individual across \ac{PV} and palm print modalities are both adjacent and consistent. Simultaneously, it pulls these co-hash codes towards a predetermined identity-specific hash centroid shared by both modalities. This enables the generation of two palm-based co-hash codes for each individual, suitable for deployment due to their compact form for storage and quick matching, making them \ac{IoT} compliant. PalmCohashNet can be deployed in many operation modes: single-modality matching (e.g., print versus print or vein versus vein), multimodality matching using both print and vein data, and cross-modality matching (print versus vein), depending on the specific requirements of the \ac{IoT} service. Empirical results on four publicly available palm databases demonstrate that the proposed method consistently outperforms existing \ac{SOTA} techniques.

\subsubsection{DTL-based  methods}

The authors in \cite{lefkovits2019applications} presents a \ac{PV} recognition system utilizing \acp{CNN}. Six \ac{CNN} models were constructed and evaluated for biometric identification using palm images. Four models were adapted from existing architectures through \ac{DTL}, namely, AlexNet, VGG-16, ResNet-50, and SqueezeNet, while two new architectures were proposed. Experimental comparisons were conducted on identification accuracy and training convergence using the PUT \ac{PV} database. Various image quality enhancement methods were explored to improve accuracy, which include modifying contrast and normalizing, applying Gaussian smoothing, utilizing \ac{CLAHE}, and employing coarse vein segmentation based on the Hessian matrix. The ResNet-50 pre-trained model yielded the highest performance. 
Cross-modality serves as a form of \ac{DTL}, transductive \ac{DTL}, leveraging information from two distinct domains to enhance the accuracy of one of them  \cite{kheddar2023deep,kheddar2023deepIDS}. Li et al. in  \cite{li2021bpfnet}, the accuracy reaches impressive levels through the fusion of multi-modal information, incorporating both palmprint and \ac{PV} \acp{ROI} into the proposed FNet network. The final descriptor E in FNet is produced via cross-modal \ac{DTL}, resulting in robust anti-falsification capabilities. Likewise, Wang et al. \cite{wang2022multispectral} introduce a framework for both palm print and palmar vein recognition. They conduct feature-level image fusion after extracting features, using wavelet transform, from palm print and palmar vein images at various wavelengths. Their experiment yields a more practical fusion scheme. Building on these fusion results, they suggest enhancing \ac{CNN} models for identity recognition using multispectral palm print and palmar vein images. The \ac{SPP} layer addresses the challenge of fixed input image sizes in \acp{CNN}, with the ResNet18 architecture demonstrating superior performance in terms of both accuracy and recognition time compared to other methods examined. Similarly, \cite{kaddoun2021convolutional} explores the utilization of \ac{DL} in \ac{PV} recognition for personal identity authentication. Various \ac{CNN} architectures are examined and contrasted, with particular focus on the ZFNet architecture, which is fine-tuned using optimal parameters. A comparative analysis is conducted with other existing \ac{CNN} models like LeNet, AlexNet, and ResNet, revealing ZFNet's superior performance, followed closely by ResNet. Moving on, the study \cite{hernandez2023synthetic} presents a novel approach to \ac{PV} recognition through \ac{DTL}, exploring the use of two synthetic databases to pre-train \ac{DL} models. The research evaluates the performance of these models on real databases. Two end-to-end \ac{CNN} architectures, based on fine-tuned SingleNet and Resnet32, are implemented for feature learning. Results from experiments on prominent public datasets indicate the efficacy of utilizing synthetic \ac{PV} images for \ac{DTL}, exceeding previous \ac{SOTA} outcomes.

Wu al. \cite{wu2021outside} have created a wavelet denoising ResNet scheme, WD Resnet, comprising two components: the wavelet denoising model and the pre-trained ResNet18 model. The fine-tuned model is designed to eliminate noise arising from skin scattering and optical blurring in \ac{PV} images. Leveraging residual learning technology, it amplifies low-frequency features to integrate them into the \ac{DL} framework, thus reinforcing the significance of handcrafted features within the network. Meanwhile, the ResNet18 model tackles challenges such as rotation, positional shifts, and scaling variations by selectively enhancing classification features while diminishing less pertinent ones. Marattukalam et al. \cite{marattukalam2021n} introduce a design employing a Siamese neural network framework for \ac{PV} identification in few-shot scenarios. The suggested model utilizes images from both left and right palms for two users, comprising two sub-networks that share user weights to distinguish individuals, as depicted on Figure \ref{fig:siamese}. The effectiveness of this architecture was evaluated on the PolyU multispectral \ac{PV} database, which has a constrained number of samples.

The work presented in \cite{kuzu2023gender} explores the utilization of hand vein patterns for biometric recognition, specifically investigating differences in vein patterns between genders and how these variances affect verification performance. The study assesses the possibility of recognizing male and female individuals based on their hand vein patterns. Transfer \ac{DL} approach is adopted to analyze hand vein features, taking into account gender-specific characteristics. Specifically, a modified Densenet-161 architecture is applied for processing hand vein patterns. This \ac{CNN} architecture is fine-tuned to analyze vein patterns and extract relevant features for gender recognition and biometric verification tasks. Results show that gender-specific models can significantly enhance recognition rates. The research conducted by Babalola et al. \cite{babalola2023boosting} focuses on enhancing hand vein recognition through the integration of various color spaces within \ac{DL} frameworks. The authors examined five distinct color spaces (RGB, XYZ, LAB, YUV, HSV) and identify, each time, the most impactful channels for recognizing patterns in \ac{PV} images. By integrating these color spaces within \ac{CNN} models such as AlexNet, VGG-19, and ResNet-50, the proposed architecture is able to achieve good accuracy for different datasets. The research demonstrates the effectiveness of considering color space in improving the performance of \ac{PV} recognition.

\renewcommand{\arraystretch}{1.25}
\begin{table}[h]
\caption{Summary of proposed works in \ac{DL}-based \ac{PV} recognition. In cases where multiple scenarios are examined, only the top-performing outcome is mentioned. }
\scriptsize
\label{tab:PalmVein}
\resizebox{\textwidth}{!}{ 
\begin{tabular}{m{0.5cm}m{0.5cm}m{2cm}m{2cm}m{2.5cm}m{1.7cm}m{3cm}} \hline
Ref. & Year & DL approach & \ac{DTL} &Data augmentation & Dataset &  Results \\ \hline
 \cite{obayya2020contactless} & 2021 & CNN &  Fine-tuning & -- & CASIA & Acc= 99.4\%, EER= 0.0683\% \\
 
 \cite{chen2021contactless} & 2021   & CNN & --& -- & CASIA, PUT & EER= 0.0556\%,  0\% \\

 \cite{lefkovits2019applications} & 2019 & ResNet-50 & Fine-tuning & Random
rotation & PUT & Acc= 99.83\% \\

 \cite{wang2022multispectral} & 2021 & CNN & \ac{DA} & -- & Self-built dataset & EER= 0.38\%; Time= 715 ms \\

  \cite{ma2023focal} & 2023 & CNN & -- & -- & PolyU, Tongji & EER= 0.28\%, 1.07\% \\
 
  \cite{qin2021multi}  & 2021 & GAN+CNN & -- & Multi-direction GAN &CASIA, PUT & Acc= 98\%, 98.82\% \\
 
  \cite{qin2023label} & 2023 & GCNLE+Attention & -- & -- &PolyU, VERA & Acc= 99.82\%, 98.27\% \\
 
\cite{thapar2019pvsnet} & 2019 & Encoder-decoder +  Triplet Siamese & Fine-tuning & Image augmentor tool& CASIA, IITI, PolyU & EER= 3.71\%, 0.93\%, 0.66\%; CIR= 85.16\%, 97.47\%, 98.78\%\\
  
\cite{zhong2018palm} & 2018 & CNN & -- & -- &PolyU & EER= 0.0222\% \\\
 
\cite{kaddoun2021convolutional}  & 2021 & CNN & -- & -- & MS-PolyU & Acc= 94.65\% \\
 
\cite{li2023transformer}  & 2023& LST+GAN & Fine-tuning & Local
Transformer-based GAN& PolyU, Tongji, MMCBNU6000 & Acc= 93.6\%, 88.7\%, 96.2\% \\
 
\cite{hernandez2023synthetic} & 2023 & ResNet-32, \newline SingleNet & Fine-tuning& ROI samples augmentation & CASIA, PolyU & Acc= 99.20\%, 100\% \newline Acc= 98.42\%, 100\%\\
 
\cite{wu2021outside} & 2021 & WD ResNet & \ac{DA} & Randomly rotate, and flip the
image horizontally or vertically & Tongji & Acc= 99.70\%, EER= 0.41\% \\
 
\cite{wulandari2019performance} & 2019 & AlexNet & \ac{DA} & -- & PUT & CIR= 93.92\% \\
 
 
\cite{kuzu2020vein} & 2020 & DenseNet-161 & \ac{DA} & -- & PolyU & CIR= 99.69\% \\
 
\cite{jia2021performance} & 2021 & MobileNet v3  \newline EfficientNet & \ac{DA} & -- & PolyU & CIR= 100\% (For both models)\\
\cite{qin2019iterative} & 2019 & DBN & -- & -- & CASIA, PolyU & EER= 0.33, 0.015\\
\hline
\end{tabular}
 }
 \begin{flushleft}
Abbreviation: Local swin Transformer (LST)   
\end{flushleft}
\end{table}

\subsection{\ac{DHV}-based methods}


\ac{DHV} recognition is a biometric authentication method that utilizes the unique vein patterns on the back of the hand for identification purposes. This method involves capturing an image of the dorsal hand using infrared or near-infrared light. Subsequently, advanced algorithms analyze the vein characteristics, including their structure, color, distribution, and geometry. \ac{DHV} recognition is particularly advantageous over other biometric methods such as \ac{FV} and \ac{PV} due to the larger surface area of the dorsal hand, which provides more detailed and distinct vein patterns. This enhances the accuracy and reliability of the system.

The robustness of \ac{DHV} biometrics is further increased as the dorsal hand is less susceptible to wear, tear, or dirt. This method offers several benefits: high accuracy, non-invasiveness, and the difficulty of replicating vein patterns, making it less prone to fraud compared to external features like fingerprints. Additionally, \ac{DHV} recognition does not require physical contact, which is ideal for applications requiring high levels of hygiene and minimal user interaction, such as secure access control, time attendance systems, and identity verification. Despite its numerous advantages, \ac{DHV} biometrics also faces challenges such as higher cost and potential privacy concerns. To facilitate a clearer understanding of the existing approaches, Table \ref{tab:DorsalmVein} summarizes the proposed schemes for \ac{DL} and \ac{DTL}-based \ac{DHV} recognition, and highlighting their respective performance. 
\color{black}

\subsubsection{\ac{DL}-based methods}
The \ac{DL}-based methods in \ac{DHV}  identification focus on leveraging \ac{DL} architectures, primarily CNN, to automatically extract and learn features from raw vein images. These methods enhance the accuracy and robustness of \ac{DHV} identification by capturing complex patterns that are difficult to detect with traditional feature extraction techniques. The primary advantage of \ac{DL}-based approaches is their ability to model intricate relationships within the data, leading to higher identification accuracy and better scalability with larger datasets.
\color{black}
\acp{CNN} are widely used for \ac{DHV} recognition due to their ability to accurately extract and learn features from complex hand patterns and textures. For example, the paper in \cite{nour2024improved} presents an unsupervised \ac{DL} method for hand vein recognition. The proposed system involves preprocessing mechanisms, \ac{DL} models based on \ac{CNN}, and an alternative approach based on \ac{RBM} to enhance feature recognition and reduce recognition time. \ac{CNN} architectures, specifically designed for image recognition tasks, are well-suited for vein recognition applications. The suggested method starts by guessing vein patterns for biometric identification. Next, \ac{DL} models like Inception and ResNet50 can automatically pull out distinguishing features from preprocessed vein images, making it easier to find \acp{DHV}. Benaouda et al. in \cite{benaouda2021cnn} introduce a methodology for \ac{DHV} recognition leveraging \ac{CNN}. To start the process, the \ac{ROI} is taken from images of the \ac{DHV}s. Next, \ac{CLAHE} is used to prepare the images. Subsequently, a database is constructed. Finally, the identification task is performed via \ac{CNN}. Hyperparameter optimization refines the experimental results, leading to a high \ac{DHV} recognition rate.
{Moving on, the} research paper \cite{li2018hand} investigates the use of \acp{CNN} for hand-dorsa vein detection to improve performance using a database expansion technique that relies on \ac{PCA} reconstruction. The paper aims to overcome the constraints of \ac{DL} in vein detection caused by a lack of sufficient dataset samples. It proposes a new image-generating method employing double \ac{PCA}. The \ac{PCA} reconstruction method makes the hand-dorsa vein database better by combining features that are similar. This makes the dataset more representative and varied for training \ac{CNN} models. By expanding the database and enhancing feature representation, the approach attains a high identification rate of 99.61\% on datasets of different scales. The \ac{PCA} reconstruction approach significantly improves the performance of \ac{CNN}, surpassing other current methods and proving the usefulness of \ac{DL} in biometric identification. Moving forward, in \cite{li2021fusion}, the proposed research integrates \ac{PLBP} with \ac{CNN} through three schemes (Serial Fusion, Decision Fusio, and Feature Fusion). This combination of traditional \ac{PLBP} features and \acp{CNN} has proven to be effective in \ac{DHV} recognition. The \ac{PLBP} feature extraction process involves scaling vein images, dividing them into non-overlapping regions, and extracting features from each sub-image. For the NCUT data set, the fusion process, with specific weights assigned, achieved an \ac{SOTA} recognition rate. Wan et al. \cite{wan2017dorsal} introduces a method for \ac{DHV} recognition using \ac{CNN}-based \ac{DL} approaches. The research involves training \ac{CNN} models like AlexNet, Reference-CaffeNet, VGG-16, and VGG-19. Before training the \ac{CNN} models for \ac{DHV} recognition, several preprocessing techniques were applied to the images. These techniques involve extracting the \ac{ROI} from \ac{DHV} images, enhancing the contrast of the images using the \ac{CLAHE} algorithm, and applying a Gaussian smoothing filter to reduce noise and enhance the quality of the \ac{CNN} models. Based on the experimental results, the size of the dataset has a significant impact on the recognition rate for \ac{DHV} recognition. Kuzu et al. \cite{kuzu2022intra} look into how similar hand vein patterns are between people using a modified Densenet-161 architecture for biometric recognition. This research employs a \ac{DL}-based method to analyze vein patterns in different areas of the hands. It builds upon prior studies by considering palm- and dorsal-vein patterns in addition to \ac{FV}. For training, the network architecture includes a Custom Embedder and an Additive Angular Margin Penalty. Experiments are conducted on various vein databases, including the SDUMLA, PolyU-P, Bosphorus, and R3VEIN datasets. The results indicate that the vein patterns of one hand of a subject show higher similarity with the corresponding traits of the other hand of the same subject. Overall, the authors illustrate how their study contributes to improving the biometric recognition systems' accuracy by understanding the similarities within hand vein patterns. Another recent study, \cite{zhang2023fast}, addresses the challenge of accurately extracting \ac{ROI} from non-contact \ac{DHV} images in complex backgrounds. It proposed an improved U-Net model to solve the problem. It proposes an improved U-Net model with a residual module to enhance feature information extraction and help to avoid gradient disappearance and explosion, thereby solving network degradation issues. It also employs the Jensen-Shannon divergence loss function to improve feature map distribution. And finally, it implemented Soft-Argmax for keypoint detection, making it suitable for practical deployment on low-resource platforms. 

The research in \cite{alashik2021human} investigates the verification of human identity through the analysis of biometric \ac{DHV} images by utilizing a methodology that incorporates \ac{DL} and \acp{GAN} (DL-GAN). An approach involving multiple steps is implemented for the selection of features and the preprocessing of images. The \ac{DL}-GAN demonstrates an efficient capability in recognizing individuals based on \ac{DHV} images, thereby improving the process of authentication. The use of deep \acp{GAN} outperforms traditional techniques such as \ac{LBP}, \ac{LPQ}, Gabor filter, and \ac{SIFT} in the realm of \ac{DHV} recognition, highlighting the superior performance of this method in biometric authentication.

\subsubsection{\ac{DTL}-based methods}
Building on the strengths of \ac{DL}-based methods, \ac{DTL}-based methods in \ac{DHV} identification aim to further enhance performance by adapting pre-trained \ac{DL} models to the specific task of \ac{DHV} recognition. By utilizing models initially trained on large, related datasets, these methods improve performance in \ac{DHV} identification, particularly in scenarios with limited \ac{DHV}-specific data. The key advantages of \ac{DTL}-based approaches include reduced training time, improved accuracy with smaller datasets, and flexibility in adapting to various \ac{DHV} identification tasks.
One of the earliest studies to investigate the application of \ac{DL} and \ac{DTL} in \ac{DHV} was conducted by \cite{li2016comparative}. In this work, the authors evaluated and compared four different \ac{CNN} architectures: Caffe AlexNet, Caffe Reference Net, VGG-16 Net, and GoogLeNet. \color{black}
These models are built using a large-scale dataset named ImageNet to extract features and identify \ac{DHV} patterns. The study shows that deep features are better than traditional methods at representing and recognizing \ac{DHV} patterns. This leads to a lot more accuracy on the NCUT database. The authors also stress the importance of fine-tuning deep models for this specific task, which leads to promising outcomes for biometric identification systems. Indeed, the fine-tuning process is crucial in \ac{DL} models for \ac{DHV} recognition as it enables the models to adjust to the unique attributes of \ac{DHV} patterns. This adaptation enhances the models' ability to differentiate individuals based on their vein images.
Another work rely on pre-trained model suggested in \cite{li2018dorsal}, introduces a technique to improve the accuracy of \ac{DHV} recognition by employing multi-bit planes and the SqueezNet network. By exploiting the intrinsic relationships among bit planes within images, the proposed algorithm achieves higher recognition rates when compared to conventional methods such as \ac{PCA} and \ac{LBP}. The incorporation of the SqueezeNet network additionally improves performance by enabling automatic feature extraction and keeping the model size efficient. The SqueezeNet network is essential in the \ac{DHV} recognition method for its efficient feature extraction and compact model size. In fact, the squeeze layer condenses the input channels, reducing computational complexity, while the expand layer helps capture detailed features. Therefore, this specific architectural design of the SqueezeNet network enables effective feature extraction while maintaining a lightweight design, making it suitable for small sample sizes and preventing overfitting in the recognition process. 
The work  \cite{li2022recognition} also employed a pretrained model for \ac{DHV} recognition,  it combines ResNet and \ac{HOG} features to improve \ac{DHV} recognition and overcome the limitations of small-scale \ac{DHV} datasets. The fusion method involves adding \ac{HOG} features and shallow semantic information (obtained by convolution) to the residual structure of ResNet for classification. This gets around the problem of \ac{DL} algorithms having small sample sizes. Additionally, the fusion approach integrates the \ac{PLBP} algorithm for comparison and feature fusion. Experimental results indicate that the proposed feature fusion method outperforms traditional methods like \ac{PLBP} and \ac{HOG} alone, emphasizing the effectiveness of the fusion approach.

In \cite{tian2022improved}, Tian and their colleagues suggested a YOLO Nano-Vein model for finding veins on the back of the hand. They wanted it to be more accurate and find veins faster than older models like YOLO Nano and YOLOv3. The authors optimize the proposed model (YOLO Nano-Vein) for embedded systems. It achieves its improvements through architectural modifications, including reducing of unnecessary modules, adjusting output scales, and the addition of an atrous \ac{SPP} module. These modifications improve the efficiency of vein detection while simplifying the network architecture. Comparative analysis with other models like faster recurrent \ac{CNN} (RCNN), YOLOv3, and YOLO Nano confirms the effectiveness of these modifications.

\renewcommand{\arraystretch}{1.3}
\begin{table}[]
\caption{Summary of proposed works in \ac{DL}-based \ac{DHV} recognition. In cases where multiple scenarios are examined, only the top-performing outcome is mentioned. }

\scriptsize
\label{tab:DorsalmVein}
\resizebox{\textwidth}{!}{ 
\begin{tabular}{m{0.5cm}m{0.5cm}m{1.5cm}m{2cm}m{2.5cm}m{2cm}m{2.5cm}}
\hline
Ref. & Year  & DL approach &DTL&Data augmentation & Dataset &  Results \\ \hline

\cite{nour2024improved} & 2023 & CNN+RBM & -- & Resizing, flipping,
rotating, cropping, padding & Badawi, DHVI & Acc= 99.76\%, 99.5\% \\

\cite{benaouda2021cnn} & 2022& CNN & -- & -- & Self-built dataset & Acc= 99\%\\

\cite{li2016comparative} & 2016 & CNN& Fine-tuning & -- & NCUT & Acc= 99.31\% \\

\cite{li2018hand} & 2018& CNN& -- & Double
PCA & NCUT & Acc= 99.61\%\\

\cite{li2021fusion} & 2021& CNN& -- & Double PCA & NCUT & Acc= 99.95\%\\

\cite{wan2017dorsal} & 2017& CNN& Fine-tuning& --& Self-built dataset & Acc= 99.7\%\\

\cite{tian2022improved} & 2022 & CNN (YOLO)& \ac{DA} & -- & Self-built dataset & Precision= 93.61\%\\

\cite{li2018dorsal} & 2018 & SqueezeNet & Feature fusion & -- & Self-built dataset & Acc= 99\%\\

\cite{zhang2023fast} & 2023& CNN (U-Net) & -- & horizontal and vertical flipping,   brightness random adjustments & Self-built dataset & Acc= 98.6\%\\

\cite{alashik2021human} & 2021 & DL-GAN & -- & GAN & Jilin Univ, 11K & Acc= 98.36\%, 96.43\%; EER= 2.47\%, 3.55\%  \\\

\cite{kuzu2022intra} & 2022 & Modified Densenet-161 & -- & -- & Bosphorus  & EER= 2.33\%\\

\cite{li2022recognition} & 2022 & ResNet+HOG & Feature fusion & Fusion of many datasets & SDUST, FYO, NCUT & Acc= 93.27\%, 95.36\%, 93.40\% \\

\cite{wang2018spatial} & 2018 & VGG-16 & Selective  feature fusion & -- & Self-built dataset & EER= 0.068\%\\
\hline
\end{tabular}
 }
\end{table}

\subsection{Multi-modal vein-based}

Biometrics has emerged as a vital form of identification as they carry rich personal identifying data. However, the shortcomings of single-modal biometric identification technology in terms of recognition accuracy is now becoming more obvious. To address this, multi-modal biometric recognition technology combines multiple complementary biometric modalities, therefore extending the range of biometric identification and enhancing accuracy and security. The growing interest in multimodal biometric systems originates from their ability to provide a more secure and precise authentication solution compared to unimodal systems. Even with this potential, performance improvement may sometimes be challenging since current biometric fusion techniques aren't robust enough to handle the correlations and redundancies of several variables at once. Several  \ac{DL}-based multi-modal vein recognition systems have been proposed in the literature. Table \ref{tab:MultiMod}  provides an overview of these schemes.

In order to address the mono-mode biometric challenges, the authors \cite{chaudhary2022pcanet} proposed a biometric system that combines both palmprint and \ac{DHV} modalities. The proposed method is primarily based on the use of \ac{CNN} \ac{DL} architecture to extract features, particularly the pre-trained function PCANeT for evaluating performance in a fusion scheme. The fusion techniques, both at the feature level and score level, have the objective of enhancing the accuracy of the biometric system.   In particular,  feature-level fusion is used to merge various features from different biometrics. On the other hand, score-level fusion enables the combination of the corresponding scores obtained from different biometric modalities. Moreover, the study \cite{bharath2023optimal} discusses a multi-modal biometric system using \ac{DL} principles. The research aims to enhance the precision of biometric identification by integrating several biometric modalities, including palm print, \ac{DHV}, \ac{WV}, and \ac{PV}. The research emphasizes the significance of multi-modal biometric technologies in improving security and authentication procedures. Furthermore, it highlights the need to use \ac{DL} techniques and optimization methodologies to improve the efficiency of biometric systems. The proposed approach consists of three primary stages: preprocessing, feature extraction, and recognition based on \ac{DL}. Preprocessing methods, such as median filtering and \ac{ROI} selection, are used to improve the quality of biometric images. Preprocessed images are used to extract features, and a \ac{DCNN} architecture is used to identify if the image belongs to an impostor or a legitimate user. An ensemble classifier is constructed of NN1 (trained using characteristics taken from the \ac{WV}, dorsal vein, and palm print) and NN2 (trained using \ac{PV}). The findings are combined using the score-level fusion technique. Next, the combined results are utilized as input for the \ac{DCNN}. The \ac{DCNN} is used for distinguishing between imposters and real individuals. An optimization model called butterfly combined Tunicate swarm optimization, named BTSA, is used to adjust the weights of \ac{DCNN} in order to enhance detection accuracy. Besides, the research \cite{daas2020multimodal} introduces a robust multimodal biometrics identification system that uses \ac{DL} to combine \ac{FV} and finger knuckle print data for enhanced security. The study presents two multimodal architectures, each including different fusion levels ({feature-level fusion} and score level fusion (weighted product, weighted sum, or Bayesian rule)), using \ac{CNN} architectures such as AlexNet, VGG16, and ResNet50. In the proposed system, the fully connected layer computes a fusion of new characteristics derived from \ac{FV} and finger knuckle print.  The ResNet50 network demonstrates superior accuracy in unimodal identification systems for \ac{FV} and finger knuckle prints. {Moreover, when comparing unimodal and multimodal techniques, the ResNet50-Softmax model combined with weighted sum fusion strategy achieves the highest recognition accuracy.} 
The study used open databases, specifically the FV SDULMA-HMT and the PolyU FKP databases, to test how well and accurately the suggested recognition algorithms worked.  Moving on, Deshmukh et al. \cite{deshmukh2022dcca} aim to illustrate how \ac{DL} and \ac{CCA} may improve biometric recognition accuracy. The authors present two frameworks for multimodal biometric systems: Deep Multiset \ac{CCA} and Deep \ac{CCA}. These frameworks use \ac{DL} approaches to acquire knowledge of the complex transformations of biometric modalities, resulting in the creation of highly correlated feature sets. These two frameworks outperform traditional fusion methods by enhancing feature fusion, minimizing irrelevant information, and enhancing recognition performance. Tests on the SDUMLA-HMT dataset show that the suggested frameworks get low \ac{EER} for both two-modal and three-modal biometrics (\ac{FV}, iris, and face).

The work in \cite{el2021efficient} introduces a \ac{MBCS} that uses \ac{DL} models to combine fingerprint, \ac{FV}, and iris biometrics, resulting in high quantitative results. The proposed work makes two primary contributions: developing a multi-exposure deep fusion module to create a combined representation of several biometric modalities; and implementing a deep dream module to construct a cancellable template from the fused biometric images. A comprehensive assessment showed favorable results. The authentication procedure of the \ac{MBCS} requires just a few sequential stages, making it very efficient for practical application. The research shows how successful the suggested \ac{MBCS} is by comparing it positively with \ac{SOTA} techniques. Later,  \ac{MBCS}  has been employed in \cite{abd2023efficient} as an efficient solution for cybersecurity-based applications. It is based on \ac{DL} and bio-hashing. The proposed system uses cascading-style transfer processes on different modalities, like fingerprint, \ac{FV}, and face images, and then applies a fusion method. This process has the advantage of facilitating the generation of cancellable templates. The system functions are the registration of input biometrics, the extraction of features, the fusion, the reconstruction, style transfer, and the authentication procedure. The overall work is validated on a dataset that includes fingerprint, \ac{FV}, and face images. The overall performance is assessed by visual and statistical analysis, which shows good results in terms of \ac{AUC} and encryption quality evaluation criteria such as \ac{SSIM}.

The study \cite{el2022multimodal} introduces a novel multimodal biometric authentication system that integrates \ac{ECG} and \ac{FV} modalities using deep fusion methods. The system consists of many steps. First, each biometric characteristic is pre-processed to ensure consistency. Next, a \ac{CNN} model extracts deep features for analysis. Finally, feature-level fusion and score-level fusion approaches are used to integrate the extracted features. The authentication procedure involves using a \ac{KNN} classifier on specifically chosen deep features. The system performs better than existing multimodal techniques and unimodal systems, as shown by extensive testing on many datasets. Huang et al. \cite{huang2023multimodal} introduces a novel end-to-end multimodal finger recognition system that addresses limitations in existing biometric fusion methods. It proposes a finger asymmetric backbone network to efficiently extract discriminative features from fingerprint and \ac{FV} images. A new attention-based encoder fusion network, so called AEF-Net, is also introduced to combine fingerprint  and \ac{FV} features and get rid of unnecessary ones, with a focus on how the different types of features affect each other. The proposed method is evaluated on three multimodal finger databases, demonstrating improved recognition performance compared to \ac{SOTA} methods. By incorporating attention mechanisms and \ac{AE}-based training, the system aims to enhance feature fusion effectiveness and achieve better recognition accuracy and speed. The paper also highlights the importance of considering feature fusion effectiveness in biometric recognition methods. Likewise, The study \cite{tyagi2022multimodal} presents an accurate multimodal biometric identification and recognition system that integrates the face and \ac{FV} modalities. Deep \acp{CNN}, inspired by AlexNet, are used to extract features from both facial and \ac{FV} modalities. These features are then fused at the score level. The combination of facial and \ac{FV} modalities leads to improved performance measures, including higher accuracy and lower equal mistake rates. This technique successfully addressed the issue of interclass dependencies, enabling accurate identification outcomes for input samples that are cross-paired. The experimental findings demonstrate a significant increase in accuracy for both identification and recognition tasks in comparison with existing systems.

The authors, in \cite{toygar2020fyo}, present a multimodal vein database, known as FYO, which includes biometric data from the \ac{PV}, dorsal vein, and \ac{WV}. The article emphasizes the usefulness of multimodal systems in attack defense, as well as the strict security requirements of vein biometrics. The objective is to improve research on multimodal authentication systems. Hand-crafted feature extractors such as \ac{BSIF}, Gabor filter, and \ac{HOG} are used, in addition to a suggested \ac{CNN} architecture for vein detection. The \ac{CNN} architecture suggested, which incorporates decision-level integration of biometric features, outperformed hand-crafted techniques and showed greater performance in vein identification. The research conducts a comparison between FYO and other databases, demonstrating the effectiveness of integrating different biometric features for improved security. Wange et al. \cite{wang2017bimodal} address gender and identity detection using hand vein data. Indeed, optimizing a task-specific \ac{DNN} model from VGG-face offers numerous benefits. Utilizing a face database to enhance hand vein detection entails using acquired information, identifying relevant features, and fine-tuning a \ac{DL} model to improve its generalization ability. This approach aims to use the advantages of face databases and \ac{DL} methods to improve the precision and effectiveness of gender and identity identification by employing hand vein data. Initially, the model uses \ac{DTL} to benefit from the knowledge acquired during a large-scale face recognition task, which may improve its performance for the current task. In addition, \ac{DNN} models are very proficient in extracting specific information from hand vein images, which is essential for achieving correct identification. Furthermore, fine-tuning enhances the model's ability to optimize, particularly for gender and identity identification using hand vein data, resulting in improved performance in comparison to conventional models. The experiments include two main databases, with an emphasis on the face and hand vein. The PolyU \ac{NIR}-face is one. The lab-made hand-dorsa vein database is another. Similarly, Yang et al. \cite{yang2018novel} introduce an approach for identifying multimodal biometrics (face and \ac{FV} images) using stacked \acp{ELM} and \ac{CCA} techniques. At first, it employs \acp{ELM} to learn the hidden-layer representation. Then, the \ac{CCA} transforms the representation into a feature space, which then reproduces the multimodal image features. Next, the obtained features serve as input for a supervised classifier. The obtained results indicate that the proposed approach outperforms some existing approaches when evaluated on hybrid datasets that merge face and \ac{FV} images. They also demonstrate the effectiveness of multimodal biometric fusion as an alternative way to enhance biometric performance. Moving on, the paper \cite{zhong2019hand} introduces a multi-biometrics algorithm that combines palmprint and \ac{DHV}  images. \Ac{DHN} and biometric graph matching are investigated for recognition. In addition, different fusion strategies are used to combine palmprint and \ac{DHV} images. \ac{DHN} is used to encode images into 128-bit codes, which are then used for similarity comparison purposes. Thus, the authors show that the fusion of palmprint and \ac{DHV} recognition at different levels (pixel, feature, score, decision) can significantly improve accuracy. The proposed scheme also enhances the score level for multi-modal fusion by incorporating \ac{SVM} for authentication. Consequently, the proposed method demonstrates superior performance in fusion recognition compared to unimodal biometrics.

The suggested work in \cite{alay2020deep} introduces a multimodal-based \ac{DL} framework for biometrics identification. Its main objective is to combine three different features from the iris, facial, and finger modalities in order to improve overall identity recognition accuracy. Two fusion methods are used in this paper to combine the data from the different biometric traits,  feature-level and score-level fusions. Feature-level fusion involves combining the extracted features from each biometric trait before classification. The score-level fusion involves a combination of similarity scores obtained from individual trait classifiers throughout the recognition process. These strategies seek to use the advantages of each biometric feature to enhance the final performance. This work evaluates the performance of both unimodal and multimodal models on the SDUMLA-HMT dataset in terms of feature extraction and classification. The obtained results show that a multimodal strategy outperforms an unimodal one, resulting in higher accuracy. Additionally, the authors suggest exploring other recognition features and fusion methods as further research directions to improve the performance. In \cite{jiang2022finger2}, the authors developed a dual-branch-Net method to carefully identify both the \ac{IKP} and \ac{FV}. To address the challenges of feature representation and multimodal fusion, they suggested a framework based on \ac{CNN}, \ac{DTL}, and the triplet loss function.  In order to address the problem of a limited training set, the augmentation technique is applied to increase the dataset size. Experiments validate the effectiveness of the proposed fusion technique compared to other approaches.\\

\begin{table}[h!]
\caption{Summary of proposed works in \ac{DL}-based multi-modal hand vein recognition. In cases where multiple scenarios are examined, only the top-performing outcome is mentioned. }

\scriptsize
\label{tab:MultiMod}
\resizebox{\textwidth}{!}{ 
\begin{tabular}{m{0.5cm}m{0.5cm}m{0.3cm}m{0.3cm}m{0.3cm}m{0.3cm}m{0.5cm}m{1.5cm}m{2cm}m{2.5cm}m{2cm}m{2.5cm}}
\hline
Ref. & Year& \multicolumn{5}{c}{Fused modalities}  & DL  &DTL &Data  & Dataset &  Results \\ \cline{3-7}
& & FV & PV& DHV& WV & Other & approach &  & augmentation & & \\\hline

\cite{toygar2020fyo} & 2020 & \xmark & \cmark & \cmark& \cmark& \xmark &CNN & -- & -- &  FYO & Acc= 100\% \\

\cite{chaudhary2022pcanet} & 2022  & \xmark &\xmark & \cmark & \xmark &\cmark &CNN+PCANet & -- & -- & IITD palmprint, Bosphorus  & Acc= 98.86\%, EER= 0.93\% \\


 \cite{daas2020multimodal} & 2020 & \cmark& \xmark & \xmark & \xmark & \cmark&AlexNet, VGG16,  ResNet50 & Fine-tuning & Translation and cropping & FV, FKP  & Acc= 98.84\%, 99.89\%; EER= 0.0142\%, 0.005\% \\
 
\cite{deshmukh2022dcca}  & 2022 & \cmark & \xmark& \xmark& \xmark& \cmark & DL & -- & -- & SDUMLA-HMT  & Acc= 99.98\%; EER= 0.7177\% \\

\cite{el2021efficient} & 2021 & \cmark & \xmark& \xmark& \xmark& \cmark & CNN & -- & -- & Created, CASIA & NPCR= 99.158\%; PSNR= 24.523 dB; SSIM= 0.079.\\

\cite{el2022multimodal} & 2021 & \cmark& \xmark& \xmark& \xmark& \cmark & CNN & -- & -- & VeinPolyU, MWMHIT
and ECG-ID & EER= 0.12\%\\

\cite{huang2023multimodal} & 2022 &\cmark& \xmark& \xmark& \xmark& \cmark & CNN &--& Rotation and translation & HDPR-310, FVC-HKP, CAS-FVU & f1= 99\%, 99.3\%, 99.4\%\\

\cite{tyagi2022multimodal} & 2022 & \cmark& \xmark& \xmark& \xmark& \cmark & CNN & -- & Translation, rotation and illumination & FV-USM, SDUMLA & Acc= 99.85\%, 94.87\%\\
 
\cite{wang2017bimodal} & 2018  & \xmark&\xmark & \cmark & \xmark& \cmark & DNN+LDM & Task-driven fine-tuning& Each 
sample is increased to 100 & PolyU NIR-face, created  HDV  & Acc= 91.6\% \\
 
\cite{alay2020deep} & 2020 & \cmark& \xmark& \xmark& \xmark& \cmark& VGG-16 & Fine-tuning & Rotation, shearing, zooming, width shifting, and height shifting & SDUMLA-HMT & Acc= 99.39\% \\ 
 
\cite{jiang2022finger2} & 2022 &\cmark& \xmark& \xmark& \xmark& \cmark &  CNN & \ac{DA} &  Translation, clipping, 
and conversion & PolyU-DB & EER= 0.422\% \\
 
\cite{haouam2021s} & 2021 & \xmark& \cmark& \xmark& \xmark& \cmark & DCTNet & -- & -- & PolyU & GAR= 100\%\\

\hline
\end{tabular}
}
\begin{flushleft}
\end{flushleft}
\end{table}

\section{Research challenges and future directions}
\label{sec6}

The field of vein biometrics, particularly with the integration of \ac{DL} techniques, holds immense potential for enhancing security and authentication methods. However, several challenges are not sufficiently considered, which often appear in real world application scenarios.  Meanwhile, future research directions should focus on developing robust solutions to these challenges while exploring new methodologies and technologies to further advance vein biometrics.

\subsection{Research challenges}
Biometric hand vein recognition, like any technology, it faces several challenges. Here are some key challenges associated with \ac{DL} and  \ac{DTL}-based biometric hand vein recognition:

\subsubsection{Data acquisition and characteristics of existing datasets}
Acquiring high-quality vein images presents several challenges that significantly impact the performance of vein biometric systems. Variations in lighting conditions can lead to inconsistent image quality, making it difficult to extract reliable vein patterns. For instance, inadequate or excessive lighting can obscure vein details or introduce shadows that distort the vein structure. Additionally, differences in skin tone among individuals can affect the visibility and contrast of the veins, complicating the image acquisition process. The presence of hair, scars, or other skin imperfections can further obscure vein patterns, leading to incomplete or noisy images that are challenging for feature extraction and subsequent recognition tasks. 

Current image preprocessing methods can improve the recognition performance of vein biometric systems, but selecting the appropriate model to enhance image quality remains challenging. \ac{DL} techniques have the potential to significantly enhance image quality by learning complex patterns and corrections that traditional methods might miss. Despite this potential, \ac{DL} methods have been cautiously applied in this context, primarily focusing on \ac{FV} \cite{bros2021vein, choi2020modified, yang2020finger, jiang2022finger, hong2024deep, gao2023drl, lei2019finger, du2021fvsr}. More research is needed to fully harness their capabilities for all types of hand veins. Introducing more powerful \ac{DL}-based methods in image preprocessing could address these inconsistencies and artifacts, improving the performance and reliability of vein biometric systems.


The field of hand vein biometrics faces a significant challenge due to the lack of large, publicly accessible datasets. Existing vein biometric datasets are often limited in size and suffer from class imbalance, which constrains  the development and evaluation of robust \ac{DL} models. The efficacy of \ac{DL}-based methods, which have shown promising performance in vein biometrics, is contingent on the availability of extensive training data. Moreover, most current \ac{FV} databases are collected using a single capture device. This uniformity fails to account for the variability introduced by different capture devices, leading to evaluations of image preprocessing and feature extraction methods that do not accurately reflect real-world conditions.

\subsubsection{Variability in vein patterns and environment}
Variability in vein patterns presents a significant challenge for vein biometric systems. Factors such as age, health conditions, and physical activities can cause substantial intra-class variations, making it difficult to consistently recognize an individual's vein patterns. As people age, changes in blood vessel structure and skin elasticity can alter the appearance of veins. Health conditions, such as vascular diseases or diabetes, can also impact vein patterns by causing blockages or changes in blood flow. Physical activities, particularly those involving extensive use of the hands or arms, can temporarily or permanently change the visibility and shape of veins. These variations complicate the feature extraction and matching processes, necessitating the development of more robust algorithms that can account for and adapt to these changes to ensure accurate and reliable vein recognition.

Current hand vein biometric systems are often designed under controlled conditions, neglecting the wide-ranging variations encountered in practical use. Factors like temperature changes, skin stretching, and fatigue can degrade accuracy. Furthermore, for \ac{DHV} recognition, challenges such as tattoos on the dorsal hand, characteristics of elderly hands, and hands with long hair remain understudied despite their impact on real-world performance \cite{jia2021survey}. Applying \ac{DTL} to hand vein biometrics can be challenging due to the lack of pre-trained models specific to this domain, necessitating the adaptation of models trained on different tasks.

\subsubsection{Multi modal biometric hand vein recognition system} 
Multi-modal biometric vein recognition systems face several challenges that stem from integrating multiple sources of biometric data . Firslty, integrating data from different modalities, such as vein patterns, finger geometry, and palm prints, requires careful consideration. Determining the optimal fusion strategy (score-level, feature-level, etc.) for each modality combination demands thorough research and evaluation.  Also, variations in image quality across modalities necessitate robust pre-processing and noise handling techniques to ensure reliable fusion. Secondly, the limited availability of datasets containing multiple modalities for training and testing poses a significant hurdle for algorithm development and validation. Thirdly, The computational complexity of multimodal systems can be a bottleneck, demanding careful optimization of algorithms and hardware resources to achieve efficient processing \cite{noh2020finger, bhilare2018single}. Ensuring that the system can process data quickly without compromising accuracy is essential for user acceptance and practical deployment. Finally, multimodal systems may require users to perform multiple actions, potentially impacting user experience and acceptance. Addressing privacy concerns and ensuring user-friendly interactions are crucial for widespread adoption \cite{bhilare2018single}. Simplifying the user interaction process while maintaining high security and accuracy is a key challenge for developers.

\subsubsection{Biometric template protection and aging consideraion}
Several traditional \ac{BTP} schemes have been developed to enhance privacy. However, these schemes can negatively impact recognition performance. \ac{DL} approaches have been proposed to mitigate this issue, but their potential has not been fully explored. To date, only a few studies have focused on this area, primarily relying on \ac{FV} data \cite{yang2019securing, liu2018finger, shahreza2021deep}. More research is needed to extend these \ac{DL}-based protection methods to other types of hand veins.

Aging significantly affects vein patterns, posing a challenge for biometric vein recognition systems. Changes in skin elasticity, collagen, and blood vessels over time can alter vein patterns, impacting the accuracy of long-term recognition technologies. Capturing these changes accurately is essential for maintaining system performance. However, current datasets, often limited to relatively short intervals (up to 106 days as shown in Table \ref{tab:VeinDatasets}), fail to fully capture the gradual aging process.  This hinders a comprehensive study of aging effects on hand vein biometric systems. \ac{DL} offers a promising solution by extracting features from vein images, potentially enabling robust and accurate recognition despite aging-related variations.

\subsubsection{PAD systems}
Despite the promising results demonstrated by \ac{DL}-based \ac{PAD} approaches in the limited studies on \ac{FV} biometrics, numerous challenges persist. The primary challenge is the limited availability of evaluation data that includes both genuine and forged samples. Only three publicly accessible \ac{PAD} datasets for \ac{FV} exist: SCUT-SFVD \cite{qiu2018finger}, ISPR \cite{nguyen2013fake}, and VERA Finger Vein \cite{tome20151st, tome2014vulnerability}. Additionally, the interaction between hand vein recognition and \ac{PAD} within the overall system requires further investigation. Although most \ac{PAD} systems can independently verify the authenticity of a hand vein sample, understanding their impact on the entire recognition system is crucial, as \ac{PAD} systems can increase the FNMR of hand vein recognition systems due to potential failures \cite{shaheed2024deep}. 


\subsubsection{Multi view vein recognition } 
Most existing databases and algorithms predominantly focus on single-view vein recognition. This method projects a 3D network topology onto a 2D plane, resulting in inevitable 3D feature loss and topological ambiguity in the images. Additionally, single-view methods are sensitive to finger rotation, translation, and hand position in practical applications, which can lead to performance degradation. Currently, there are few dedicated studies and public databases on multi-view vein recognition \cite{qin2022local, zhao2024vpcformer}. More research and development of multi-view databases and algorithms are needed to overcome these limitations and enhance recognition accuracy and robustness.

\subsubsection{Accelerating \ac{DL} techniques for hand vein biometrics} 
\ac{DL} models have shown impressive performance in hand vein recognition, but their computational demands, especially during training and for complex architectures, pose a significant challenge for real-time performance and deployment on resource-constrained devices. Achieving efficient and fast hand vein recognition requires careful consideration of hardware limitations and algorithmic optimizations.

Current \ac{DL} models often rely heavily on powerful \acp{GPU}, both for training and achieving acceptable inference speeds. While specialized hardware like \ac{FPGA} offer potential advantages in terms of efficiency, their deployment for hand vein biometrics necessitates further research and development \cite{chang2023design, janaki2024fpga}. Furthermore, deploying complex \ac{DL} models on devices with limited memory, processing power, and battery life, such as mobile devices or embedded systems, presents a major hurdle \cite{roth2020resource}.

Several model optimization techniques can help mitigate these challenges. Model compression techniques like pruning \cite{han2015learning}, quantization \cite{jacob2018quantization}, and low-rank factorization \cite{denton2014exploiting} can significantly reduce model parameters and operations, leading to smaller models and faster inference. Knowledge distillation, where a smaller "student" model learns to mimic the behavior of a larger pre-trained "teacher" model, can also reduce computational complexity while maintaining performance \cite{hinton2015distilling}. Additionally, employing lightweight \ac{CNN} architectures specifically designed for efficient computation, can enhance performance on resource-constrained platforms \cite{fang2018novel, XIE2019148, tang2019finger,  shen2021finger, zheng2020new, zhao2020finger, zhang2022convolutional, chai2023vascular, zhang2023convolutional}.

Parallel and distributed computing offer further avenues for acceleration. Utilizing multi-node and multi-core architectures, \ac{GPU} clusters, or cloud computing platforms can significantly speed up both training and inference \cite{dean2012large}. However, effectively leveraging these technologies for hand vein biometrics necessitates addressing challenges related to data partitioning, communication overhead, and resource management \cite{li2014scaling}. Researchers are encouraged to investigate the capabilities of real-time \ac{DL}-based hand vein recognition by implementing their techniques on embedded platforms like \ac{FPGA}, and by assessing real-time metrics such as inference time, latency, resource utilization, and algorithmic complexity. These assessments offer critical insights for users and developers, aiding in the selection of the optimal \ac{DL} for particular hardware configurations and specific types of hand veins.

\color{black}

\subsection{Perspectives}
Considering the challenges faced by hand vein biometrics, from our humble perspective, the following lines of investigation can be considered by the
research community:

\subsubsection{Creating synthetic large datasets} 
Collecting large-scale real-world training data for hand vein recognition has proven challenging due to noise and irregular variations during acquisition. One promising future direction is the creation of synthetic large datasets for hand vein biometrics. Developing realistic and diverse synthetic datasets can address the limitations of existing datasets, which are often small and lack the variability needed for robust training and testing. Advanced generative models, such as \ac{GAN}, can generate synthetic vein patterns that mimic real-world variations in skin tone, age, health conditions, and environmental factors. These synthetic datasets can provide a valuable resource for training \ac{DL} models, enabling researchers to explore new algorithms and approaches without the constraints of limited real-world data. Additionally, synthetic datasets can facilitate the study of the aging problem by simulating longitudinal changes in vein patterns over extended periods, thus offering a comprehensive platform for testing the robustness and adaptability of biometric recognition systems.
The review paper of Salazar-Jurado et al. \cite{salazar2023towards} discusses the challenges, insights, and future perspectives in generating synthetic \ac{PV} images, highlighting the potential benefits and ongoing advancements in this field.

\subsubsection{3D hand vein recognition} 
3D hand vein biometric systems offer enhanced accuracy and security by capturing the depth and spatial characteristics of vein patterns, providing more detailed biometric data than traditional 2D imaging. Current vein verification systems typically use a monocular camera to acquire a single-view 2D image, which limits vein pattern information and causes inconsistencies due to positional variations. These limitations adversely affect system performance, particularly in contact-free modes, where pitch and roll movements create significant variability in the samples. This concern remains a challenge despite some efforts to address it in recent years \cite{qin2022local, zhao2024vpcformer, de20173d, jia20212d}.  Future research should focus on designing software and hardware platforms to capture a comprehensive view of vein patterns, developing methods for 3D reconstruction to construct a complete 3D hand vein image or point cloud \cite{sohail2024advancing}, advancing 3D feature extraction and matching strategies, and creating diverse 3D datasets.

\subsubsection{Gender and aging recognition based hand vein} 
Exploring gender recognition using hand vein patterns presents a novel and promising direction in biometric research. Hand vein patterns, influenced by physiological and anatomical differences between genders, can potentially serve as distinctive biometric traits for gender classification. Advanced \ac{ML} techniques, particularly \ac{DL} models, can be employed to extract and analyze features from hand vein images, enabling accurate gender prediction. This approach can enhance the robustness and security of biometric systems by adding an additional layer of verification. This problem has been preliminarily studied \cite{damak2017age, wang2018gender, sellami2019palm, hernandez2022cnn, damak2023pyramid, kuzu2023gender}. However, significant challenges remain, such as the need for large, diverse datasets that capture gender-specific variations in hand vein patterns. Addressing these challenges requires developing sophisticated algorithms capable of handling intra-class variability and ensuring high recognition accuracy. Thus, in the future, more effort should be invested in promoting the research of this problem, focusing on creating comprehensive datasets, refining feature extraction methods, and optimizing classification algorithms to fully realize the potential of gender recognition based on hand vein patterns. \ac{DTL} can be used to transfer knowledge learned from a source domain (e.g., a dataset of younger individuals) to a target domain (e.g., a dataset of older individuals). This helps in extracting relevant features from vein patterns that are less affected by aging. Besides, fine-tuning pre-trained models on datasets that include both younger and older individuals can help adapt the model to recognize vein patterns across different age groups more effectively.

\subsubsection{\ac{FL} in hand vein recognition} 
\ac{FL} offers a promising approach to enhancing hand vein recognition systems while preserving user privacy. Unlike traditional methods that collect all data centrally, \ac{FL} trains models locally on individual devices, sharing only model updates. This decentralized method improves data security and privacy by keeping sensitive vein patterns on users' devices. Additionally, it can leverage the diversity of vein patterns from various devices, enhancing model robustness.

\ac{FL} has been preliminarily studied in recent hand vein recognition research \cite{lian2023fedfv, mu2024pafedfv, mu2024federated}. However, several challenges remain, such as ensuring consistent model updates across devices with different resources and network conditions, and developing robust aggregation methods for diverse, \ac{non-IID} data sources. Future research should focus on optimizing \ac{FL} algorithms, enhancing communication efficiency, and managing the variability in hand vein data across devices. Addressing these challenges will help create more secure and effective biometric recognition technologies

\subsubsection{DRL and GNN in hand vein recognition}

The perspective of using \ac{DRL} in hand vein recognition offers intriguing possibilities. \ac{DRL} can optimize biometric systems by enabling adaptive learning based on feedback from the environment \cite{gao2023drl}. For instance, \ac{DRL} algorithms could dynamically adjust feature extraction parameters or model hyperparameters to improve recognition accuracy over time. \ac{DRL} could also enhance system robustness by optimizing decision-making processes in challenging conditions such as varying illumination or hand posture. However, applying \ac{DRL} to biometric recognition involves challenges like defining appropriate reward functions and ensuring data privacy and security. Future research may focus on developing \ac{DRL} frameworks that integrate seamlessly with hand vein recognition systems, optimizing performance while addressing ethical and technical considerations.

Employing \acp{GNN} for hand vein modularization offers promising opportunities in biometric classification. \acp{GNN} can model intricate relationships between vein patterns as graphs, enhancing feature extraction and classification accuracy. \ac{DRL} can further optimize \acp{GNN} by guiding node and edge updates to refine graph representations based on classification feedback. \ac{DRL} enables \acp{GNN} to dynamically adjust to varying vein patterns, improving adaptability and robustness. This synergy allows for more effective integration of complex spatial and temporal dependencies in hand vein biometrics, paving the way for advanced biometric authentication systems with enhanced accuracy and resilience to environmental variations.

\subsubsection{DL-Based integrity assurance for hand vein images}

\textit{Hand vein} image datasets benefit from \ac{DL}-based steganalysis to detect alterations, ensuring data integrity \cite{kheddar2024deep}. \ac{DL} models can effectively analyze subtle changes in vein patterns caused by steganography or watermarking techniques \cite{rebahi2023image}, verifying authenticity and preventing unauthorized modifications. Watermarking adds embedded data to images, ensuring traceability, while steganography hides imperceptible data within dataset samples without detection \cite{kheddar2022speech}. \ac{DL} enhances these techniques by robustly identifying alterations in hand vein patterns, crucial for maintaining the reliability and security of biometric data in applications requiring stringent integrity checks, such as healthcare and secure access systems. Further research is necessary to enhance the capabilities of \ac{DL} in accurately detecting and mitigating sophisticated forms of image tampering in hand vein biometrics.

\subsubsection{IoT authentication and biometrics} 

Although the co-hash code proposed for \ac{PV} is poised for use as an authenticator in \ac{IoT}  services due to its stable, compact, and straightforward nature, it can also serve as an intermediate representation for \ac{BTP} schemes, as suggested in references \cite{ebrahimi2021lightweight,yin2021iot}. This opens avenues for privacy-enhanced \ac{IoT}  authentication systems with rigorous requirements such as non-invertibility, unlinkability, and revocability. Another significant downstream \ac{IoT}  application of PalmCohashNet is in biometric cryptosystems, where biometrics and cryptography are integrated for purposes like \ac{BTP}, safeguarding secret keys, and key generation, as outlined in reference.

It is important to note that PalmCohashNet is also applicable in large-scale biometric identification, indexing, and retrieval tasks commonly encountered in \ac{IoT}-related applications. These applications necessitate that entities (biometric templates) are identified, indexed, or retrieved in a compact and simple format, such as binary representation. Our experiments demonstrate that PalmCohashNet achieves \ac{SOTA} top-1 identification accuracy on various benchmark datasets, indicating its significant potential for such applications. Although PalmCohashNet utilizes a lightweight \ac{CNN}, specifically MobileNetV2, as its backbone, a more efficient lightweight \ac{CNN} backbone, like MPSNet \cite{horng2021recognizing}, could be integrated to meet the requirements of \ac{IoT}  devices.

\subsubsection{Transformers and LLM for hand vein biometric} 

Looking ahead, Transformers and \acp{LLM} offer promising avenues for advancing hand vein biometric recognition beyond traditional attention mechanisms. Future research might explore novel transformer architectures tailored to capture intricate vein patterns effectively. This could involve designing Transformer variants that integrate convolutional or recurrent layers optimized for processing spatial and temporal features in 3D vein data.
\acp{LLM}, known for their ability to learn complex patterns from vast datasets, could enhance feature representation and extraction in hand vein recognition. By pre-training \acp{LLM} on diverse biometric data, they could learn discriminative features across various demographics and conditions, improving robustness and generalization. Additionally, fine-tuning \acp{LLM} on specific vein recognition tasks could enhance their ability to extract relevant features from vein images, potentially outperforming traditional feature extraction methods. These advancements hold potential for developing more accurate, adaptable, and secure hand vein biometric systems, applicable in healthcare, security, and other fields requiring precise and reliable identification methods. Specificaly,  \acp{LLM} such as \ac{GPT} , \ac{BERT}, and others \cite{kheddar2024transformers}, can be fine-tuned on hand vein datasets to extract discriminative features. Feasibility lies in training a specialized BiometricChat \ac{LLM}, tailored for vein recognition, by pre-training on a diverse set of vein images to capture complex patterns. This would require careful dataset curation, algorithm optimization for biometric tasks, and addressing privacy concerns in biometric data usage.

\section{Conclusion} \label{sec7}

In conclusion, hand vein biometrics, including \ac{FV}, \ac{PV}, \ac{DHV}, and multi-modal-based  recognition, have emerged as a highly secure, accurate, and non-intrusive method for identity verification. The distinctiveness and complexity of vein patterns within the hand provide a robust basis for biometric identification, offering significant advantages over other modalities. Hand vein recognition is contactless, which enhances user convenience and hygiene, and the internal location of veins makes them less susceptible to damage or tampering, thereby improving the overall security and reliability of biometric systems.

This comprehensive review has delved into the latest advancements in \ac{DL} techniques applied to hand vein biometrics. It has covered essential fundamentals of hand vein biometrics and summarized publicly available datasets, providing a valuable resource for researchers in the field. Additionally, the review has discussed \ac{SOTA} evaluation metrics for finger, palm, and dorsal vein recognition, highlighting the best performance achieved and identifying optimal methods and effective transfer learning approaches.

The review also addressed several research challenges, such as the need for more extensive and diverse datasets, the development of more sophisticated preprocessing techniques, and the refinement of \ac{DL} models to improve accuracy and efficiency. By outlining future directions and perspectives, this review encourages researchers to build upon existing methods and propose innovative techniques that further enhance the capabilities of hand vein biometrics.

In light of these findings, it is clear that hand vein biometrics hold significant promise for the future of secure and efficient identity verification. The ongoing advancements in \ac{DL} and related technologies are expected to drive further improvements in the performance and reliability of these systems. As researchers continue to address current challenges and explore new possibilities, hand vein biometrics are poised to play a critical role in the evolution of biometric authentication, offering a highly effective solution for a wide range of applications.

\printcredits

\section*{Declaration of competing interest}
The authors declare that they have no known competing financial interests or personal relationships that could have appeared to influence the work reported in this paper.

\section*{Data availability}
No data was used for the research described in the article.


\bibliographystyle{elsarticle-num}
\bibliography{references}

\end{document}

%% file: acro_list.tex
\begin{acronym}[AAAAA] 
\acro{AHE}{adaptive HE}
\acro{Acc}{accuracy}
\acro{ACER}{average classification error rate}
\acro{AE}{autoencoder}
\acro{APCER}{attack presentation classification error rate}
\acro{AUC}{area under curve}
\acro{BERT}{bidirectional encoder representations from Transformers}
\acro{BIM}{basic iterative method}
\acro{BPCER}{bona fide presentation classification error rate}
\acro{BSIF}{binarized statistical image features}
\acro{BTP}{biometric template protection}
\acro{CAE}{convolutional autoencoder}
\acro{CBAM}{convolutional block Attention module}
\acro{CCA}{canonical correlation analysis}
\acro{CCD}{charge-coupled device}
\acro{CLHE}{contrast-limited HE}
\acro{CLAHE}{contrast-limited adaptive histogram equalization}
\acro{CMH}{cross-modality hashing}
\acro{CNN}{convolutional neural network}
\acro{CIR}{correct identification rate}
\acro{DA}{domain adaptation}
\acro{DBN}{deep belief network}
\acro{DCNN}{deep convolution neural network}
\acro{DHN}{deep hashing network}
\acro{DHV}{dorsal hand vein}

\acro{DL}{deep learning}
\acro{DNN}{deep neural network}
\acro{DRL}{deep reinforcement learning}
\acro{DTL}{deep transfer learning}
\acro{DWT}{discrete wavelet transform}
\acro{ECG}{electrocardiogram}
\acro{EER}{equal error rate}
\acro{ELM}{extreme learning machine}
\acro{F1}{F1-score}
\acro{FAR}{false acceptance rate}
\acro{FGSM}{fast gradient sign method}
\acro{FPGA}{field programmable gate array}
\acro{FDR}{false discovery rate-correction}
\acro{FID}{Frechet inception distance}
\acro{FMR}{false match rate}
\acro{FNMR}{false non-match rate }
\acro{FL}{Federated learning}
\acro{FRR}{false rejection rate}
\acro{FV}{Finger vein}
\acro{FVI}{finger vein image}
\acro{FFN}{feature fusion network}
\acro{GAN}{generative adversarial network}
\acro{GCNLE}{graph convolutional network-based label enhancement}
\acro{GAR}{genuine acceptance rate}
\acro{GNN}{graph neural networks}
\acro{GPT}{generative pre-trained Transformer}
\acro{GRU}{gated recurrent unit}
\acro{GPU}{graphics processing unit}
\acro{HE}{histogram equalization}
\acro{HOG}{histograms of oriented gradients}
\acro{HSJA}{hop-skip-jump-attack}
\acro{HTER}{half total error rate}
\acro{IoT}{internet-of-things}
\acro{IKP}{inner knuckle print}
\acro{IRI}{infrared image}
\acro{KNN}{k-nearest neighbors}
\acro{LBP}{local binary patterns}
\acro{LDA}{linear discriminant analysis}
\acro{LED}{light emitting diode}
\acro{LLM}{large language model}
\acro{LPQ}{local phase quantization}
\acro{LSTM}{long short-term memory}
\acro{LTGAN}{local transformer-based GAN}
\acro{ML}{machine learning}
\acro{MLP}{multi-layer perceptron}
\acro{ML-ELM}{multilayer extreme learning machine}
\acro{MBCS}{multi-biometric cancellable system}
\acro{MTL}{multitask learning}
\acro{NIR}{near-infrared}
\acro{non-IID}{non-independent and identically distributed}
\acro{NPCER}{normal presentation classification error rate}
\acro{PAD}{presentation attack detection}
\acro{PCNN}{pulse coupled neural network}
\acro{PCA}{principal component analysis}
\acro{PLBP}{partition local binary patterns}
\acro{Pre}{precision}
\acro{PSNR}{peak signal-to-noise ratio}
\acro{PV}{palm vein}
\acro{Rec}{recall}
\acro{RAB}{residual attention block}
\acro{RBM}{restricted Boltzmann machines}
\acro{RNN}{recurrent neural network}
\acro{ROI}{region of interest}
\acro{SIFT}{scale-invariant feature transform}
\acro{SPP}{spatial pyramid pooling}
\acro{SVM}{support vector machine}
\acro{SSIM}{structural similarity index measure}
\acro{SOTA}{state-of-the-art}
\acro{TProt}{template protection}
\acro{T2T}{tokens-to-token}
\acro{TL}{transfer learning}
\acro{ViT}{vision Transformer}
\acro{WD}{Wasserstein distance}
\acro{WV}{wrist vein}
\end{acronym}